\documentclass{book}
\usepackage{amsmath}
\usepackage{amsfonts}
\usepackage{amssymb}
\usepackage{graphicx}
\begin{document}

\frontmatter

\begin{titlepage}
\begin{figure}[t]
\begin{center}
\includegraphics*[width=0.2\textwidth]{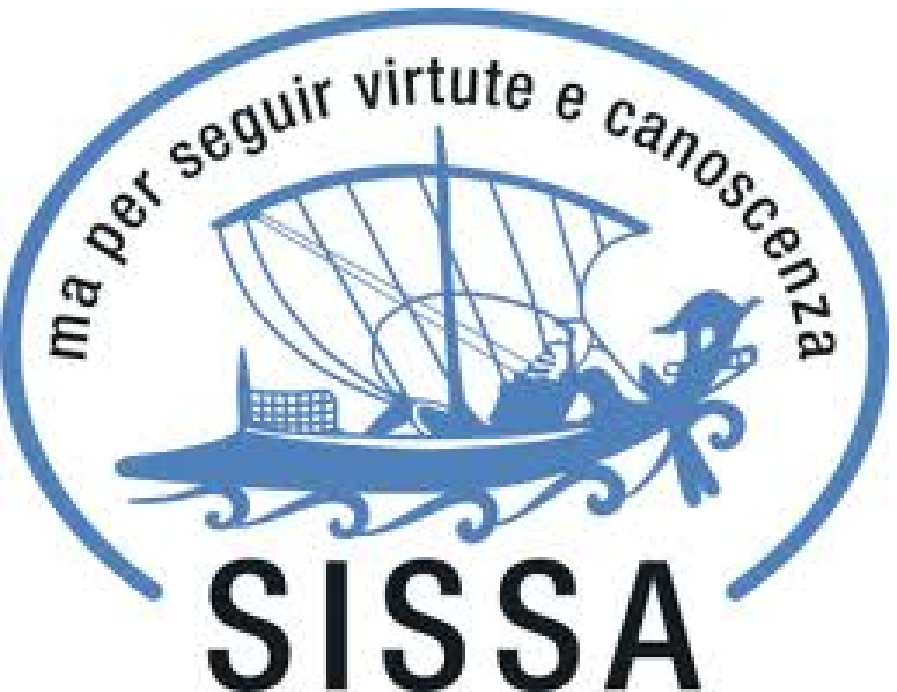}
\end{center}
\end{figure}

\begin{center}
SCUOLA INTERNAZIONALE SUPERIORE DI STUDI AVANZATI\\
INTERNATIONAL SCHOOL FOR ADVANCED STUDIES\\
{\bf Elementary Particle Theory Sector\\
Statistical Physics Curriculum\\}
\vspace{1.5cm}
{\Huge\bf Phase transitions on heterogeneous\\ random graphs: \\some case studies }\\
\vspace{2cm}{\large
Thesis submitted for the degree of\\
\emph{Doctor Philosophi\ae}}
\end{center}
\vspace{2cm}
\begin{tabular*}{\textwidth}{@{\extracolsep{\fill}}lcr}
\large ADVISOR: &  & CANDIDATE:\\
{\large\textbf{Prof. Matteo Marsili}} &  & {\large\textbf{Daniele De Martino}} \\
\end{tabular*}
\vspace{2cm}
\begin{center}
{\large 30$^{\rm th}$  September 2010}
\end{center}
\end{titlepage}




\tableofcontents

\chapter{Introduction}
\emph{It is possible to say that the research in statistical mechanics is in an historical phase akin to that one in quantum mechanics at the beginning of the last century\cite{galla}.} 
There is a lot to be studied and discovered, both in fundamentals and in  applications. 
One fundamental point is in fact  the range of application of the theory itself.  
Methods and concepts from statistical mechanics are starting to be currently used in scientific fields as different as e.g. 
biology\cite{sneppen}, economy\cite{stanley}, information theory\cite{mertens} and traffic engeneering\cite{helbing}.  
Statistical mechanics, in fact, naturally proposes itself as a general framework to connect microscopic mechanisms and macroscopic collective behaviors. 
In condensed matter physics, quantitative physical laws can be seen as emerging out of a statistical description 
of the dynamics of the microscopic units that form the system, 
or even out of that one of a simpler, coarse grained version of it.
Nowadays, simple lattice models are widely used to gain a qualitative and often deeper understanding of physical phenomena.

However, when a statistical mechanics  perspective is adopted  
in fields different from physics, an interesting point comes out. 
In many contexts, the structure of the interactions among the microscopic units can be  
often heterogeneous nor embedded in a real dimensional space, and moreover, it can evolve in time. 
Thanks to the recent development of the numerical calculus power and of the memory resources in information technology, 
recent analysis show that the topology of graphs as different as social networks 
(friendship patterns, scientific collaboration networks, etc.) food webs in ecology, 
critical infrastructure like the Internet and so on, 
is truly heterogeneous and very complex (see \cite{complexnet1} and ref. therein).
If the analysis of the structure of such complex networks requires statistical methods, the study of the dynamical 
processes occuring {\em on} them can get useful insights from statistical mechanics\cite{complexnet2}. 

A good example is provided by the study of epidemic models in heterogeneous networks\cite{epidemic}.
The real networks on which these processes are taking place are in fact very heterogeneous, i.e. they are {\em scale free}.
The dynamics of these processes in heterogeneous graphs can be ruled by the tails of the degree distribution. 
They can be in practice always in the infectuos phase 
and this is true e.g. for the spreading of viruses in large scale informatic systems.  
Interestingly, the paradigmatic Ising model has a dependence of this kind on the heterogeneity of the graph\cite{ising}. 

The general study of how the underlying topology affects the collective statistical behavior of model systems
is at the core of research in statistical mechanics.
However, up to recent times this study  was almost limited to homogeneous, 
or at least symmetrical structures of interactions, tipically d-dimensional or bethe lattices.
The heterogeneity calls at identifying general mechanisms and unifying schemes in the dynamics
of cooperative models on general heterogeneous graphs, 
since they can show truly different behaviors with respect to regular lattices.
The focus of this thesis is about statistical mechanics on heterogeneous random graphs, i.e. how such heterogeneity 
can affect the cooperative behavior of model systems, but it is not intended as a general review on it. 
Rather, I will show more practically how this question emerges naturally and can give new insights for specific instances, 
in both physics and interdisciplinary applications, for equilibrium and out of equilibrium issues as well.

The first chapter is devoted to the study of the congestion phenomena in networked queuing systems, 
like sub-networks of the Internet. After a brief introduction on the workings of the Internet, 
we will review the classic results of the queuing network theory, 
and the recent numerical results on congestion phenomena on complex networks.
Then, I will show how to combine them within a minimal model that in practice extends queuing network theory 
in the congested regime\cite{conge}. With the use of network ensemble calculation techniques, it is possible 
to study the dependence of the traffic dynamics on the topology of the graph and on the level of traffic control as well, 
up to the possibility of drawing a mean field phase diagram of the system. 
We find many results. In particular, we find that traffic control is useful only 
if the network has a certain degree of heterogeneity, 
but, in any case, it can trigger congestion in a discontinuous way.
Then, the second chapter is about the nature of the dynamical crossover in glass forming systems.
After a brief review on the experimental phenomenology of the glass transition, we will do a short review on the
theoretical perspectives on it. Then, I will show, within the framework of a simple facilitated spin model,
how the question of the heterogeneity of the underlying spatial structure is crucial\cite{ourfred}.
The dynamical arrest can change from a bootstrap percolation scenario to a simple one considering  an heterogeneous lattice (e.g. diluted).
This helps to shed lights on analogies and differences between the jamming of supercooled liquids and more heterogeneous
systems, like polymer blends or confined fluids.
The third chapter is on a general relationship between models and the underlying topology: 
how some specific features of the graph can induce inverse phase transitions in tricritical model systems.
After a brief review on inverse phase transitions, we will discuss the simplest model that reproduces this behavior, 
i.e. the Blume-Capel model with high degeneracy of the interacting states. 
I will show that tricritical model systems have this behavior if
sparse subgraphs are crucial for the connectivity\cite{nostopo}. 
Within this framework, I will work out many results for the Blume-Capel model and give
some insights about the fact that the random field ising model shares the same phenomenology.
Finally, the subject of the fourth chapter is on the co-evolving models of social networks.
We will give a brief introduction to the field of social networks. The interesting point is that here the graph itself is subject
to a dynamical evolution that can lead in turn to different states, with different connectivity properties.
The evolution of the network can be coupled to the dynamics defined on top of it, i.e. a so called co-evolution mechanism.
I will show how the volatility, i.e. the rate at which nodes and/or links disappears, affects this evolution with a simple model\cite{nostro}. 
Many results are found, in particular high node volatility can definitively suppress the emergence of an ordered, connected phase.

In the conclusions there is a review of the results and 
I will point out a general insight about the statical mechanics of models on heterogenous random graphs, 
supported by specific examples took out from the cases we dealt with.

\mainmatter

\chapter{Statistical mechanics of queuing networks}
The Internet\cite{wiki} is perhaps the most complex engineering system created in the human history.
Its exponential growth has played a pivotal role in the recent surge of interest in the study of complex networks.
It is not a static system, rather it evolves according to a self-organized and decentralized dynamics.
The structure\cite{vespi} and the properties of traffic dynamics\cite{regi} it supports show a very rich phenomenology.
The basic theory to analize traffic dynamics of information processing
networks, queuing network theory, relies on the simplifying hypothesis of stationarity.
This theory is mainly used to investigate single cases of small systems whose structure does not change.
Statistical mechanics can extend this theory,
allowing the investigation of  congested states and
the general study of the effects of topology and of traffic control on the traffic dynamics.
In this chapter, after a brief introduction
on the structure and traffic dynamics of the Internet,
I will review the main results of queuing network theory\cite{bolch} and
the recent results in congestion phenomena
on complex networks(see table $6$ in\cite{regi}).
Then, in the final paragraph,  I will show how to combine them togheter within the framework
of a minimal model, that allows the study of congestion in queuing network systems
up to the possibility of drawing mean-field phase diagrams\cite{conge}.
We find many general results, both theoretical, e.g. I will show analitically the presence of a dynamical phase transition in queuing networks ,
and of practical importance, e.g. I will  show that traffic control is useful only in heterogeneous networks.

\section{The Internet}
The Internet was originally conceived for experimental reasons within a military project in the '60 of the last century.
This network of networks  of computers has nowadays a worldwide extension, connecting hundreds of millions of \emph{hosts}
\footnote{An host is a device connected to the Internet, with its own  address, that can inter-operate with other hosts},
through which users can communicate in real time, sell and buy goods, exchange and share music,videos, informations, etc.
The handling of the information flows is the result of a complex interplay of different rules, protocols and devices acting at different levels.
These levels go from the physical one (the transport of electrical signals along wires or optical fibers) to that one of applications
(e.g. the standard SMTP protocol to forward e-mail), usually without common standards all along the network.

The Internet is a network of networks:
hosts  are joined together by \emph{switches} in  LAN (local area network)
or WAN (wide area network), and the exchange of information
among these networks is provided by specific devices called \emph{routers},
forming a network that represents
the physical connectivity of the Internet.
Routers are themself grouped together in \emph{autonomous systems} (AS),
i.e independently administered domains.
The traffic at the network (routers) level is ruled by the TCP/IP 
(transmission-control protocol/Internet protocol), 
perhaps the only common standard protocol in the Internet:
\begin{itemize}
\item The information is framed in discrete units, called IP \emph{packets}.
      The packet has a part devoted to the addresses of source and destination.
      There is a common address space for all the network.
\item All the packets are routed independently by the routers.
      Each router has a  list of paths,
      i.e. a kind of coarse-grained map of the network.
      It sends packets to its neighbors along the shortest path.
\item Routers exchange continuosly informations on the topology of the network,
      signaling damages, outages, etc.
\item Each single transmission between neighbouring routers is ruled
      by the TCP protocol through the exchange of check and confirmation signals 
      (ACK acknowledgements signals).
      A delay of ACK signals induces an  halving of the packets' sending rate along
      that line of transimission (window-based congestion control mechanism).
\end{itemize}
The overall structure, at each level, is self-organized and evolving.
The network of routers changes continuosly,
the nodes and the links being removed  or added according to the reasons (mainly of economical nature)
of single \emph{providers} and not by a central authority.
Therefore, it is hard to monitor the topology of such a graph, that is still partially unknown.

\begin{figure}[h]
\begin{center}
\includegraphics*[width=0.75\textwidth]{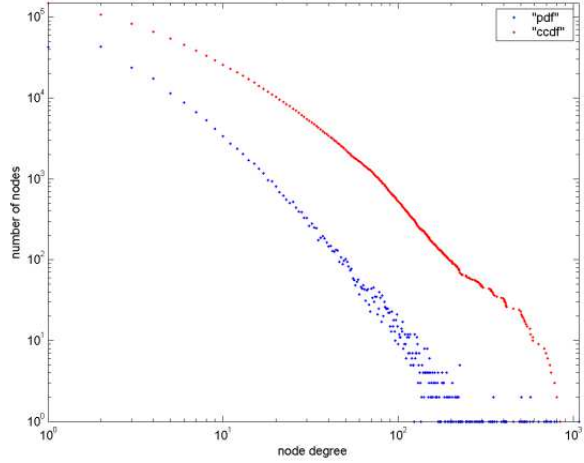}
\caption{Degree  distribution and complementary cumulative distribution 
of a subnetwork of the Internet at routers level monitored within the CAIDA project\cite{caida}}
\label{caida}
\end{center}
\end{figure}

Fig.\ref{caida} shows the degree distribution of a sub-network of the Internet monitored within the CAIDA project\cite{caida}.
The curve is well fitted by a power-law with an exponent between $2$ and $3$.
For a power law distribution $P(x) \propto x^{-\gamma}$, we have $\frac{d \log P(x)}{d(\log x)} \propto -\gamma$,
independently of $y$, and moreover, if $\gamma<3$ the error on $\langle x \rangle$ is not defined.
The networks  with a power law degree distribution with an exponent $\gamma<3$ are thus called {\em scale free}.
Dynamical processeses defined on them can show qualitative change with respect to homogeneous networks\cite{complexnet2},
as we will point out later in this chapter about congestion phenomena.
It seems that the scale-free degree distribution characterizes the Internet graphs at many scales,
from the routers to the AS level\cite{vespi}. This finding has attracted many  research efforts on the Internet structure\cite{workshop}.

There are few stilizyed facts about traffic dynamics,
the main being the \emph{self-similarity} of inter-arrival time signals\cite{willy}.
Looking at the temporal evolution $y(t)$
of the time spent by a signal to travel along a given path in the network under controlled conditions it is found that
the self correlation function 
\begin{equation}
C(\tau) = \frac{\langle (y(t) y(t+\tau) \rangle -\langle y \rangle^2}{ \langle y^2 \rangle - \langle y \rangle^2}:
\end{equation}

\begin{itemize}
\item is unsummable $\sum_i  \vert C(i) \vert \to \infty$
\item and has a power-law tail $C(\tau) \propto \tau^{-\nu}$.
\end{itemize}
Moreover:
\begin{itemize}
\item The scaling of the variance of the coarse-grained signal over intervals $M$ times larger
      $y_M (t) = \frac{1}{M} \sum_{i=t-M}^{t+M} y(i)$ is not normal, i.e.
      $\sigma_M^2 = \frac{\sigma^2}{M^\beta}$
\end{itemize}

There are many different ways to define and measure these features.
They seem to be independent of the path, the time of measure and the level of traffic.
There are many ways to interpret them.
The robustness of these features has attracted several modeling efforts
to interpret it as a signature of the fact
that the  network is working at criticality between a free and a congested phase
through a self-organized mechanism of some kind, as we shall see in the next paragraph.
But self-similarity is even  too robust for this mechanism at work:
it is still present even for low level of traffic load, far out from the congested regime.
Therefore, the most accepted explanation for the self-similarity of inter arrival times signals relies
on the heterogeneity and strong correlation in time of the demand itself\cite{willy2}.
In fact, many packets can belong to the same request,
with a distribution of the flows' size
\footnote{A flow is a group of packets within the same request} that is heterogeneous itself.

Apart from this, even if this network mostly works in the free regime,
time delays and packets' loss continue to threaten Internet pratictioners,
because some parts of the network can be sometimes in the congested regime.
However, congestion events are difficult to monitor and study,
and a clear phenomenological picture is still missing.
This calls for a theoretical understanding of what happens above the threshold at which a queuing network system can work.

\section{Models of network traffic dynamics}
\subsection{queuing network theory}
The classical framework used to study performances of information processing and/or service delivering networked systems
is queuing network theory (QNT)\cite{bolch}.
Its applications range from the study of costumers forming queues in banks and offices, to the study of
data traffic in packet switching networks of routers in communication systems.

The main model  is the  Jackson or open queuing network\cite{jack}, consisting  of $N$ nodes such that:
\begin{itemize}
\item each node $i$  is endowed with a FIFO (first-in first out) queue
      with unlimited waiting places (it can be arbitrary long).
\item The delivery of a packet from the front of $i$ follows a poisson process with a certain frequency $r_i$(service rates), and
      \item[-] the packet exits the network with some probability $\mu_i$, or
      \item[-] it goes on the ``back'' of another queue $j$ with probability $q_{ij}$.
\item Packets are injected in each queue $i$ from external sources
      by a Poisson stream with intensity $p_i$.
\end{itemize}

\begin{figure}[h]
\begin{center}
\includegraphics*[width=0.75\textwidth]{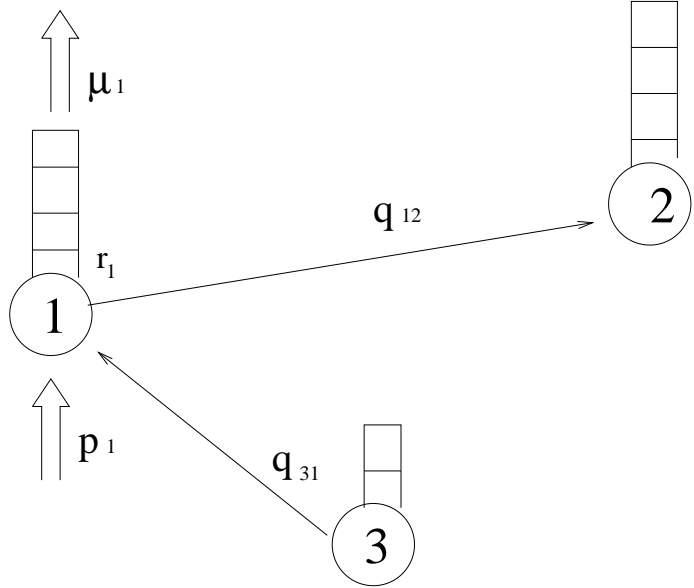}
\caption{queuing network: $p_i$ creation rates,$r_i$ service rates, $\mu_i$ absorbing rates, $q_{ij}$ routing probabilities}
\label{net}
\end{center}
\end{figure}

The state of the system is specified by the vector $\mathbf{n} = (n_1,...,n_N)$, where $n_i$ is the $i$'s queue length.
If we indicate with $\mathbf{i}$ the vector with all components equal to zero, apart from the $i^{th}$ that is equal to one,
we have the expressions for the transition rates\footnote{In the second and third equations $n_i>0$}:
\begin{eqnarray}
W(\mathbf{n} \to \mathbf{n}+\mathbf{i}) = p_i      \\
W(\mathbf{n} \to \mathbf{n}-\mathbf{i}) = r_i \mu_i     \\
W(\mathbf{n} \to \mathbf{n}-\mathbf{i} + \mathbf{j}) = r_i (1-\mu_i) q_{ij}.     
\end{eqnarray}
The master equation for the probability distribution of states $P(\mathbf{n})$ reads:
\begin{equation}
\dot{P}(\mathbf{n}) = \sum_{\mathbf{n'}} W(\mathbf{n'}\to\mathbf{n}) P(\mathbf{n'}) - \sum_{\mathbf{n'}} W(\mathbf{n}\to\mathbf{n'}) P(\mathbf{n}).   
\end{equation}
QNT studies the stationary state, assuming it exists: $\dot{P}=0$. How we see next, 
the probability distribution factorizes $P(\mathbf{n}) = \prod_{i} p_i(n_i)$.
By using this form of the distribution  as an ansatz to solve the master equation, we find:
\begin{equation}
p_i(n) = (1-x_i) x_i^{n},
\end{equation}
where $x_i = \lambda_i/r_i$. Here the coefficients $\lambda_i$, the average packet flow towards node $i$,
can be found on specific networks solving the set of linear equations:
\begin{equation}
\lambda_i = p_i + \sum_{j} q_{ji} (1-\mu_j) \lambda_j .
\end{equation}

The framework of QNT can easily accomodate modifications.
For instance, it is possible to think of  finite size capacity for the queues,
queue length state dependent service rates and/or transition probabilities.
It is possible to study closed
instances, with a given number of packets $K$ that are not generated neither absorbed.
QNT has many practical applications in very different contexts, from telecommunications networks to the scheduling design
of factories, hospitals, ecc, and, moreover, it can give interesting insights to theoretical research.

For instance, there was a recent debate\cite{scaling} about the scaling of fluctuations of the Internet time series.
Looking at the time series of the amount of bytes $x(t)$ processed by  a router,
it is found that $\sigma \propto \langle  x \rangle ^{\gamma}$,
with $1/2<\gamma<1$, depending on the aggregation time, and/or crossover between these limits.
This can find a nice and natural explanation within QNT.
In fact, if $x$ follows an exponential distribution, as is the case for most of the queuing networks,
it is $\sigma^2 = \langle  x \rangle + \langle  x \rangle ^2$, and $\langle  x \rangle$ can vary by aggregating times,
mimicking exponents between $1/2$ and $1$.

The main limitation of this theory is in the stationarity assumption.
It gives for guaranteed that,
given a certain external demand $\mathbf{p} =(p_1,...,p_N)$, we are always able
to build a network such that $\lambda_i<r_i$. It basically
avoids completely the study of \emph{congested} states, in which queues can grow out of stationarity.
It should be noticed that in self-organized evolving networks, like the Internet, 
the external demand may change on times faster than our capacity to modify the network to mantain
stationarity.
This can trigger \emph{congestion phenomena}, that are interesting
to study from a theoretical point of view.

\subsection{The congestion phase transition}
Apart from this theory, the recent years have witnessed
the proposal of several models of interacting particles hopping on graphs,
to study the interplay of  topology and routing strategy on the performances
of networked systems in processing information (see table $6$ in\cite{regi}).

In all these models packets are injected into the network with some rate $P$,
they have to travel between  given sources and destinations, where they exit the network, and they interact by forming queues.
Can all the packets reach their destinations, or, alternatively, can the network process all the incoming information (quantified by $P$)?
If it can do it,  the total number of packets $N(t)$ will be stationary in time, if it cannot, $N(t)$ will be growing in time.
A good parameter to distinguish these two different phases is the average queues' growth rate
divided by the average rate of incoming packets, i.e. the percentage of packets trapped in queues\cite{arenas}:
\begin{equation}
\rho = \frac{\langle \dot{N} \rangle}{P}. 
\end{equation}
By studying the curves $\rho(P)$, once the network and the routing strategy  are given,
it is possible to distinguish two phases: $\rho=0$ (free flow) and $\rho>0$ (congestion), clearly divided by a point $P_c$.
Upon approaching this point from the free phase, the self-correlation in time of the queues' length starts to develop fat tails.
This was seen as an elegant explanation of the self-similar character of real time series.
However, as it was previously stated, self-similarity in a real network is present even far out of the congested regime.
Anyway, all these works show the inherent presence in queuing networks of a dynamical phase transition towards congestion.

\begin{figure}[h]
\begin{center}
\includegraphics*[width=0.75\textwidth]{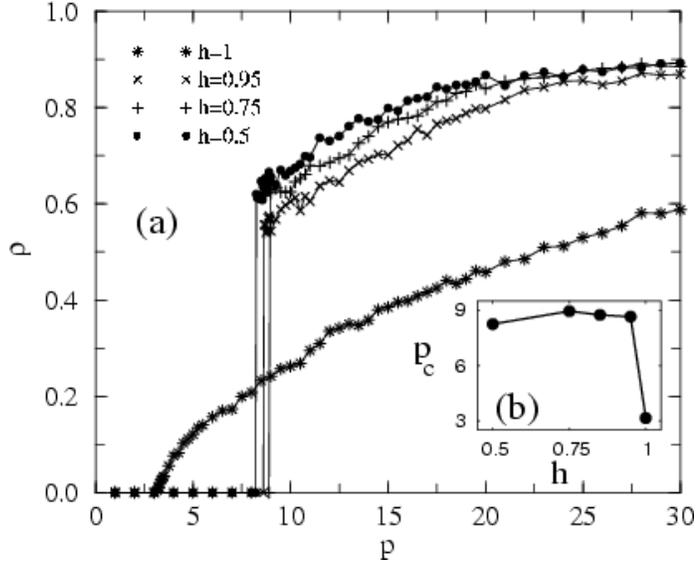}
\caption{Congestion parameter curves $\rho(P)$ from simulations of a particle hopping model on the Internet at level of AS graph, from\cite{eq}.}
\label{eche}
\end{center}
\end{figure}

The  numerical investigation of the dependence of such a transition on the structure
of the graph and on the routing strategy, shows interesting phenomena.
One of the most interesting is  the apparent \emph{tricritical} character of the congestion transition in  queuing network systems.
In\cite{eq} the authors propose the following model:
onto a given network, packets arrive from external source with rate $P$ on random nodes, each packet has
its destination $d$, and it hops from the current node $i$ to the neighbor $j$ such that the quantity:
\begin{equation}
w_j  = d_{jd}h + n_j (1-h),
\end{equation}
is minimum. There $d_{jd}$ is the distance between $j$ and $d$, $n_j$ is the number of packets sitting on $j$,
$h$ is a parameter that quantifies the level of congestion control ($h=1$, no congestion control, shortest path routing).
This mimick an attempt to minimize travelling times instead of distances with the use of
local information. The authors  did simulations on a realistic instance, i.e. the Internet network at autonomous system level.
They found that a certain level of traffic control can avoid the transition up to a certain point, after which congestion
is triggered in a discontinuous way, i.e. upon decreasing $h$, $P_c$ is growing, but exactly at $P_c$,
$\rho$ jumps from $0$ to a finite value (see fig.\ref{eche}).

This approach to network traffic based on the  dynamics of individual particles
has the problem of not being amenable to analytic approaches.
These models become analitically tractable  considering a randomization of the trajectories.
The reach of a destination by a particle should be mimicked in a probabilistic way,
i.e. during the hoppings the particle can be absorbed with some probability.
This defines a framework very similar to the QNT, that I will analize in detail in the next paragraph.

\section{Statistical mechanics of congestion}
I will review the model in ref.\cite{conge}. It consists of particles hopping randomly among the nodes of a graph such that:
\begin{itemize}
\item They form queues,
\item They are created with a certain rate.
\item They have a certain probability of being absorbed during the hoppings.
\end{itemize}
Then we will mimick a protocol of congestion control in the following way:
\begin{itemize}
\item The node $j$ starts to reject particles with probability $\bar{\eta}$ once its queue is longer than $n^*$.
\end{itemize}

\begin{figure}[h]
\begin{center}
\includegraphics*[width=0.75\textwidth]{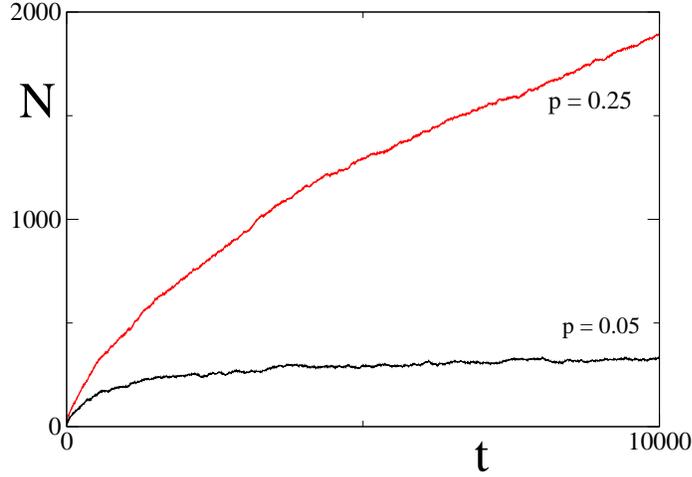}
\caption{Number of packets as a function of time for an homogeneous network of $10^3$ nodes, degree $K=4$, without routing procol ($\bar\eta=0$),
with homogeneous condition $r_i=1$, $\mu_i=\mu=0.2$, in the free ($p_i=p=0.05$) and congested phase ($p=0.25$).}
\label{phases}
\end{center}
\end{figure}

This model of particles corresponds to a queuing network such that:
\begin{itemize}
\item The hopping probability from node $i$ to node $j$ is
      $q_{ij} = \frac{1-\bar{\eta}\theta(n_j-n^*)}{k_i}a_{ij}$\footnote{$\theta(x)$ is the step function, i.e. it is $1$
      for $x\geq0$, $0$ otherwise}
\item A certain set of values ${r_i,p_i,\mu_i}$ for the service rate, demand and adsorbing probability of the node $i$, respectively, is given.
\end{itemize}
Here $a_{ij}$ is the adjacency matrix of the graph
\footnote{$a_{ij}$ is $1$ if $i$ and $j$ are connected by a link, $0$ otherwise} and  $k_i$ the degree of the node $i$.
The important difference from the Jackson framework
is that packets have to \emph{move} to be absorbed.
They are absorbed during the hoppings and  not when they are stored in the queues.
Within this new framework, it is possible to extend QNT beyond stationarity.
There are two phases (see fig.\ref{phases}):
as the demand increases the system pass from a free phase, in which the number of particles is stationary,
to a congested phase, where it is growing.

\begin{figure}[h]
\begin{center}
\includegraphics*[width=0.6\textwidth]{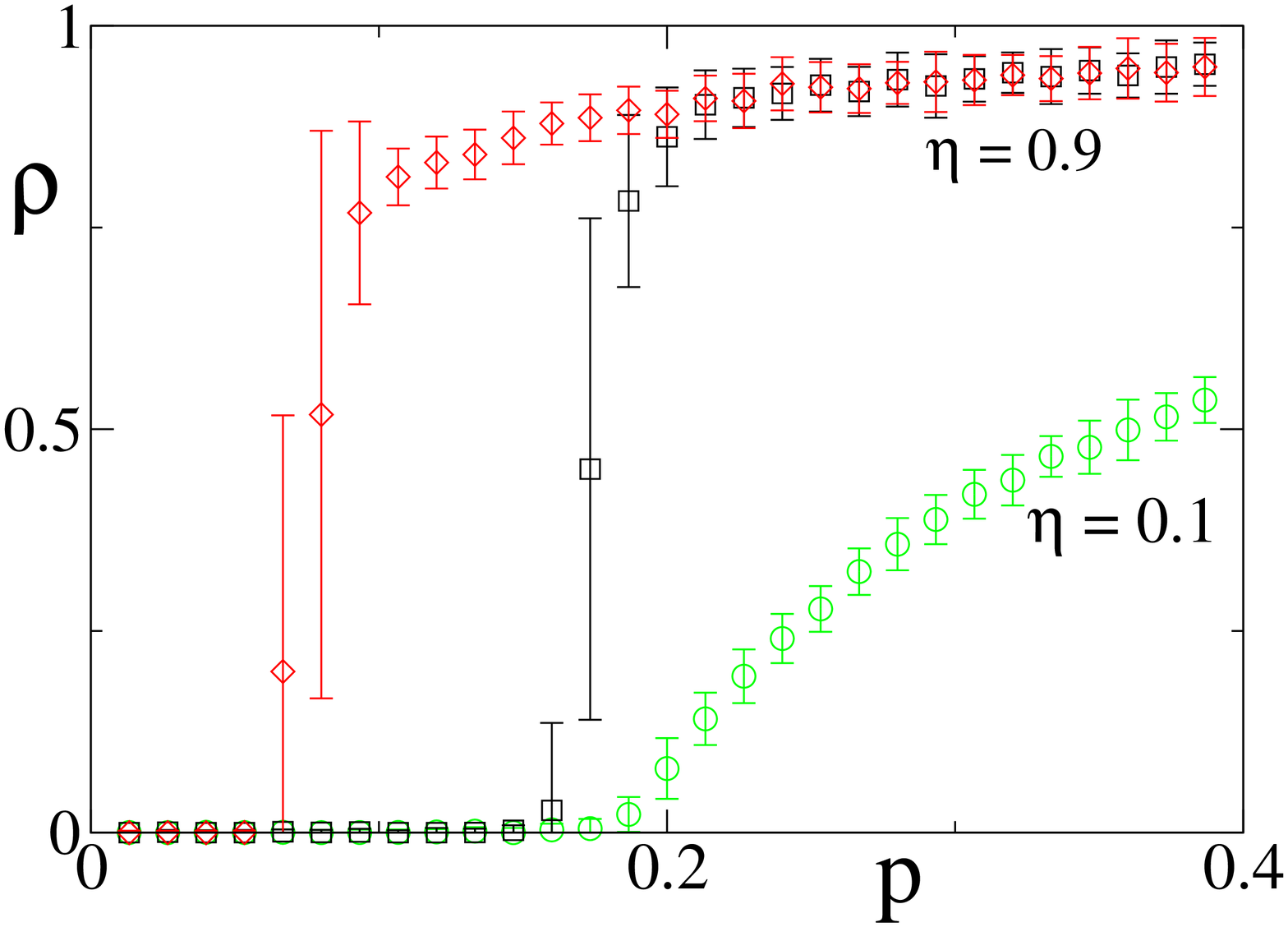}
\includegraphics*[width=0.6\textwidth]{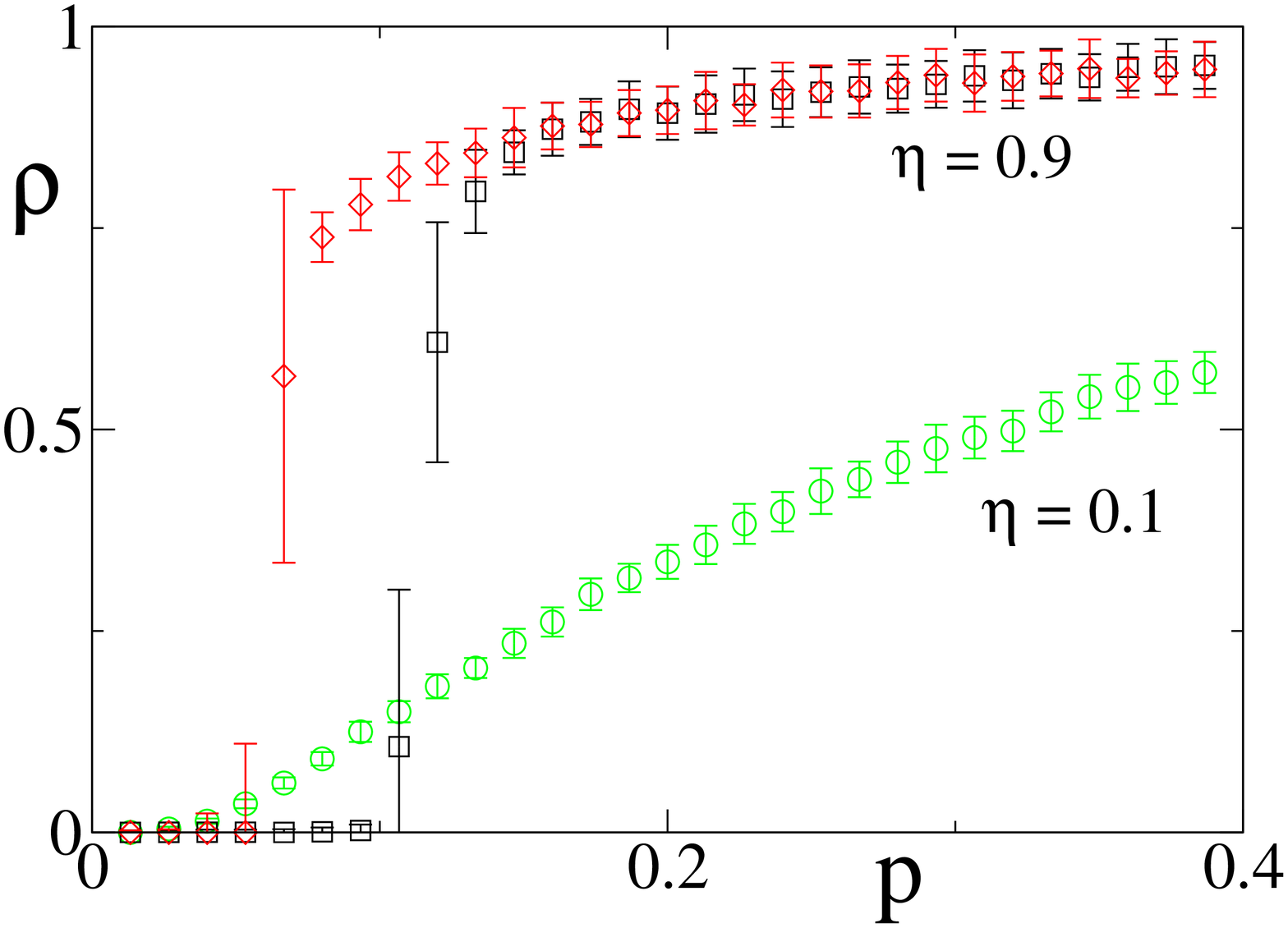}
\caption{Top: Transition curves $\rho(p)$ for a random regular graph of size $N=10^4$,
$\mu=0.2$, $\bar{\eta}=0.1$, $\bar{\eta}=0.9$.  Bottom:
Transition curves $\rho(p)$ for an uncorrelated scale free graph with $\gamma=3$, $k_{min}=2$
of $N=3 \centerdot 10^3$ nodes, $\mu=0.2$, $\bar{\eta}=0.1$, $\bar{\eta}=0.9$ for $n^*=10$.
For $\eta=0.9$ the system shows \emph{ hysteresis} in both the homogeneous and heterogeneous case.
}
\label{rhop}
\end{center}
\end{figure}

There is a phase transition between them, whose nature depends on the topology of the graph and on the level of traffic control.
This is shown in fig.\ref{rhop}, which reports simulations  on homogeneous and heterogeneous graphs, with low and high level of traffic control.
The curves $\rho(p)$ suggest that an high level of traffic control trigger the transition in a discontinuous way
and can displace the transition point to higher values of $p$ only in the heterogeneous case.

In order to study the generality of such results, I will exploit the QNT formalism, combined with the use
of techniques from statistical mechanics.
The transition rates  are:
\begin{eqnarray}
W(\mathbf{n} \to \mathbf{n}+\mathbf{i}) = p_i      \\
W(\mathbf{n} \to \mathbf{n}-\mathbf{i}) =  r_i \sum_j \frac{\mu_j (1-\bar{\eta}\theta(n_j))}{k_i} a_{ij}     \\
W(\mathbf{n} \to \mathbf{n}-\mathbf{i} + \mathbf{j}) = \frac{r_i (1-\mu_j) (1-\bar{\eta}\theta(n_j))}{k_i} a_{ij} .
\end{eqnarray}

We work with  the approximation of a factorized form for the probability distribution function
$P(\mathbf{n}) \simeq \prod_{i} p_i(n_i)$
\footnote{It works exactly for $\bar{\eta}=0$ because this is equivalent to the Jackson network.
See appendices in the second reference of ref.\cite{conge} for a discussion about the extension of its validity.}.

Imposing detailed balance $p_i(n_i) W(n_i \to n_i+1) = p_i(n_i+1) W(n_i+1 \to n_i)$,
we can express the single point distributions $p_i(n_i)$ in terms of the
the two local quantities $q_i = p_i(0)$ and $\chi_i = Prob(n_i\ge n^*)$.
We have
\begin{eqnarray}
W(n_i \to n_i+1) = p_i + (1-\mu_i) \sum_{j} r_j q_j \frac{1-\bar{\eta}\theta(n_i-n^*)}{k_j} a_{ij} \\
W(n_i+1 \to n_i) = \frac{r_i}{k_i} \sum_j (1-\bar{\eta}\chi_j)a_{ij}.
\end{eqnarray}
Then, the average growth of the queue lenght of node $i$ is:
\begin{equation}
\langle \dot{n}_i \rangle = p_i + (1-\mu_i)(1-\bar{\eta}\chi_i)\sum_{j} a_{ij} \frac{r_j(1-q_j)}{k_j} -\frac{(1-q_i)r_i}{k_i}\sum_{j} (1-\bar{\eta}\chi_j) a_{ij} . 
\end{equation}

The equations for the ${q_i,\chi_i}$ come from:
 \begin{itemize}
\item The normalization conditions $\sum_{n_i} p_i(n_i) =1$.
\item The stationarity of queues'length  $\langle \dot{n_i} \rangle =0$.
\end{itemize}
If the second condition gives not physical results, we have that $q_i=0$, $\chi_i=1$, from which we can calculate $\langle \dot{n_i} \rangle$.
This can be summarized in terms of the linear set of equations
\begin{eqnarray}
\chi_i = \max\{0,\min[1,C_i(\vec{\chi},\vec{q})]\} \\
q_i = \max\{0,\min[1,Q_i(\vec{\chi},\vec{q})]\},
\end{eqnarray}
where
\begin{eqnarray}
Q_i(\vec{\chi},\vec{q}) = 1 -\frac{p_i+(1-\mu_i)\sum_{j} a_{ij} r_j(1-q_j)/k_j}{p_i-r_i/k_i \sum_{j} a_{ij} \bar{\eta}\chi_j} \\
C_i(\vec{\chi},\vec{q}) = \frac{1}{\bar{\eta}}\Big[  1 +  \frac{p_i-r_i/k_i \sum_{j} a_{ij} \bar{\eta}\chi_j}{(1-\mu_i)\sum_{j} a_{ij} r_j(1-q_j)/k_j} \Big].
\end{eqnarray}
For sake of simplicity we consider from now on $p_i=p$, $r_i=1$ and $\mu_i=\mu$.

\subsection{Network Ensemble calculations and results}
We consider an uncorrelated random graph
with a given degree distribution $P(k)$.
This is a graph taken from the ensemble
of all the graphs with a given degree distribution,
all with equal statistical weights.
If we have $N$ nodes $i=1 \dots N$, it is possible to build a graph of this kind along these lines (configuration model\cite{bollobas}):
\begin{itemize}
\item We extract the degree $k_i$ of each node $i$ randomly according to the desired distribution $P(k)$.
\item Each node $i$ has $k_i$ {\em stubs} dangling from it. We randomly match these stubs, taking  care
      to avoid tadpoles and double links\footnote{This can introduce undesired correlation, see\cite{bollobas}}.
\end{itemize}
A typical network of this ensemble has locally the structure of a tree,
such that dynamical processes defined onto it
can be successfully approximated with the use of mean field techniques.
In particular, for our model,  we will make the hypothesis
that all the nodes with the same degree have the same statistical dynamical features.

The mean field rates for the queue length of a node with degree $k$ are
\footnote{We have absorbed for sake of simplicity the $\bar{\eta}$ in the definition of the $\chi$}:
\begin{eqnarray}
w_{k}(n \to n+1) &=& p + (1-\mu)(1-\bar{q})\frac{k}{z}(1-\bar\eta\theta(n-n^{*}))\nonumber \\
w_{k}(n \to n-1) &=& \theta(n)(1- \bar{\chi}),
\end{eqnarray}
where $z$ is the average degree,  $\bar{q} = \sum_{k} q_{k} P(k)$ and $\bar{\chi} = \sum_{k} \frac{k}{z} \chi_{k} P(k)$.
The average queue length $\langle n_{k}\rangle$ follows the rate equation
\begin{equation}\label{dotn}
\langle\dot{n}_{k} \rangle = p +(1-\mu) (1-\bar{q}) \frac{k}{z} (1 - \chi_{k})-(1-q_{k}) (1-\bar{\chi}) .
\end{equation}
Note that summing over $k$ and dividing by $p$ we obtain a measure of the order parameter $\rho (p)$.

Since $\dot{n}_{k}$ depends linearly on $k$, high degree nodes are more likely to be congested,
therefore, for every $p$, there exists a real valued threshold
$k^{*}(p)$ such that all nodes with $k > k^{*}$ are congested whereas nodes with degree less than $k^*$ are not congested.
Congested nodes ($k > k^{*}$) have $q_{k} = 0$ and $\chi_{k} =\bar\eta$.
The probability distribution for the number of particles in the queue of
free nodes with degree  $k<k^{*}$ can be extracted by calculating
the generating function $G_{k}(s) = \sum_{n} \mathcal{P}_k(n_k=n)s^n$ from the
detailed balance condition $w_k(n_k+1 \to n_k) \mathcal{P}_k(n_k+1) = w(n_k \to n_k+1) \mathcal{P}_k(n_k)$.
 The generating function takes the form
\begin{equation} \label{genfun}
G_{k}(s) = q_{k}\left\{ \frac{1-{(a_{k} s)}^{n^{*}}}{1-a_{k} s} +
  \frac{{(a_{k} s)}^{n^{*}}}{1-(a_{k} - b_{k})s}\right\}
\end{equation}
 corresponding to a double exponential, where $a_{k} = [p+(1-\mu) \frac{k}{z} (1-\bar{q})]/[1-\bar{\chi}]$
and $b_{k} = \bar\eta [(1-\mu)\frac{k}{z}(1-\bar{q})]/[1-\bar{\chi}]$.
From the normalization condition $G_{k}(1)=1$ and the condition $\dot{n}_{k} = 0$,
we get expressions for $q_k$, $\chi_k$,
\begin{eqnarray}
q_k & = & \left[  \frac{1-a_{k}^{n^{*}}}{1-a_{k}} +
  \frac{a_{k}^{n^{*}}}{1-a_{k} + b_{k}} \right]^{-1} \\
  \chi_k & = & 1 + \frac{p-(1-q_k)(1-\bar\chi)}{(1-\mu)(1-\bar{q})\frac{k}{z}}
\end{eqnarray}
and, finally, for $\bar{q}$, $\bar{\chi}$.

The value $k^{*}$ is  self-consistently determined imposing that nodes with $k=k^{*}$
are marginally stationary, i.e. $\dot{n}_{k^{*}}=0$ with $q_{k^{*}} = 0$, $\chi_{k^{*}} =\bar\eta$, that translates into the equation
\begin{equation} \label{kstar}
k^{*} = \frac{1-p-\bar{\chi}}{(1-\mu)(1-\bar\eta)(1-\bar{q})}z . 
\end{equation}
The set of closed equations for $\bar{q},\bar\chi$ can be solved for any degree distribution
$P(k)$ and $\rho(p)$ can be computed accordingly.

\begin{figure}[h]
\begin{center}
\includegraphics*[width=0.75\textwidth]{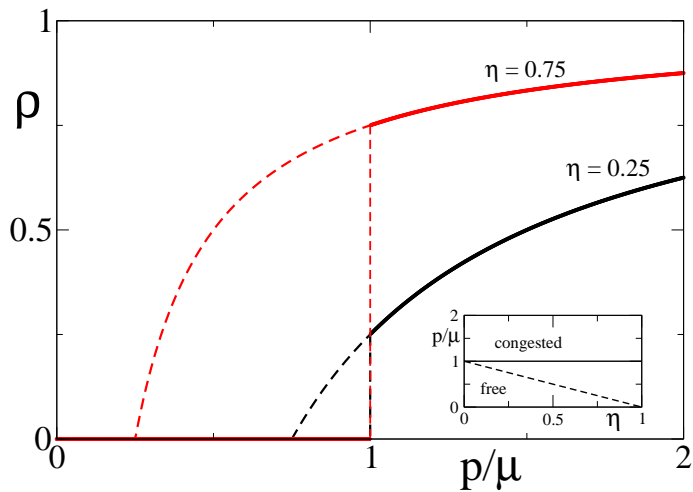}
\caption{Behavior of the congestion parameter
$\rho(p/\mu)$ for a random regular network
for $\eta = 0.25$,  $0.75$. Inset: phase diagram for the same graph.
}
\label{figreg}
\end{center}
\end{figure}

\subsubsection{Homogeneous networks}
The equations for $\bar{q}$ and $\bar\chi$ simplifies to a single equation when all nodes have the same properties, and in particular the same degree  ($k_i=K, ~\forall i$).
On these networks, the mean-field behavior can be trivially studied for any value of $n^*$, but we consider as an illustrative example the limit $n^{*}\to \infty$.
Only two solutions of the equation relating $\bar{q}$ and $\bar\chi$ are possible: the free-flow solution ($\rho=0$) with $\bar q=1-p/\mu$
and $\bar\chi=0$ that exists for $p\le \mu$, and congested-phase solution, where all nodes have
$n_i\to\infty$, i.e. $\bar\chi=\bar \eta$ and $\bar q=0$. The latter solution has $\rho=\dot n/p=1-(1-\bar\eta)\mu/p$
and exists for $p\ge (1-\bar\eta)\mu$.
The behavior of the congestion parameter with both the continuous and discontinuous
transitions to the congested state  is plotted in Fig. \ref{figreg} for $\bar\eta=0.25, 0.75$.
The corresponding phase diagram, reported in the inset of Fig. \ref{figreg}, shows that in the interval
$p\in [(1-\bar\eta)\mu,\mu]$ both a congested- and a free-phase coexist. We find  an hysteresis cycle, with  the system that turns from a free phase into
 a congested one discontinuously as $p$ crosses $\mu$.
It reverts back to the free phase only at $p=(1-\bar\eta)\mu$ as $p$ decreases.
It is interesting to observe that in the homogeneous case
the transition is always discontinuous until there is traffic control $\bar{\eta}>0$.

\subsubsection{Heterogeneous networks}
\begin{figure}[h]
\begin{center}
\includegraphics*[width=0.75\textwidth]{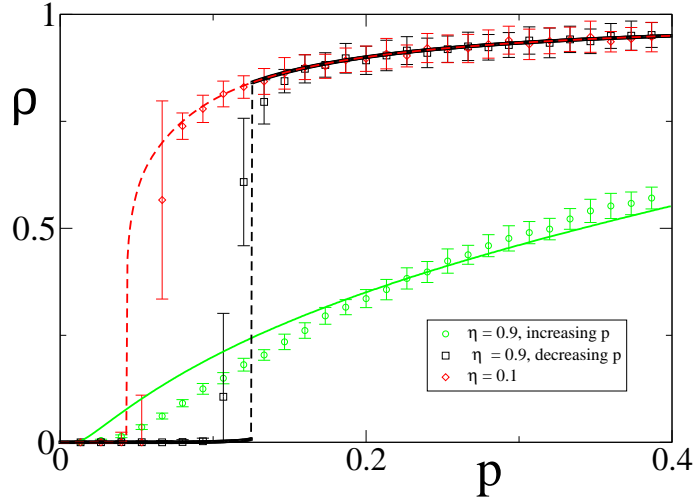}
\caption{
$\rho(p)$ for an uncorrelated scale-free graph
($P(k)\propto k^{-3}$, $k_{min}=2$, $k_{max} = 110$, $N = 3000$),
$\mu=0.2$, $n^{*}=10$ and $\bar\eta=0.1$
and $\bar\eta=0.9$, from both simulations (points) and theoretical predictions (lines). Hysteresis is observed increasing (black curve and points) and then decreasing (red curve and points) $p$ across the transition.
}
\label{fig1}
\end{center}
\end{figure}

In the case of heterogeneous networks the equations for $\bar{q}$ and $\bar\chi$ have to be solved numerically.
For instance, in Fig. \ref{fig1} we compare the theoretical
prediction (full line) for $\rho(p)$ in a scale-free network  with results of  simulations (points).
The agreement is good, the theoretical prediction at the ensemble level  confirming the scenario already observed in the simulations.
The curves are obtained for $\mu = 0.2$ and $n^{*} = 10$, but the behavior does not
qualitatively change for different values of these parameters. The dependence on $\bar\eta$ brings instead qualitative changes.
Increasing $\bar\eta$ from $0.1$ to $0.9$, the transition becomes discontinuous and $p_{c}$ increases.

The main difference with respect to homogeneous networks is that not all nodes become congested at the same time. The rate $p$ at which a node becomes congested depends on its degree, the hubs being first. The process governing the onset of congestion and the effects of the rejection term can be understood in the limit $n^{*} \to \infty$, that  simplifies considerably the calculations without modifying the overall qualitative behavior for sufficiently large $n^*$.
We have to solve in the limit $n^* \to \infty$ the self-consistent equations for $\bar\chi$ and $\bar{q}$. In this limit, uncongested nodes have $a_{k} <1$, hence  $\chi_{k} \to 0$ and $q_{k} = 1- a_{k}$. All nodes with degree $k<k_F$, where  $k_{F} = \max (k^{*} (1-\bar\eta), k_{min})$, are free from congestion.  Congested nodes have $q_{k} \to 0$ and $\chi_{k} = \bar\eta$ (for $k \geq k^*$). In addition there are
also fickle nodes, which are those with $k_F \leq k < k^{*} $ and $\chi_{k} = 1-\frac{k_F}{k}$.
Using this classification, we get a first expression for $\bar\chi$, i.e.
\begin{equation}
\bar{\chi}_{1}  = \sum_{k = k_{F}}^{k^{*}} \left[ 1 - \frac{k_F}{k}\right]\frac{k}{z} P(k) 
+ \bar\eta \sum_{k=k^{*}}^{k_{max}} \frac{k}{z} P(k) .
\end{equation}
Eq. (\ref{kstar}) provides a further relation between $\bar{q}$, $\bar{\chi}$ and $k^*$. We eliminate $\bar{q}$  using its definition which leaves us with another expression for $\bar{\chi}$,
\begin{equation}
\bar{\chi}_{2} = 1 - \frac{1}{2A}\left\{ 1+ A p -B + {\left[ (1+ A p -B)^2 + 4 A B p \right]}^{1/2}\right\} 
\end{equation}
where $A = z/[k^{*}(1-\bar\eta)(1-\mu)]$ and $B = \sum_{k = k_{min}}^{k_{F}} \left[ 1- \frac{k}{k_F}\right] P(k)$. To determine $\bar{\chi}$ we have to solve the implicit equation $\bar\chi_1 = \bar\chi_2$.\\
\begin{figure}[h]
\begin{center}
\includegraphics*[width=0.6\textwidth]{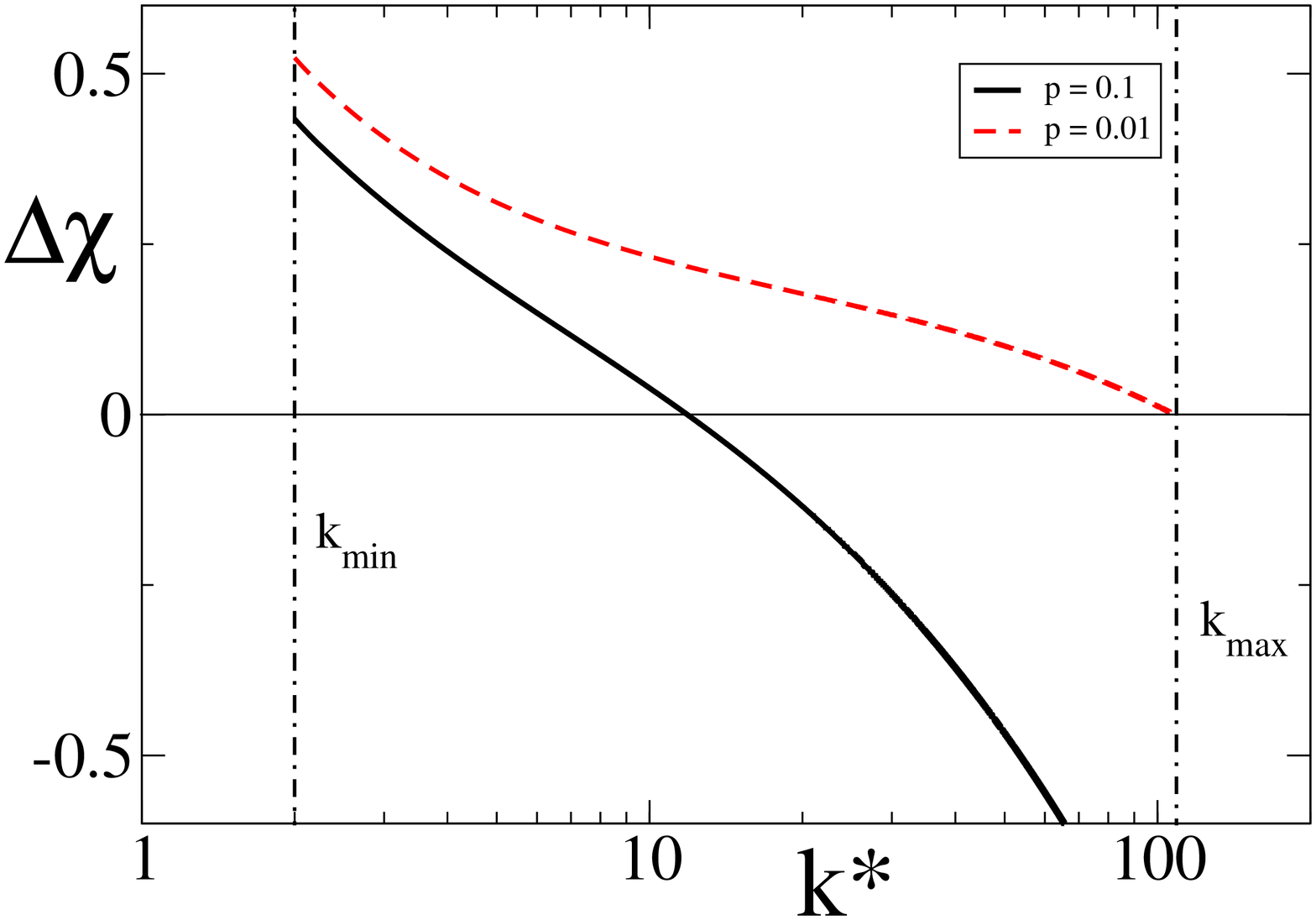}
\includegraphics*[width=0.6\textwidth]{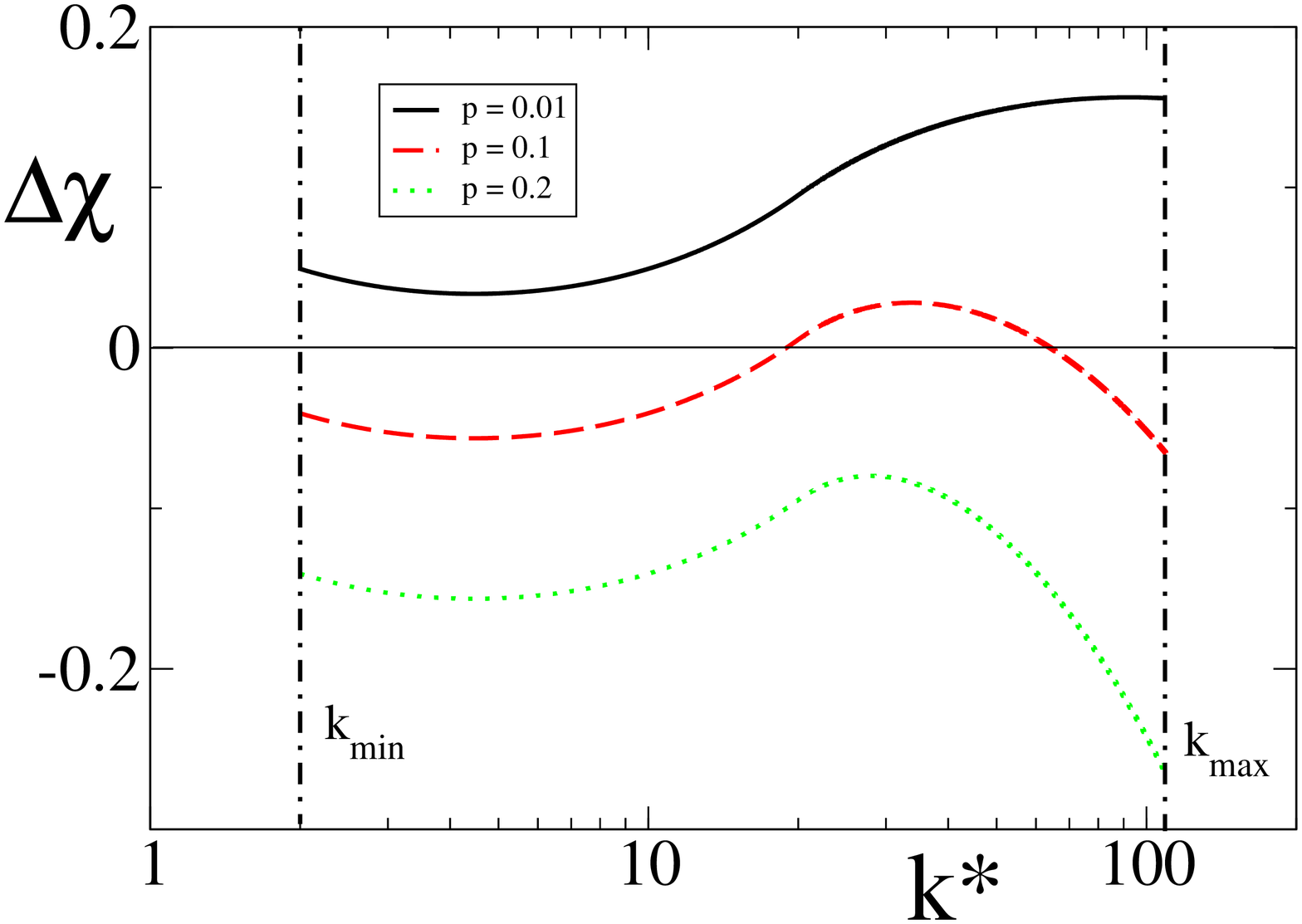}
\caption{
The zeros of $\Delta \chi (p)$ vs. $k^{*}$ define the threshold degree for the onset of congestion in a network.
The picture refers to a scale-free random network with $\gamma = 3.0$, $k_{min}=2$ and $N=3000$
($k_{max}= 110$), and different values for $\bar\eta = 0.1$ (left) and $0.9$ (right) and $p$. The
solution $k^*_1(p)$ in the right panel falls outside the plot.
} \label{fig2}
\end{center}
\end{figure}

In Fig. \ref{fig2} we plot the difference $\Delta \chi = \bar\chi_{1} - \bar\chi_{2}$ vs. $k^{*}$, for $\bar\eta=0.1$ (left) and $0.9$ (right) and different values of $p$ on a scale-free graph.
The zeros of $\Delta\chi(k^*)$ correspond to the only possible values assumed by $k^{*}$. For small rejection probability ($\bar\eta = 0.1$ in Fig. \ref{fig2}), there is only one solution $k^{*}(p)$, which decreases from  $+\infty$ when increasing $p$ from $0$. The value $p_{c}$ at which $k^{*}(p_{c}) = k_{max}$ is the critical creation rate at which largest degree nodes become congested. At larger $p$, $k^{*}(p)$  decreases monotonously until eventually all nodes are congested when $k^{*}(p) = k_{min}$.  Hence for low values of $\bar\eta$, the transition from free-flow to the congested phase occurs continuously at the value of $p$ for which $k^{*}(p) = k_{max}$. \\
At large $\bar\eta$ ($\bar\eta = 0.9$ in Fig. \ref{fig2}), the scenario is more complex. Depending on $p$, the equation can have up to three solutions, $k^{*}_{1} (p)\leq k^{*}_{2} (p)\leq k^{*}_{3}(p)$. It is easy to check that only $k^{*}_{1}$ and $k^{*}_{3}$ can be stable solutions. For $p\ll 1$ there is only one solution at $k^{*}_{3} (p)\gg k_{max}$, corresponding to the free phase. This is thus the stable solution for $p$ increasing from zero. As $p$ increases, another solution $k^*_1(p) < k^*_3(p)$ can appear, and $k^*_3(p)$ moves towards lower degree values. Three situations may occur:
\begin{itemize}
\item[{\em i.}] The solution $k^{*}_{3}(p)$ disappears before reaching $k_{max}$. Then $k^*_1(p)$ becomes the stable solution, and the congested phase appears abruptly.  However, given the shape of the function $\Delta \bar\chi$ (see Fig. \ref{fig2}), when this happens $k^*_1(p) \to 0$ and in particular we expect $k^*_1(p) < k_{min}$, so that  above the transition the whole network is congested and follows the law  $\rho(p) = 1-(1-\bar\eta)\frac{\mu}{p}$.
\item[{\em ii.}] The solution $k^{*}_{3}(p)$ crosses $k_{max}$ and exists until it reaches $k_{min}$. Then the congested phase emerges continuously and the network is only partially congested (i.e. only the nodes with $k \ge k^*_3(p)$). The order parameter grows until it reaches the curve of complete congestion $\rho(p) = 1-(1-\bar\eta)\frac{\mu}{p}$ ($k^*_3(p) < k_{min}$).
\item[{\em iii.}] The solution $k^*_3(p)$ crosses $k_{max}$ but disappears before reaching  $k_{min}$, and $k^*_1$ becomes the stable solution. In this case the congested phase appears continuously (only high-degree nodes are congested), but at some point another transition occurs that brings the system abruptly into the completely congested state.
\end{itemize}

In general, the exact phenomenology observed in the mean field and simulations depends strongly on the tail of the degree distribution, i.e. on the graph ensemble considered.

Note that in case of discontinuous transitions, the presence of an hysteresis phenomenon is associated to the stability of the two solutions $k^*_{1}(p)$ and $k^*_{3}(p)$. For instance, in case {\em ii} or {\em iii}, we start from the free-phase at low $p$, the system selects the solution $k^{*}_{3}(p)$ and follows it upon increasing $p$ until the solution $k_3^*(p)$ disappears. On the contrary, starting from the congested phase (large $p$) the system selects the solution $k_1^*(p)$ and remains congested until this solution disappears (see inset of Fig. \ref{fig1}).

\begin{figure}[h]
\begin{center}
\includegraphics[width=0.75\textwidth]{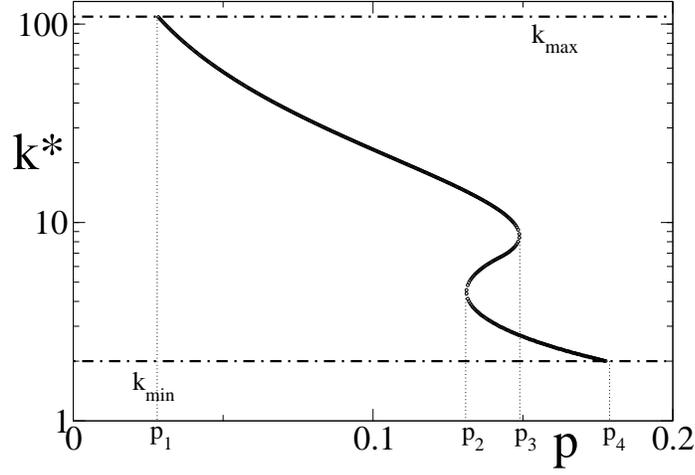}
\caption{The solution $k^*(p)$ for the scale-free graph of Fig.\ref{fig1}, with $\bar\eta=0.7$.
At $p_1$, $k^*=k_{max}$, and the system becomes partially congested in a continuous way.
Between $p_2$ and $p_3$ there are three solutions, two of them are stable.
Increasing $p$, the system jumps suddenly to a more congested state at $p_3$, whereas decreasing $p$, the system jumps to a less congested state at $p_2$. Above $p_4$ the system is completely congested.
} \label{keta}
\end{center}
\end{figure}

In Fig. \ref{keta} we can see the solution $k^*(p)$ for the same graph of Fig.\ref{fig1}, with $\bar\eta=0.7$:
at $p_1$, when $k^*=k_{max}$, the system becomes congested in a continuous way, at $p_3$ there is a discontinuous jump
to higher values of congestion, while above $p_4$ the network is fully congested and finally, coming back to $p_2$
there is a jump to a less congested state. Between $p_2$ and $p_3$ there is coexistence of high
and low congested states with hysteresis.

\begin{figure}[h]
\begin{center}
\includegraphics[width=0.75\textwidth]{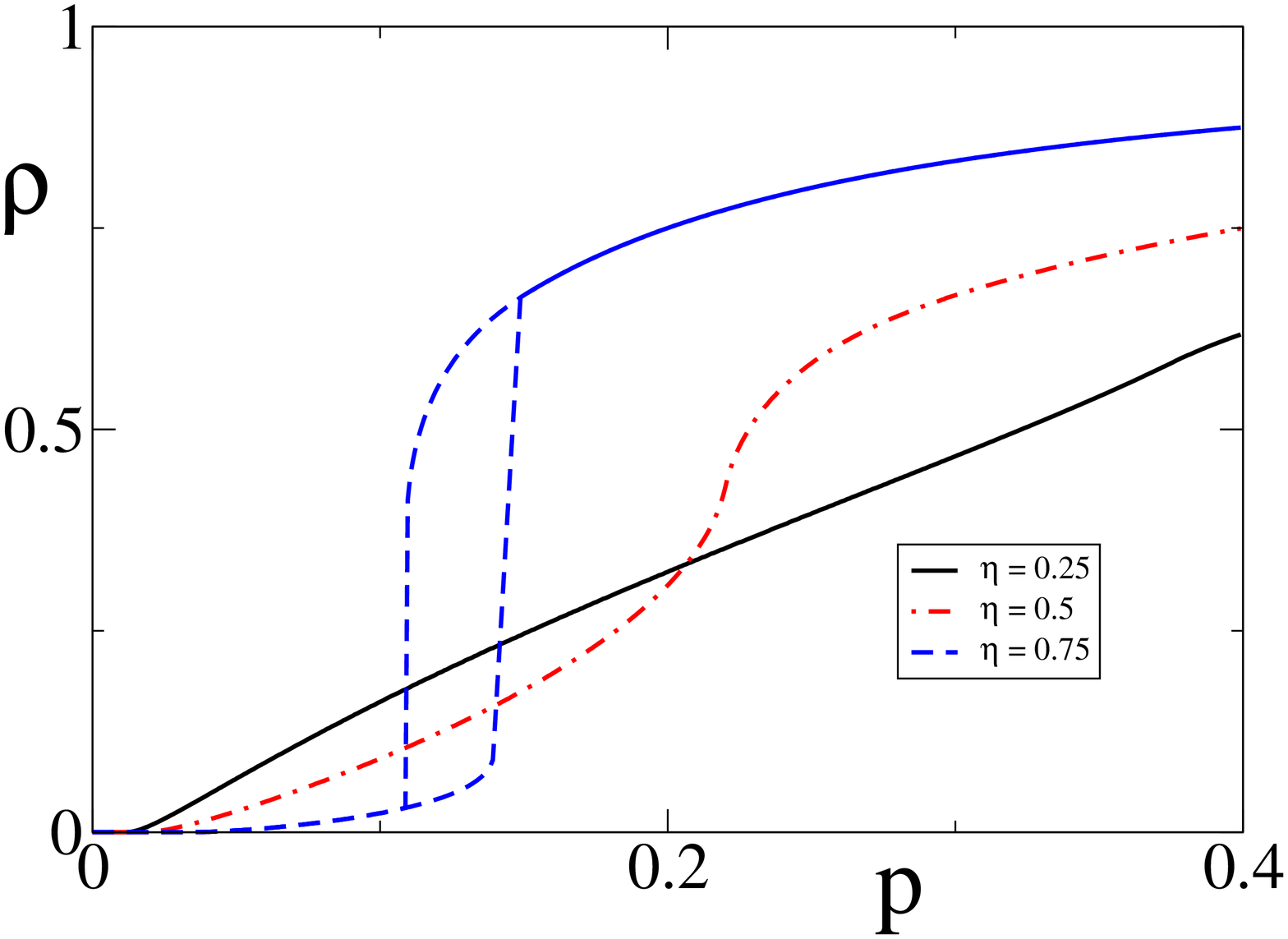}
\caption{Increasing $\bar\eta$, the congestion parameter $\rho(p)$ develops a discontinuous transition. Here we report the case of the graph of Fig.\ref{fig2}. For $\eta=0.75$, we have first a continuous, then a discontinuous transition.
} \label{fig_OP}
\end{center}
\end{figure}

In summary, the system can show a sort of hybrid transition: a continuous transition to a partially congested state followed by a discontinuous one to a (almost) completely congested one (see Fig.\ref{fig_OP}).

On heterogeneous random graphs, the behavior of the system in the plane $(\bar\eta, p)$ depends in a complex way on its
topological properties, such as the degree cut-off and the shape of the degree distribution. For this reason the precise location of the
critical lines, separating different phases, can be determined only numerically using the methods exposed in the previous section.
In the following, we give a qualitative description of the general structure of the phase diagram in the limit $n^* \to \infty$, then we substantiate the analysis reporting
an example of phase diagram obtained numerically for the same networks ensemble of Fig. \ref{fig1}.

A first important region of the space of parameters is the one in which a completely free solution exists, i.e. $k_{max} \le k_F$. This solution is characterized by $\bar{q} = 1-p/\mu$, $\bar\chi = 0$ and $\rho=0$. From the expression for $\dot{n}_k=0$ computed in $k_{max}$ we find that this happens as long as  $p\le p_{c_0}$ with
\begin{equation} p_{c_0} = \frac{\mu}{\mu + (1-\mu)\frac{k_{max}}{z}} . \end{equation}
Note that this region does not depend on the rejection probability $\bar\eta$, because rejection affects only congested nodes.

\begin{figure}[h]
\begin{center}
\includegraphics*[width=1\textwidth]{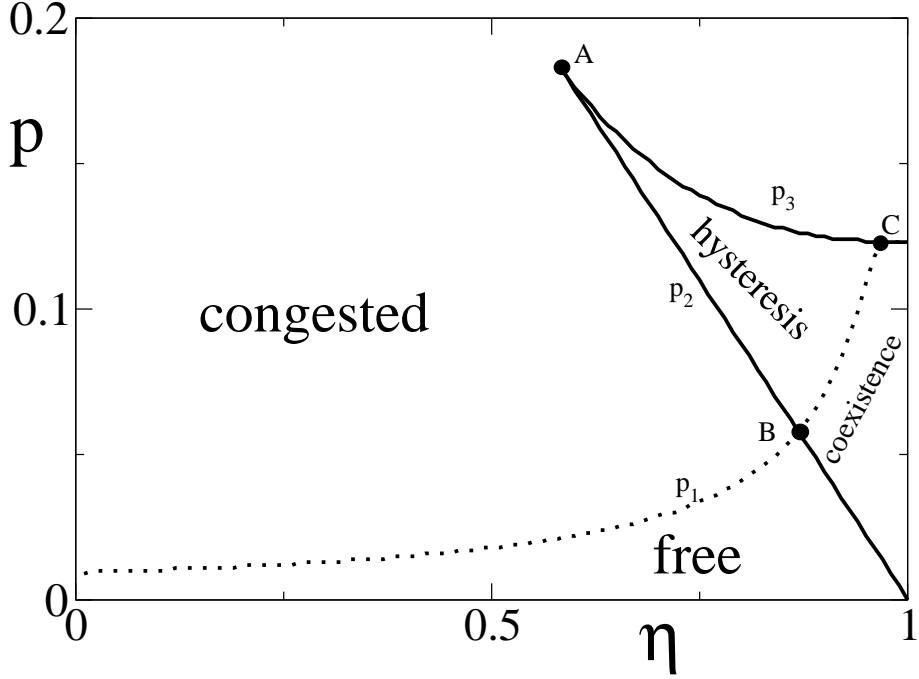}
\caption{
$(\bar\eta,p)$ phase diagram for the uncorrelated scale-free graph
of Fig. \ref{fig1}.
} \label{fig3}
\end{center}
\end{figure}

The transition takes place when the maximum degree nodes first become congested, i.e. $k^{*}=k_{max}$. Since $\dot{n}_{k^*} =0$, $q_{k_{max}}=0$ and $\chi_{k_{max}} = \bar\eta$, we get from Eq. \ref{dotn} a first expression for $p_{c} =  1-\bar\chi - \frac{k_{max}}{z}(1-\mu)(1-\bar\eta)(1-\bar{q})$. Now computing $\rho$ averaging Eq. \ref{dotn} and imposing  $\rho = 0$, we find a second expression for $p_{c} = \mu (1-\bar{q})(1-\bar\chi)$. Eliminating $\bar{q}$ from these two equations, we find the critical line
\begin{equation}
p_c(\bar\eta)=\frac{(1-\bar\chi)^2}{1-\bar\chi+\frac{k_{max}}{z}(1-\bar\eta)\frac{1-\mu}{\mu}}
\end{equation}
where $\bar\chi = \sum_{k\ge k_F} \frac{k P(k)}{z}\left(1-\frac{k_{max}(1-\bar\eta)}{k}\right)$.
Below this line (dotted line in Fig. \ref{fig3}) the system is not congested ($\rho = 0$), even if in the region $p_{c_0} \le p \le p_{c}(\bar\eta)$ higher-degree nodes are unstable ($k_F \le k_{max} \le k^*$). \\
It is possible to show that $p_c(\bar\eta)$ attains its maximum in $\bar\eta_c = 1-\frac{k_{min}}{k_{max}}$ where $p_{cmax} = \mu \frac{k_{min}}{z}$, where
$k_F=k_{min}$ and so above this point the curve is constant $p_c(\bar\eta>\bar\eta_c)=p_c(\bar\eta_c)$.

The transition line $p_c(\bar\eta)$ corresponds to the point $p_1$ in Fig. \ref{keta}, calculated for all values of $\bar\eta$.
We can calculate the two curves $p_2(\bar\eta)$, $p_3(\bar\eta)$ as well, in order to get the points at which there are discontinuous jumps in the congestion parameter $\rho(p)$.

Looking at Fig. \ref{fig3} we can distinguish three points A, B, C dividing the phase diagram into different regions:
\begin{itemize}
\item[{\em i.}] Below $\bar\eta_A$ we have a continuous transition to a congested state increasing $p$ above $p_1$.
\item[{\em ii.}] Between $\bar\eta_A$ and $\bar\eta_B$ the transition is continuous at $p_1$. Then, increasing $p$ above $p_3$,
there is a discontinuous jump to a more congested state. Coming back to lower values of $p$, there is a discontinuos jump
to a less but still congested state at $p_2$, and the system eventually becomes free below $p_1$ in a continuous way.
\item[{\em iii.}] Increasing $p$ in the region between $\bar\eta_B$ and $\bar\eta_C$, there is a continuous transition
from free-flow to a congested state at $p_1$, and a sudden jump to a more congested phase at $p_3$; but, this time, by
decreasing $p$ from the congested state, the transition to the free phase is discontinuous and located in $p_2$.
\item[{\em iv.}] For $\bar\eta > \bar\eta_C$ the transition is a purely discontinuous one with transition points $p_2$ and $p_3$.
\end{itemize}

Increasing $p$ above the transition, at some point $p_{c_1}(\bar\eta)$ the system becomes completely congested. For $p \ge p_{c_1}(\bar\eta)$, the order parameter follows the curve $\rho = 1-\mu(1-\eta)/p$. This happens for
 $p \ge p_{c_1}(\bar\eta) = (1-\bar\eta)(1-(1-\mu) k_{min}/z)$, where $k^* \le k_{min}$, $q=0$, $\chi=\bar\eta$.

These calculations show that the phase diagram crucially depends on the tail of the degree distribution.
In scale-free networks $k_{max}$ scales with the network's size $N$ as $N^{\frac{1}{\omega}}$ with $\omega = 2$
(structural cut-off) or $\omega = \gamma-1$ (natural cut-off).
Accordingly the critical line depends on the system's size, $p_{c} \propto N^{-\frac{1}{\omega}}$.
The only region that does not depend on $k_{max}$ is the one for $\bar\eta \ge \bar\eta_C$.

\subsection{Conclusions}
The model discussed above, inspired by the recent literature on congestion on complex networks,
basically extends the classic framework of Jackson queuing networks along three lines:
\begin{itemize}
\item
It goes beyond stationarity, exploring \emph{congested} regimes, where the queues can grow.
\item
It accounts for congestion control protocols and this requires that the absorption of packets
takes place during the hoppings and not within the queue.
\item
It exploits graph ensemble calculation techniques, allowing the study
of how traffic is affected by the general features of the underlying network.
\end{itemize}
Within this framework it is possible  to obtain transition curves and phase diagrams at analytical level for the
ensemble of uncorrelated networks and numerically for single instances.
We found that traffic control improves global performance, enlarging the free-flow region
in parameter space only in heterogeneous networks.
In very heterogeneous networks, e.g. with scale-free degree distribution, its role should be crucial,
since for low enough traffic control the critical packets inserction rate per node goes to zero with the system size.
Traffic control introduces non-linear effects and, beyond a critical
strength, may trigger the appearance of a congested phase in a
discontinuous manner.
This work can be extended in several interesting directions.
First, it should be interesting to study how the dynamics change once considering a bias in the random routing, e.g.
an hopping probability proportional to the betwennes centrality of the
neighbouring nodes\footnote{The betweennes centrality of the node $i$ is $\sum_{j \neq i,k \neq i} \frac{n(j,k,i)}{n(j,k)}$,
where $n(j,k)$ is the number of shortest paths between $j$ and $k$, and $n(j,k,i)$ is the number of them that pass through $i$.}, to better mimick shortest path routing.
The possibility of solving the model on a given network with realistic parameters
could provide both specific predictions for the robustness of the network to traffic overloads and important hints for the design of systems
less vulnarable to congestion. The dynamical environment created within this model could be also axploited as a framework for testing the statistical properties of single particle
dynamics under more complex routing schemes, like the study of tracking particles in hydrodynamics.
Finally, it would be interesting to model the complex adaptive behavior of human users in
communication networks, such as the Internet, by introducing variable rates of packets production in response to network performances.
It is known that users face the social dilemma of maximizing their own communiction rates, maintaining the system far from the congested state \cite{HL97}.
In such a situation, the presence of a continuous transition may allow the system to self-organize
at the edge of criticality, whereas a discontinuous transition may have catastrophic consequences.

\chapter{Dynamical arrest on disordered structures}
\emph{"The deepest and most interesting unsolved problem in solid state theory is probably the theory of the nature of glass and the glass transition. 
This could be the next breakthrough in the coming decade. The solution of the problem of spin glass in the late 1970s had broad implications in unexpected 
fields like neural networks, computer algorithms, evolution and computational complexity. 
The solution of the more important and puzzling glass problem may also have a substantial intellectual spin-off.  
Whether it will help make better glass is questionable"}.\\
P.W.Anderson, Science (1995).

Fifteen years have passed since this statement, and the dramatic slowing down of the dynamics of glass forming systems is still puzzling us\cite{biroli}.
Its study requires a deep reasoning on the fundamentals of statistical mechanics, 
and methods and concepts developed in this field are likely to  become paradigms 
for the study of complex systems in general.
In this chapter I will show how a certain degree of fixed heterogeneity, e.g. in the underlying spatial structure, 
can change the character of the jamming transition in glass forming systems.
In glass science a great deal of efforts is devoted to understanding the dynamical properties of supercooled liquids.
Relaxation and transport properties of such a state are subject to a dynamical crossover upon decreasing temperature. 
There is an anomalous relaxation with heterogeneous patterns in space and time that are the signature of strongly cooperative effects.
At a mean-field level, this crossover becomes a true phase transition.
After a brief introduction on the phenomenology of glass forming systems, 
we will review the theoretical perspectives on it, from thermodynamical to purely dynamical approaches. 
In particular I will expand on the spin facilitated model by Frederickson and Anderson.
Within the framework of this model it is possible to recast the 
dynamical jamming transition in terms of a bootstrap percolation scenario.
Then, I will show how a certain degree of fixed heterogeneity, 
being it encoded as a simple dilution of the underlying lattice, 
or as a distribution in mobilities,
can dramatically change the collective behavior from  bootstrap to simple percolation scenario. 
This can give insights on analogies and differences among the jamming of supercooled liquids and more heterogeneous systems, 
like polymer blends and confined fluids.

 \section{Main experimental features of the glass transition}
\begin{figure}[h!!!!!!!]
\begin{center}
\includegraphics*[width=0.75\textwidth]{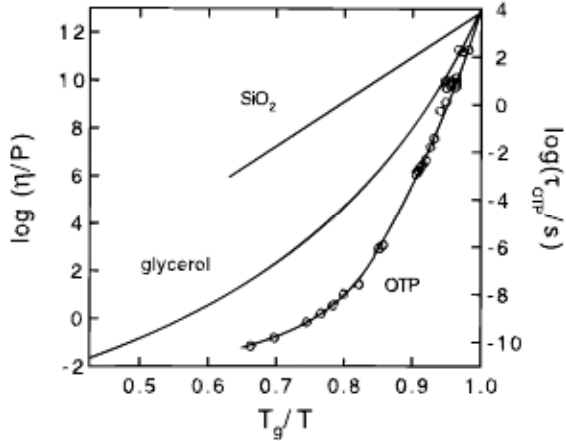}
\caption{Viscosity as a function of reduced inverse temperature for three liquids: $SiO_2$, glycerol and $o$-terphenyl.
For the $o$-terphenyl are also shown the typical time of reorientation of molecules. From\cite{angell}}
\label{angell}
\end{center}
\end{figure}

If we cool a liquid fast enough, it can avoid crystallization 
entering in a metastable, \emph{supercooled} state \cite{cavagna}.  
The typical timescales of the relaxation and transport properties 
of such a state dramatically increase once we further cool it.
Fig\ref{angell} shows the shear viscosity of some supercooled liquids as a function of the temperature, divided 
by the temperature at which it
becomes of the order of $10^{14}P$.
This temperature $T_g$ is called glass transition temperature and it is weakly dependent on the cooling rate.

All these curves are fitted well by the Vogel-Fulcher-Tamann law (VFT):
\begin{equation}
\eta \propto e^{\frac{A}{T-T_k}}
\end{equation}
It is possible to distinguish \emph{strong} and \emph{fragile} behaviors. 
The former is consistent with $T_k = 0$, and $A$ can have in this case the meaning of an energy activation barrier. 
The latter has a true divergence at $T_k>0$, and the  
typical energy scale to relax continuously increases upon decreasing temperature. 

Near $T_g$ the typical timescale to relax at equilibrium 
exceeds the experimental one and the system is practically out of equilibrium. 

\begin{figure}[h]
\begin{center}
\includegraphics*[width=0.75\textwidth]{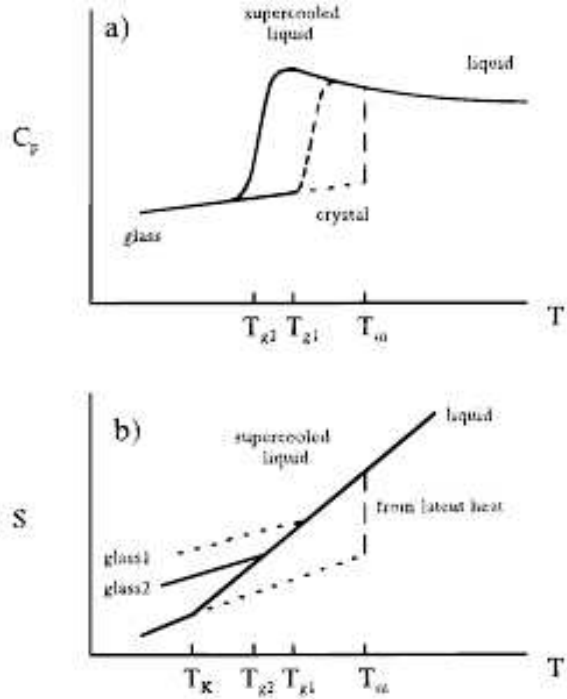}
\caption{Top: a schematic plot of the temperature dependence of the specific heat for a liquid.
Avoiding the melting point doesn't change its fate: at a certain point (cooling-rate dependent) there is a drop towards a solid-like dependence.
Bottom: The extrapolated entropy of a supercooled liquid below the glass transition point equals the entropy of the crystal}
\label{angell2}
\end{center}
\end{figure}

In these range of temperatures there is  a drop of the specific volume 
and of the specific heat towards the same value of the crystal (see fig.\ref{angell2}).
It is possible to calculate the entropy of the supercooled liquid  and
to extrapolate it below $T_g$: at a certain point $T_k'$ its value equals that one of the crystal\cite{Kauz}. 
There are strong empirical evidence in favour of the fact that $T_k = T_k'$ \cite{richangel}. 
Therefore $T_k$ should represent an infinite cooling rate limit of $T_g$, where there
should be concomitantly a thermodinamic singularity and a divergence of the relaxation time. 
Hence, is it possible to speak of the glass transition in terms of a truly thermodynamic phase transition? 
Unfortunatively, the density-density correlation function, 
or its fourier transform, the structure factor\footnote{This quantity can be directly measured through scattering experiments.}, 
$F(\vec{k}) = \langle \sum_i e^{i \vec{k} \cdot \vec{r}_i} \rangle$ 
doe not show any interesting change when decreasing the temperature. 
On the other hand, the relaxation of its time-dependent version $F(\vec{k},t)$,
the intermediate scattering function, 
shows very interesting features upon approaching $T_g$ from the supercooled phase, 
with heterogeneous patterns in space and time.
 
Therefore, even if the static structure of the system doesn't show
when decreasing the temperature any intereasting change,
from the dynamical point of view, interesting phenomena are taking place.

\begin{figure}[h]
\begin{center}
\includegraphics*[width=0.8\textwidth]{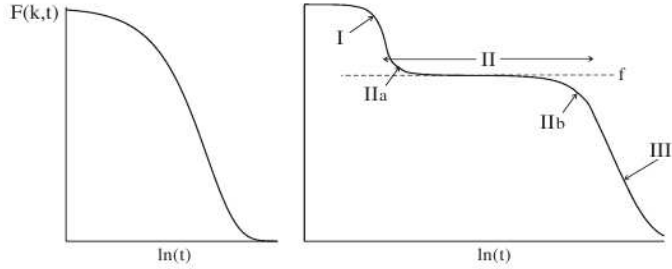}
\caption{A schematic plot of the relaxation in time of the intermediate scattering function. Left: high temperature, liquid phase, exponential relaxation.
Right: at lowering temperature the function relaxes in two steps, first towards a plateau, the $\beta$ step, and then,  
after a while, to zero, the $\alpha$ step.}
\label{relax}
\end{center}
\end{figure}

\begin{figure}[h]
\begin{center}
\includegraphics*[width=0.75\textwidth]{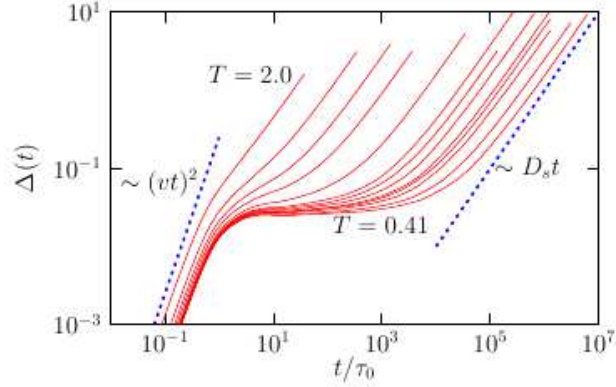}
\caption{Average displacement in time of particles by simulations of Lennard-Jones spheres 
for several temperature. Upon decreasing temperature, the division the between short-time ballistic regime
and long-time diffusive one is given by a plateau. From \cite{biroli}}
\label{displ}
\end{center}
\end{figure}

The relaxation of $F(\vec{k},t)$ at low temperatures is indeed not exponential, 
rather it procedees by two step (see fig\ref{relax}). First it approaches 
a plateau, the $\beta$ relaxation, and then it departes from it towards the equilibrium value, 
the $\alpha$ relaxation. 
The height of the plateau starts discontinuously from a value larger than zero at a certain temperature. 
This is usually related to the miscroscopic motion of the particles of the system. 
If we follow the average displacement in time (see fig\ref{displ}) of particles, 
at high temperature we a have a sharp crossover from a ballistic ($ d \propto t$) to a diffusive regime  ($ d \propto \sqrt{t}$).
At low enough temperature this two regimes are separated by a plateau. 
This means that a particle is trapped for a while by the cage formed by its neighbours, where it vibrates. 
A picture confirmed by numerics and experiments as well\cite{cage}. 
The motion within the cage is related to the $\beta$ step, the rearrangment of the cages should be related with the $\alpha$ one. 
Finally, the $\alpha$ step itself is not exponential, being fitted by a stretched exponential formula $\exp^{-(t/\tau)^\beta}$.
This last trend is usually ascribed to a certain degree of dynamical heterogeneity: 
different parts of the same system can have different relaxation patterns, 
this causing in turn a deviation from the exponential.
The microscopic resolution of the dynamics with numerical simulations has shown strong correlation patterns, like e.g. the clustering of more mobile particles (see \cite{biroli} and ref therein).
\begin{figure}[h]
\begin{center}
\includegraphics*[width=0.35\textwidth]{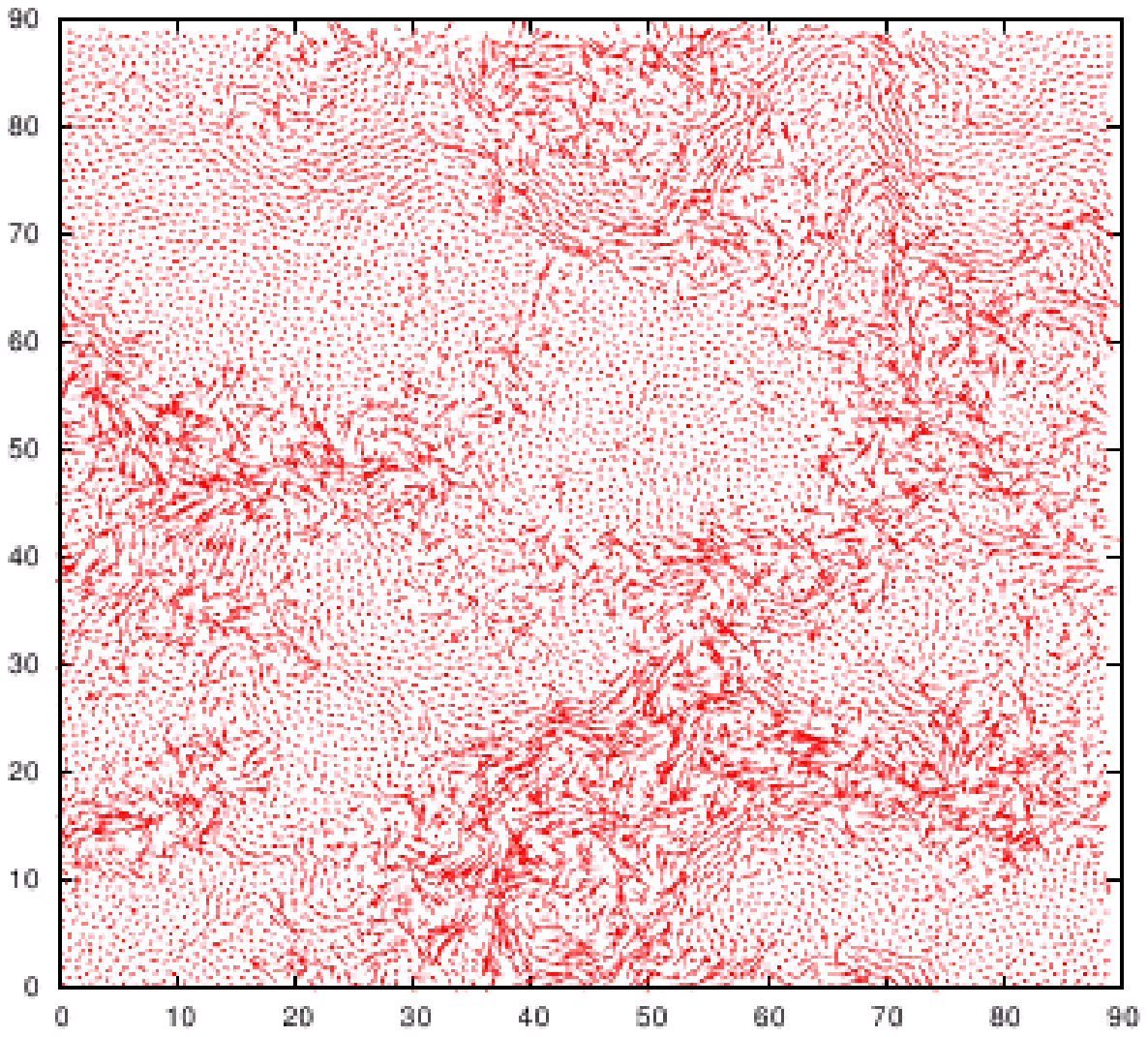}
\includegraphics*[width=0.6\textwidth]{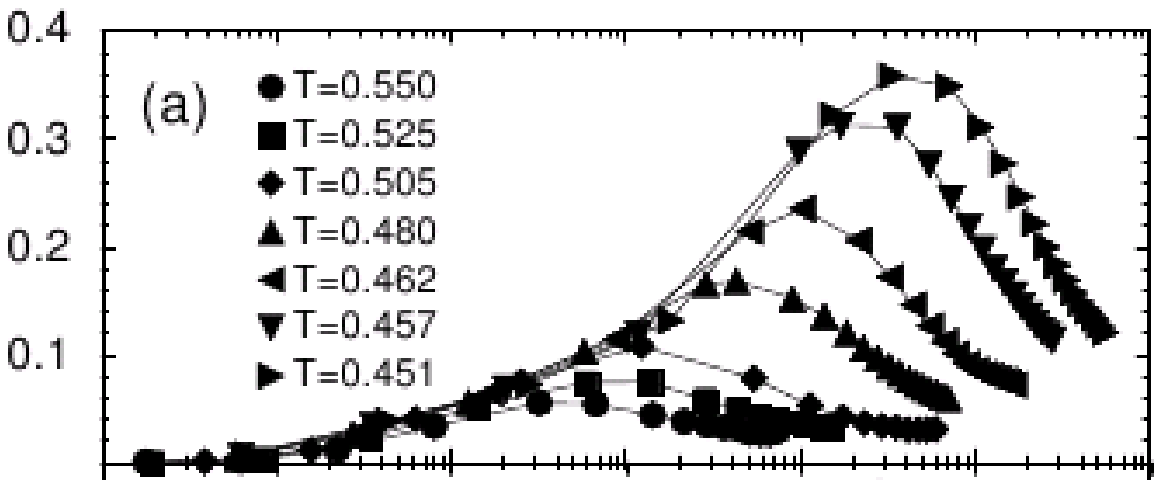}
\caption{Left: spatial map of single particle displacements from simulations of a bidimensional Lennard-jones system. It is possible to recognize qualitatively
different mobilities and the presence of non trivial correlations in the motion. From \cite{biroli} 
Right: The dynamical susceptibility 
develops a maximum in time whose height increases upon decreasing temperature, from simulations of a Lennard-Jones system in\cite{ofrat}}
\label{ofrat}
\end{center}
\end{figure}
The fluctuations of the intermediate scattering function, i.e. the dynamical susceptibility, develops a maximum in time whose height increases upon decreasing the temperature. 
This maximum is related with the typical size of the correlated regions, thus defining a dynamical correlation lenght that diverges together with the relaxation time\cite{biroli}.

\subsection{Other-than-molecular glass formers}
One interesting point, from a statistical mechanics perspective, 
is about the generality of such phenomenology.  
Interestingly enough, the same phenomenology, 
with a dramatic slowing down of the dynamics and complex patterns of relaxations in space and time, 
is shared by systems whose interacting units are of different scales 
from the molecular ones like colloidal suspensions, granular media and polymer solutions.

Colloidal suspension\cite{collozacca} consist of big particles in a solvent,  
with typical sizes of $1-500$ nm. The continuous scattering events with the much smaller particles of the solvent renders the dynamics 
of such particles brownian, 
with diffusion time of the order of $1$ ms.  
They can be modeled as hard spheres, with an interaction potential
that is infinite below a certain distance $2R$ and zero otherwise. 
In this case the temperature has the only role of rescaling times and 
what matters is the density $\rho$, or alternatively, the packing fraction $\Phi = 4/3 \pi R^3 \rho$. 
At increasing $\Phi$ the viscosity and/or the relaxation times
of the system dramatically increases, and
at $\Phi_g \simeq 0.58$ the relaxation times exceed the experimental ones and the system is jammed.
The system is now an elastic amorphous solid, the gel. 
All this remember the already seen phenomenology of the glass transition, and, in fact, it is found
that a VFT formula is a good fit for the dependence of relaxation times on density, 
the dynamics of the correlation function has a two step relaxation and there is a certain degree of spatial dynamical heterogeneity.
Another kind of systems that show jamming are granular media\cite{unify}. 
They consist of large ($N = 10^2$-$10^6$) assembly of macroscopic particles, 
from powder($10^{-5}$ m) to rocks ($10$ m).
Because of their macroscopic scales, the thermal energy has no role 
and they have to be vibrated, sheared, etc, 
by an external source to explore the phase space. 
Therefore, there should  be a continuous injection of energy that is continuously 
dissipated by friction, a force that play a key role in these systems.
Their phenomenology is very rich. 
In particular, depending on the strength of the driving force,
their properties can  be seen as similar to that one of 
the usual phases of matter, with transition among them. 
That's why it is common to speak of granular gases, liquids and solids\cite{nagel}.
An interesting point comes up: when the  grains are in the ``solid'' phase 
they are usually not arranged in a regular and/or crystalline structure.
The solid is amorphous and the transition from the fluid phase
is a dynamical arrest with a complex relaxation pattern, 
as can be seen e.g. in compaction-by-vibration experiments\cite{compaction}.
Interestingly enough, also polymer solutions have a jamming transition. 
When decreasing the temperature/increasing the density, the dynamics of these systems is slowed down with 
a dramatic increase of the viscosity, till the system becomes an elastic solid, a gel\cite{degennes}.
There is clear dynamical crossover with the intermediate scattering function 
showing a stretched exponential decay.
However, at odds with  simple liquids, this sol/gel phase transition is well known.
When decreasing the temperature/increasing the density, 
the polymers stick together, forming a network that at a certain point can span the whole
system, a process that is called \emph{percolation}. 
The study of such phenomenon opened the huge field of percolation theory,  
a kind of general and geometrical view on phase transitions\cite{stauffer}.
In the simplest percolation scenario we have a lattice, whose bonds can be either empty or occupied with some probability $p$.
At low $p$, the lattice is disconnected in clusters of finite size. 
Upon increasing $p$, at a certain point $p_c$, there is a continuous transition by which
one of the clusters span the whole lattice, that is now connected. 
The average clusters'size defines in this case naturally a correlation length\cite{abete}. 
It is interesting to notice that there are numerical evidences that 
this simple percolation scenario for a dynamical arrest seems 
to be present even in super-cooled liquids once we confine them\cite{confi}.

\section{Theoretical views on the glass transition}
As we pointed out before, an intriguing qualitative difference in the dynamics of the supercooled liquid, 
that should give insights about mechanisms underlying the dramatic slowing down, 
is the anomalous two steps relaxation of correlation functions. 
A picture by Goldstein\cite{golde} tries to explain it in terms of a dynamics in the phase space ruled by activated processes.
In a  certain range of temperatures the energy landscape visited by the system is composed of many local minima. 
The dynamics should consist of vibrations within one minimum, the $\beta$ step, and then jumps among minima, the $\alpha$ step. 

On the other hand, it is possible to write down equations for the correlation function, 
and by means of suitable approximations to solve them. 
This is the framework of the mode coupling theory (MCT)\cite{MCT1},
that gives many interesting insights and experimentally proved results for the high temperature regime of supercooled liquid\cite{MCT2}.
It predicts quantitatively the features of the relaxation of correlation functions, with a two step relaxation, 
the development of a plateau in a discontinuous way and increasing fluctuations.
But, at odds with the real liquid, at a certain point $T_c$ the correlation function sticks to that plateau.
Hence, the main drawback of this theory is the wrong prediction of a singularity in the dynamics
with a power law divergence of the relaxation time at $T_c>T_g$.

Interestingly, the approximated equations of this theory are exactly the same of a mean-field model of spin glass: 
the p-spin spherical model\cite{cavagna2}.
Spin glasses are disordered materials whose magnetic properties show interesting behaviors that rensemble very close that ones of glasses.
They are usually modeled by classical spins on lattice, whose interactions can be both antiferromagnetic and ferromagnetic.
These model systems are characterized by the phenomenon of frustration.
It is impossible or extremely hard to satisfy all the interaction terms in the Hamiltonian,
and this gives rise to a very complex energy landscape, full of minima and saddles.
This  is often a distinctive feature of complex systems in general,  
and concepts and methods used for spin glasses are currently used in fields as different as biology 
(neural networks) and information theory (algorithmic complexity)\cite{parisi}.
In the p-spin spherical model $N$ continuous spins $\sigma_i$ interact by p-body terms, the hamiltonian being:
\begin{equation}
H = - \sum_{i_1,...,i_p} J_{i_1,...,i_p} \sigma_{i_1}   \cdots \sigma_{i_p} 
\end{equation}
where the ${J}$ are quenched\footnote{The interaction terms  are slowly changing with respect to the spin variables, 
i.e. they are fixed once for all. In the thermodynamic limit, the average over all the possible configurations of interactions 
of extensive quantities, like the free energy, should give the same result of a given single instance.} 
random variables with a gaussian probability distribution of zero mean,
and the spins are subject to the spherical constraint $\sum_i \sigma_i^2 = N$. 
Most of these model systems in general, and the p-spin model in particular, are subject, upon lowering $T$, to a dynamical phase transition
and moreover they have a truly thermodynamic singularity, the replica simmetry breaking phase transition
\footnote{A replica is a copy of the system with exactly the same realization of quenched disorder, if any.
Actually replicas were first introduced as a trick for calculations.}.
The first, corresponding  exactly to the singularity of the MCT, 
it is an extreme case of the already mentioned Goldstein scenario.
The dynamics is ruled by activated processes, i.e jumps among local minima of the energy, whose number is exponential in the system size.
This crossover becomes a truly phase transition because, at mean field level, the barriers among minima are infinite in the thermodynamic limit.
The second is static and it is characterized by ergodicity breaking.

The phenomenology of the p-spin model seems to give a good metaphor for the dynamics in phase space of structural glasses.
Therefore, it has recently inspired replica-based approaches for Lennard-Jones fluids\cite{parisi2} and hard-sphere systems\cite{parisi3}
\footnote{However, in finite dimension the scenario is even more complex: 
different parts of the same system could  be in different minima, with a characteristic size $\xi$ for these domains}. 
However spin glasses are different from structural glasses, the main difference
being the presence of quenched disorder.

An alternative approach is to look at glassiness from a pure dynamical perspective, 
without recurring to a complex energy landscape scenario.
This approach is based to the study of lattice models with simple hamiltonian and trivial equilibrium behavior,
but whose dynamics is subjected to some kinetic constraints, such that they can show glassy relaxation patterns\cite{ritort}.
These models can give deep and useful insights about the miscroscopic mechanisms of the first step of the glass transition, the dynamical crossover.
It should not be forgotten that the equilibrium dynamical properties of the supercooled system around this crossover 
are accessible to experiments. 
Below it, the experimental investigation of the thermodynamic properties requires excedeengly large times.
In particular, within their framework the question of how  
the underlying spatial topology affects the dynamics can be directly addressed and easily analized,
as we shall see soon for a particular case.

\subsection{The Frederickson-Andersen model}
One of the first kinetically constrained model was introduced by Frederickson and Andersen (FA)\cite{fred}.
On top of each site $i$ of a lattice there is a classical $1/2$ Ising spin $s_i$ that can be $1$ or $-1$. 
The spins are uncoupled and there is a global magnetic field  of strenght $1$ pointing up, 
the Hamiltonian being simply $H = -\sum_i s_i$.
The static properties are very simple and the stationary probability distribution function 
of the spin configurations factorizes  $P(s_1 ,..., s_N) = \prod_i p_i(s_i)$. 
The dynamics is characterized by an additional constraint:
a spin can flip if at least $f$ of its neighbours are down. 
Down spins should represent region with high mobility such that they trigger the  relaxation of their neighbours.
This rule doesn't violate detailed balance but it can trigger a dynamical arrest.
Upon decreasing the temperature, at a certain point, 
the system cannot relax because a finite fraction
of the spins doesn't flip anymore, i.e. they are frozen.
A good parameter to characterize this transition is thus the fraction of blocked spins $\Phi$ as a function of the temperature.

It is possible to analize this model at a mean field level onto a bethe lattice of degree $z$ \cite{sellitto}. 
This lattice can be seen as the infinite size limit of  a Cayley tree, 
i.e. the graph obtained starting to branch from a seed node
with a constant branching $k=z-1$, or of a random regular graph, i.e. a graph taken 
from the ensemble of all the graphs whose nodes have degree $z$, 
all with equal statistical weight. 
The first is a tree but it has strong boundary effects, the second is locally a tree, 
having loops whose lenght scales with the logarithm of the system size. 

Let us call $B = P(A)$ the probability of the event $A$: 
the spin at the end of a random link is in the state $-1$, or it can flip to this state by rearranging the $k$ sites above it. 
$B$ verifies the iterative equation:
\begin{equation}
B = (1-p) + p \sum_{i=0}^{k-f} \binom{k}{i} B^{k-i} (1-B)^{i}
\label{bootstrap}
\end{equation}
where $p=\frac{1}{1+e^{-1/T}}$ is the probability that the spin is $+1$ at equilibrium. 
The term $1-p$ on the rhs is the probability that the spin is already in the state $-1$. 
The other term is the probability that the spin is in the state $+1$ ($p$) and that
it can flip by rearranging the neighbours (the sum). 
The sum is thus the probability that the event $A$ is not verified for at most $k-f$ out of $k$ neighbours, 
i.e. that $A$ is verified at least for $f$ of them, that is, the constraint is satisfied.
There is always a  solution $B=1$. For $f=1$ it is the only solution. 
For $f>1$, the so-called cooperative cases, we can have a fixed point $B<1$.
We define $x=1-B$ that verifies the equation:
\begin{equation}
x = p \sum_{i=0}^{f} \binom{k}{i} x^{k-i} (1-x)^i
\end{equation}
The parameter that distinguish jammed from free phases is the fraction of spins permanently frozen $\Phi$:
\begin{eqnarray}
h = \sum_{i=0}^{f-2} x^{k+1-i} (1-x)^i \\
\Phi = p \sum_{i=0}^{f-1} x^{k+1-i} (1-x)^i +(1-p)\sum_{i=0}^{f-1} (ph)^{k+1-i} (1-ph)^i 
\end{eqnarray}
The two contributions are the probability that a spin is frozen in the $+1$ or $-1$ state, respectively. 

Let's analize the case $f=2$, $k=3$. We have the trivial solution $x=0$ and  $1 = p x(3-2x)$. The critical value at which the transition takes place
is $p_c =8/9$, or $T_c \simeq 0.48$, where $\Phi$ jumps discontinuosly from $0$ to $\Phi_c \simeq 0.67$, such that $\Phi-\Phi_c \simeq (T_c-T)^{1/2}$.
The dynamics of relaxation at equilibrium is analized in terms of the persistence function $\phi(t)$, i.e. the fraction of spins that do not flip till time $t$.
This is a measure of the self correlation of the system and $\lim_{t \to \infty} \phi(t) = \Phi$. 
Looking at the temporal trends of $\phi$ in fig.\ref{selli}, left, we can see how effectively, it has an exponential behavior
at high $T$, then it deviates from it, starting to develop a plateau upon lowering $T$, till $T_c$, where it sticks to the plateau concident with the value of $\Phi$.
The integral of $\phi$ gives an estimate of the typical relaxation time, that diverges at $T_c$ with exponent $\gamma \simeq 3$ (see fig.\label{selli}, right).

\begin{figure}[h!!!!!!!!!]
\begin{center}
\includegraphics*[width=1\textwidth]{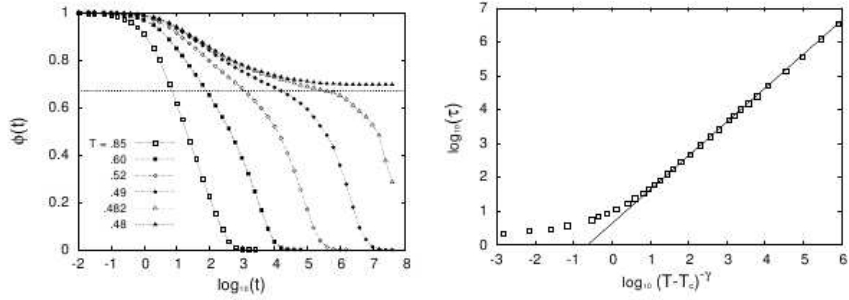}
\caption{left: Time trends of the persistence function $\phi(t)$. 
Upon decreading temperature, it has a two step relaxation, developing a plateau in a discontinuous way. The straight
line is calculated analitically. 
Right: Power law divergence of the integral relaxation time as a function of the temperature at $T_c$. Simulations from\cite{sellitto}}
\label{selli}
\end{center}
\end{figure}

\begin{figure}[h]
\begin{center}
\includegraphics*[width=0.6\textwidth]{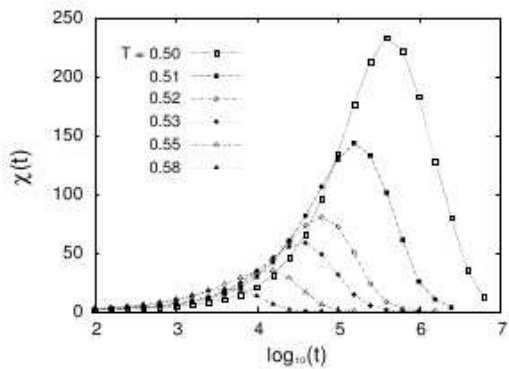}
\caption{Time trends of the dynamical suscieptibility. A maximum is developing, whose height increases at decreasing temperature. Simulations from \cite{sellitto}}
\label{selli2}
\end{center}
\end{figure}

The fluctuations of $\phi(t)$ show critical behavior upon approaching the $T_c$. 
Fig \ref{selli2} shows the dynamical susceptibility $\chi(t) = N (\langle \phi(t)^2 \rangle -\langle \phi(t) \rangle ^2)$ 
that develops a maximum whose height is diverging at $T_c$.

This dynamical phase transition is thus called {\em hibrid}, 
because the parameter $\Phi$ jumps discontinuously at the transition point to a finite value, but it has critical fluctuations 
and a well defined exponent for the value of $\Phi-\Phi_c$ upon approaching the $T_c$.

Even if an exact mapping is still missing, this jamming scenario is in very good agreement 
with the dynamical phase transition of the simplest MCT and of the p-spin spherical model.

\subsubsection*{The bootstrap percolation problem}

\begin{figure}[h]
\begin{center}
\includegraphics*[width=0.6\textwidth]{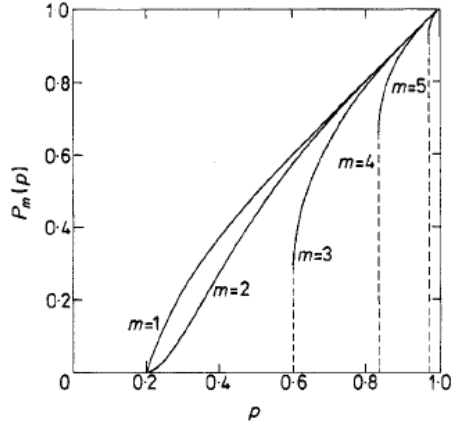}
\caption{Relative size of the BP m-cluster as a function of the density $P_m(p)$, for several $m$ for a bethe lattice of degree $z=6$.}
\label{chalupa}
\end{center}
\end{figure}

The FA model can be mapped onto the bootstrap percolation (BP)\cite{bootstrap} problem.
In BP, each site is first occupied with a particle with probability $p$, then, particles with less then $m$ neighbouring particles are removed.
Iterating the procedure, we can end up with a remaing m-cluster of particles or not, depending on $p$, the initial density.
This model can be analized exactly on a Bethe lattice of degree $z$.
We let $R$ be the probability that an occupied site $i$ is not connected to an infinite m-cluster containing its nearest neighbor $j$.
It can be so if $j$ is not occupied (w.p. $1-p$) or if less than $m-1$ of the other neighbors of $j$ are also not in the m-cluster.
Thus we find that:
\begin{equation}
R =1-p + p\sum_{i=0}^{m-2} \binom{z-1}{i} R^{k-i} (1-R)^{i}
\end{equation}
And this is the same equation of $B$ if 
\begin{equation}
m = z-f+1 
\end{equation}
There is always a  solution $R=1$.
Depending on $p$ we can have a fixed point $R<1$.
The case $m=2$ has the same equation of the ordinary percolation problem\cite{stauffer}.
The remaining cluster in the $m=1$ case is the  same of $m=2$ case with the adjoint of dangling bonds,
i.e. the chain structures connected to the 2-cluster.
Below the transition point of the $m=2$ case, in the $m=1$ case, there are still remaining clusters, but they are disconnected chains
whose relative size decreases to zero in the thermodynamic limit.
The fraction of sites in the m-cluster, or the probability that a site is part of it, $P_m $, has the form:
\begin{equation}
P_m = p \sum_{i=0}^{z-m} \binom{z}{i} R^{i} (1-R)^{z-i}
\end{equation}
These equations can be solved along the same lines of the FA model.
Fig.\ref{chalupa} shows the transition curves $P_m(p)$ for a bethe lattice with connectivity $z=6$.
We have a continuous transition $P_m \propto (p-p_c)^\beta$ for $m=1,2$ with exponents $\beta =1,2$ respectively.
For $m>2$ the transition is discontinuous $P_m -P_{mc} \propto (p-p_c)^\beta$ with exponent $\beta =1/2$.

\section{The heterogeneous FA model}
As we have seen, it is possible on a random regular graph to recast the jamming in the FA as a bootstrap percolation transition. 
In particular, depending on the facilitation parameter $f$
of the model, and on the degree $z$ of the underlying lattice, 
it is possible to have a bootstrap or a simple percolation scenario, 
if $m=z-f+1$ is larger or not than $2$ respectively.
It is possible to consider  situations 
in which $m$ is varying from site to site, i.e. considering different degrees and/or facilitation parameters\cite{branco}.
Let us consider a diluted version of the previously analized lattice, i.e. a trimodal random graph with degree distribution 
$P(k) = u \delta_{k,2} + (q_1-u) \delta_{k,3} + (1-q_1) \delta_{k,4}$, the average degree is $z = 4-q_1-u$.
Let us\cite{ourfred} put on top of such a lattice the FA model with facilitation parameter $f=2$ \cite{ourfred}.
We can extend the equation \ref{bootstrap} to general heterogeneous random graphs
considering the probability $B_k$ that a spin verifies $A$ and it has $k+1$ neighbours,
\begin{equation}
B_k = (1-p) + p \sum_{i=0}^{k-f} \binom{k}{i} B^{k-i} (1-B)^{i}
\end{equation}
And $B = \sum_k \frac{(k+1)P(k+1)}{z} B_k$ is the average of $B_k$ over the degrees, that verifies the equation:
\footnote{On a random graph with degree distribution $P(k)$, the degree distribution of a random neighbouring node is $P(k)k/z$ }
\begin{equation}
B = (1-p) + p \sum_{k=0}^{\infty} \frac{(k+1)P(k+1)}{z} \sum_{i=0}^{k-f} {k \choose i} B^{k-i}(1-B)^i 
\end{equation}

or, in terms of $x=1-B$:
\begin{equation}
x = p \sum_{k}^{\infty} \frac{(k+1)P(k+1)}{z} \sum_{i=0}^{f-1} {k \choose i} x^{k-i} (1-x)^i
\end{equation}

The equations for $f=k+1$ and $f=k$ are the same, and they are the same of ordinary percolation. 
We have, finally (apart of the $x=0$ solution):

\begin{equation}
\frac{1}{p} = a (2-x) +(1-a)x(3-2x) 
\end{equation}

where $a = \frac{3q_1-u}{z}$. Then we have:
\begin{equation}
x = \frac{ 3-4a + \sqrt{9-8a-\frac{8(1-a)}{p}}}{4(1-a)}
\end{equation}

This solution exists if $p > p_{c1} = \frac{8-8a}{9-8a}$, and, if $a > a_c = 3/4$, it is positive until $p>p_{c2} =\frac{1}{2a}$.
Below $a_c$, we can expand around $p_{c1}$, $p=p_{c1}+\epsilon$, we have 
\begin{equation}
x \simeq A_1 + A_2 \sqrt{\epsilon}
\end{equation}
where $A_1 =\frac{3-4a}{4(1-a)}$ and  
$A_2=\frac{\sqrt{2(1-a)}}{2p_{c1}(1-a)}$.
The transition is discontinuos with exponent $1/2$  and $x_c = A_1$ at the critical point.
Above $a_c$, expanding around $p_{c2}$ we end up with 
\begin{equation}
x \simeq A_3 \epsilon
\end{equation}
 where
$A_3 = \frac{4a^2}{4-3a}$, a continuos transition with exponent $1$.
The crossover is at the point $a_c=3/4$, $p_c=2/3$.

The fraction of blocked spins $\Phi$ has the general form:
\begin{equation}
\Phi = p \sum_{k}^{\infty} P(k) \sum_{i=0}^{f-1} {k \choose i} x^{k-i} (1-x)^i + 
(1-p)\sum_{k}^{\infty} P(k) \sum_{i=0}^{f-1} {k \choose i} h^{k-i} (1-h)^i
\end{equation}
where
\begin{equation} 
h = p \sum_{k}^{\infty} \frac{(k+1)P(k+1)}{z} \sum_{i=0}^{f-2} {k \choose i} x^{k-i} (1-x)^i
\end{equation}
that, in our case it is:
\begin{equation} 
h = p x \frac{2u + 3(q_1-u)x + 4(1-q_1)x^2}{4-q_1-u}  
\end{equation}
\begin{eqnarray}
\Phi = p (u x (2-x) +(q1-u) x^2 (3-2x) + (1-q1)x^3 (4-3x))  \nonumber \\
+ (1-p)(u h (2-h) +(q1-u) h^2 (3-2h) + (1-q1)h^3 (4-3h))
\end{eqnarray}
Below $a_c$ we have
\begin{equation}
\Phi \simeq  2p_c u (1+\frac{2(1-p_c)u}{4-q1-u}) A_3 \epsilon
\end{equation}
If $u=0$, the leading order in the $\epsilon$ expansion is $2$ (see the right part of fig.\ref{trans}):

\begin{figure}[h!!]
\begin{center}
\includegraphics*[width=0.7\textwidth]{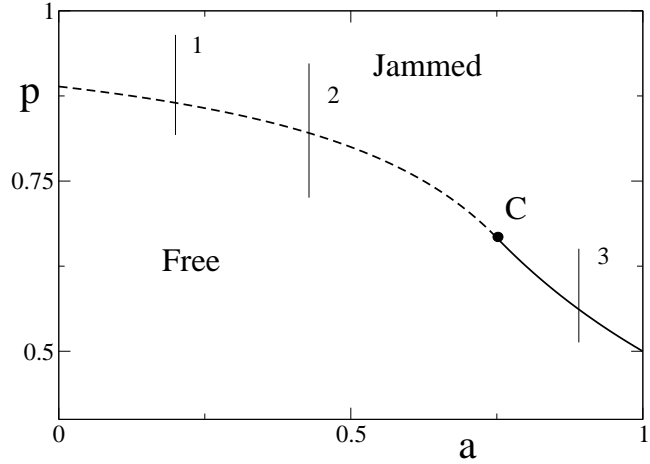}
\caption{Phase diagram of the system in the $(a,p)$ plane, at $C = (3/4,2/3)$ the transition changes character}
\label{phdi}
\end{center}
\end{figure}

Above $a_c$, we have instead:
\begin{equation}
\Phi-\Phi(x_c) \propto \sqrt{\epsilon}
\end{equation}
In synthesis, the picture is as follows:
\begin{itemize}
\item If $a<3/4$, above $ p_{c1} = \frac{8-8a}{9-8a}$
      there is a discontinuous jamming transition 
      with exponent $1/2$
\item If $a>3/4$, above $p_{c2} = \frac{1}{2a}$
      there is a continuous transition with exponent $1$
\end{itemize}
We will consider from now on three lines in the phase diagram, fig.\ref{phdi}: 
$q_{11} = 0.25$, that can be considered a perturbation with respect to a random regular graph;
$q_{12}=0.7$, where the transition is still discontinuous but closer to the point $C$, and 
$q_{13} = 0.85$, in the region with continuous transition.
Fig.\ref{trans} shows the transition curves $\Phi(T)$ along these three lines. 
We have, respectively, the critical temperatures $T_1 = 0.5386$, $T_2 = 0.9362$ and $T_3=2.0843$.
\begin{figure}[h!]
\begin{center}
\includegraphics*[width=0.4\textwidth]{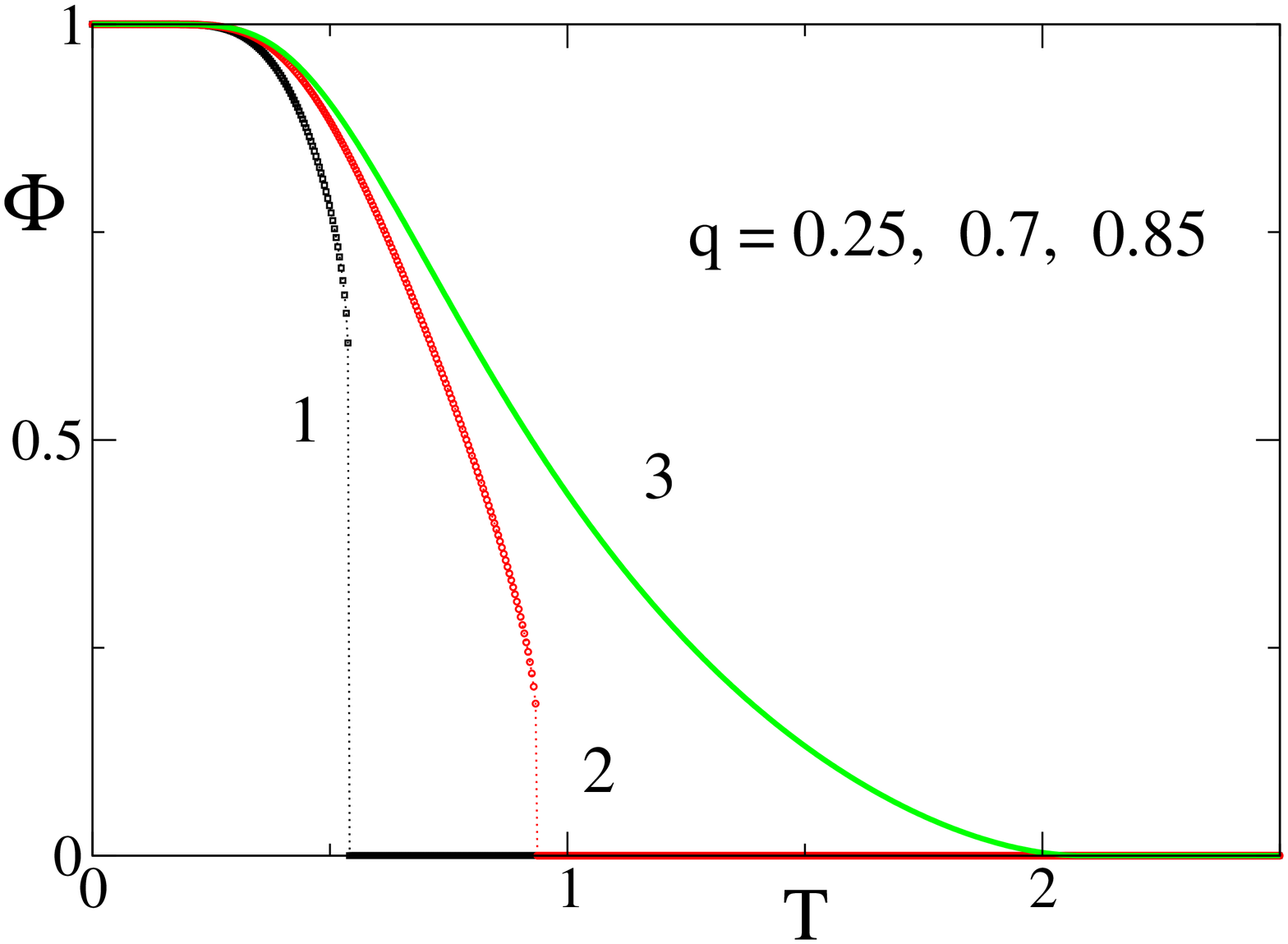}
\includegraphics*[width=0.4\textwidth]{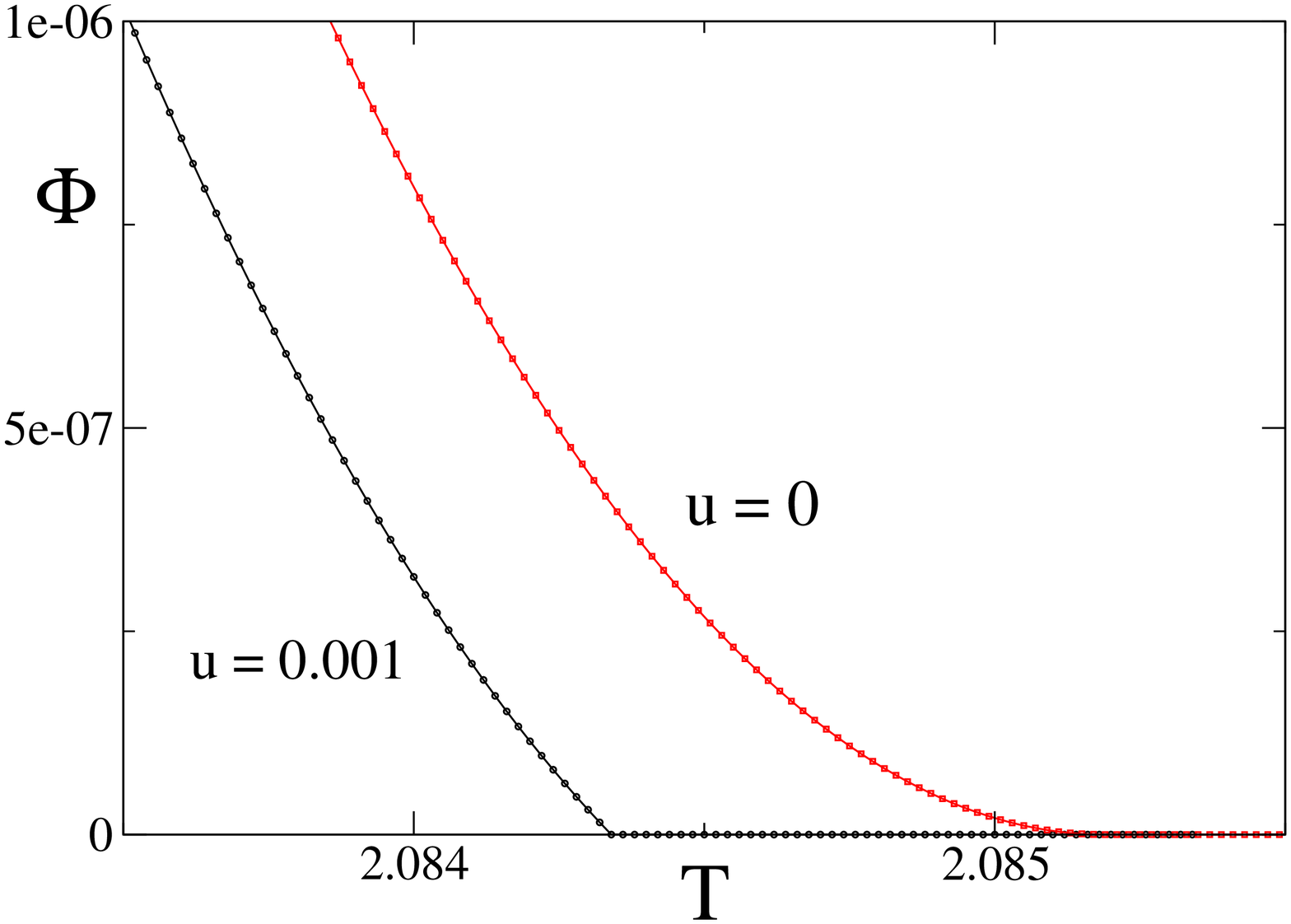}
\caption{
Left: Transition curves $\Phi(T)$ for $q_1 = 0.25, 0.7, 0.85$ , $u=0.001$   
Right:  Inserting  a small fraction of nodes with degree $2$ the transition exponent changes from $2$ to $1$ ($q_1=0.85$).
}
\label{trans}
\end{center}
\end{figure}

\begin{figure}[hb!!!]
\begin{center}
\includegraphics*[width=0.28\textwidth]{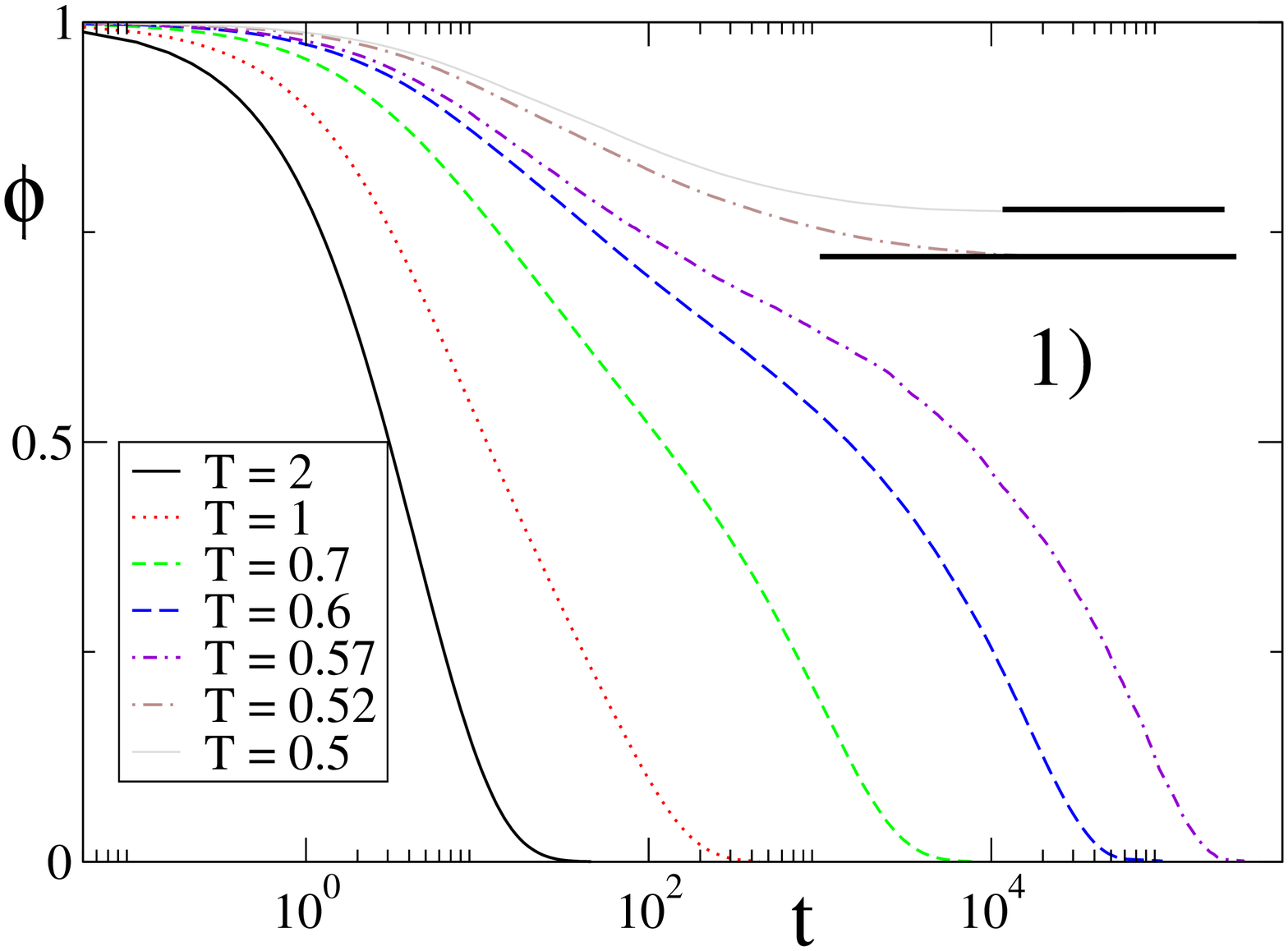}
\includegraphics*[width=0.28\textwidth]{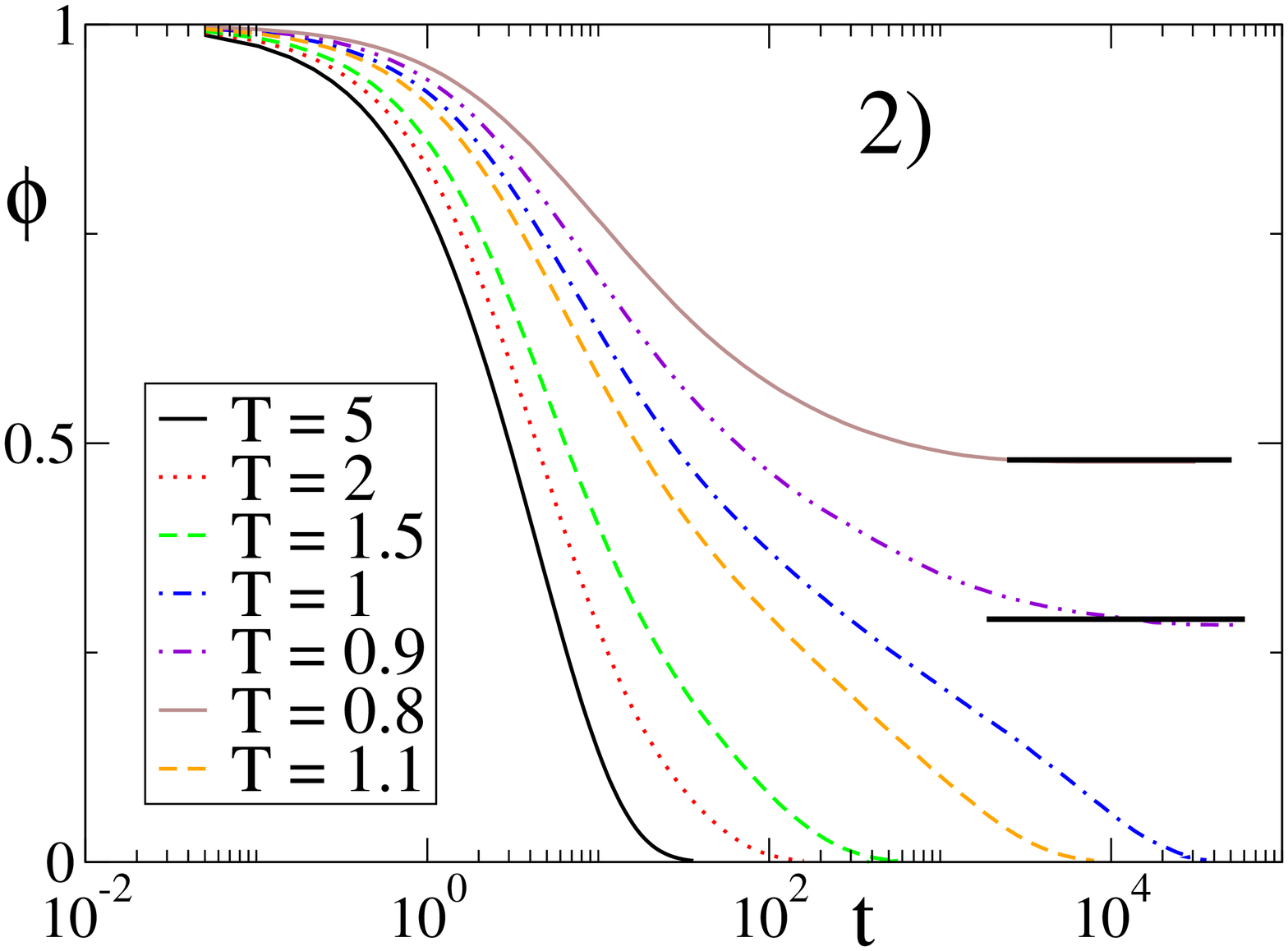}
\includegraphics*[width=0.28\textwidth]{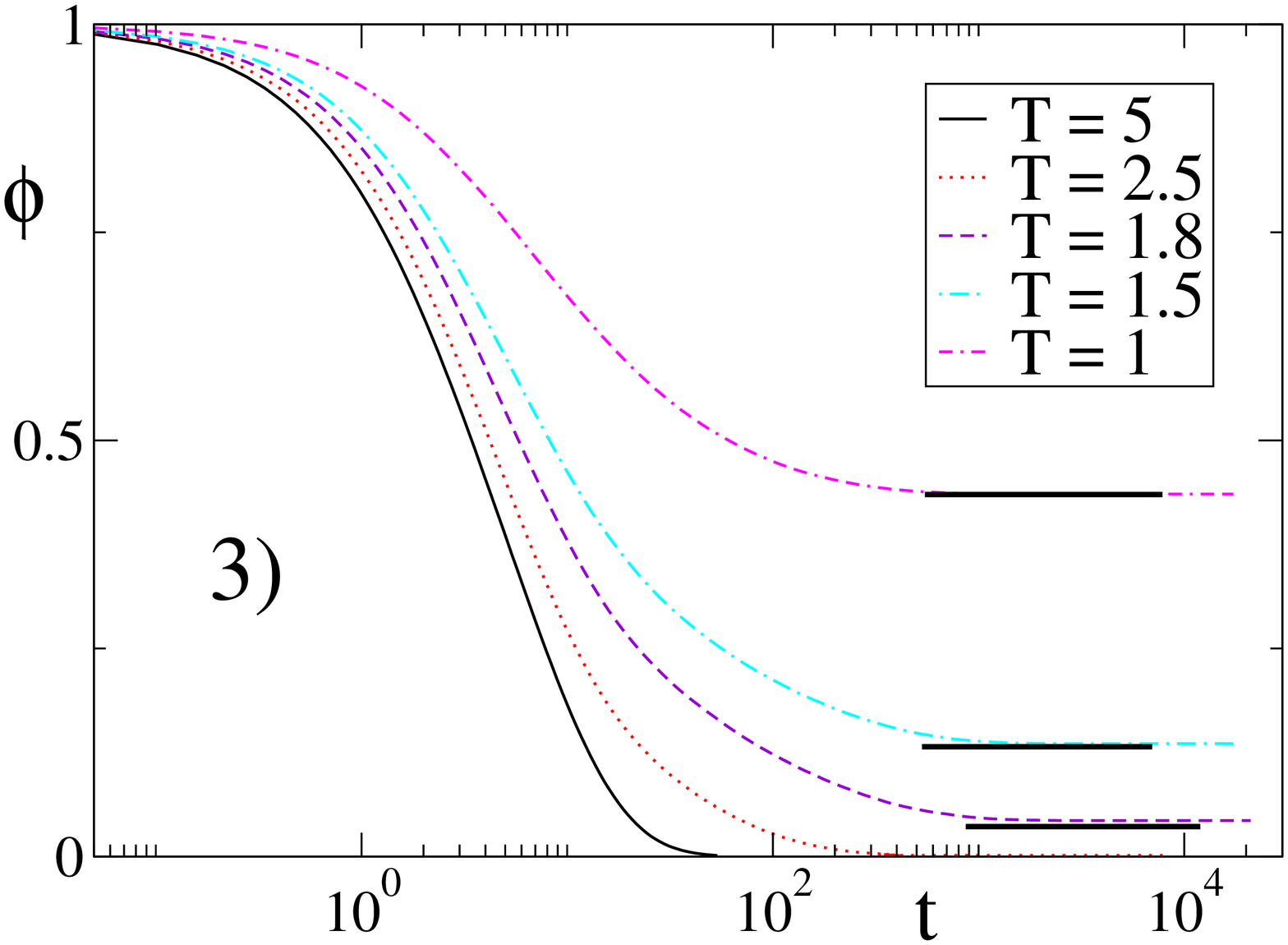}
\includegraphics*[width=0.28\textwidth]{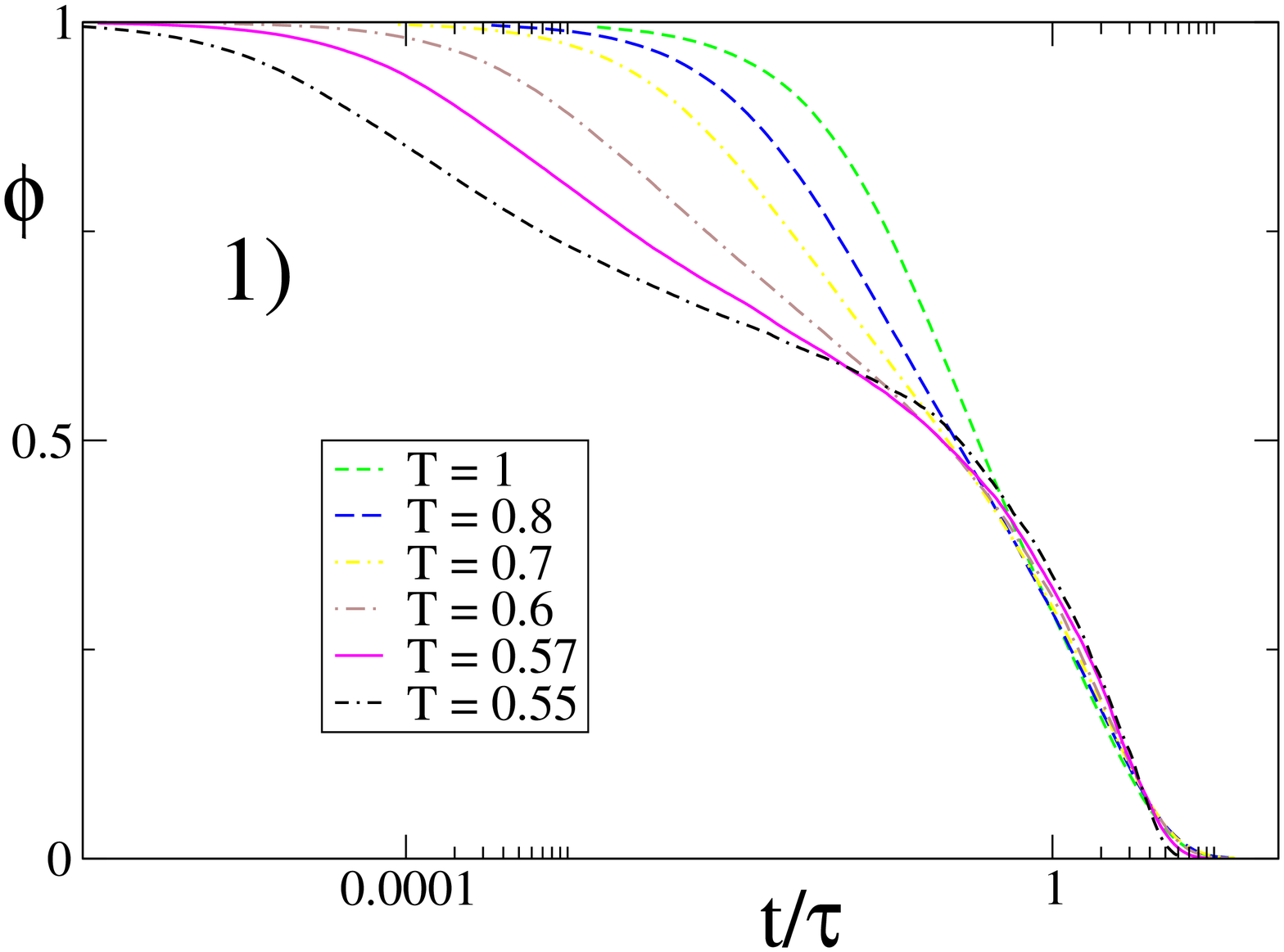}
\includegraphics*[width=0.28\textwidth]{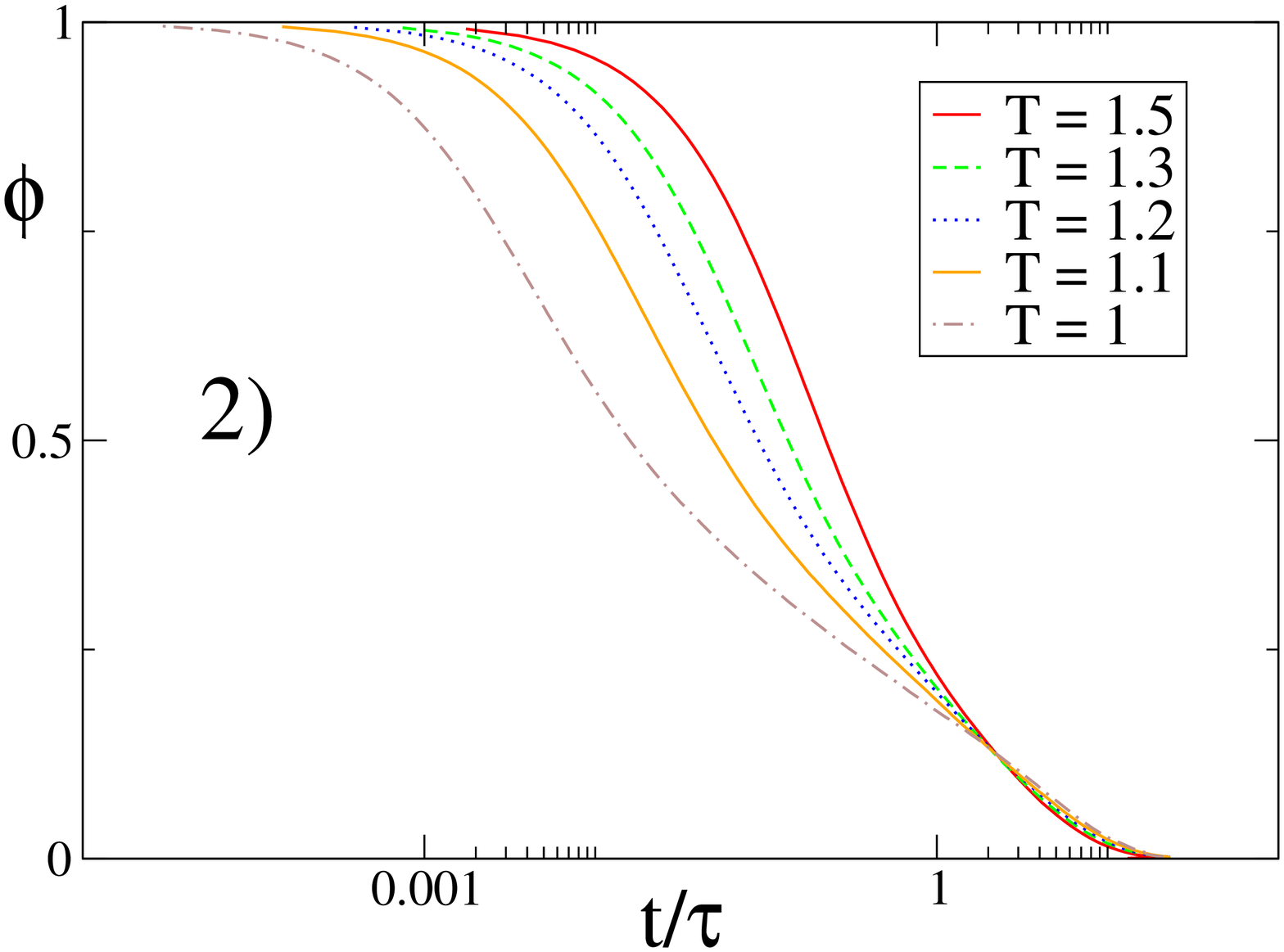}
\includegraphics*[width=0.28\textwidth]{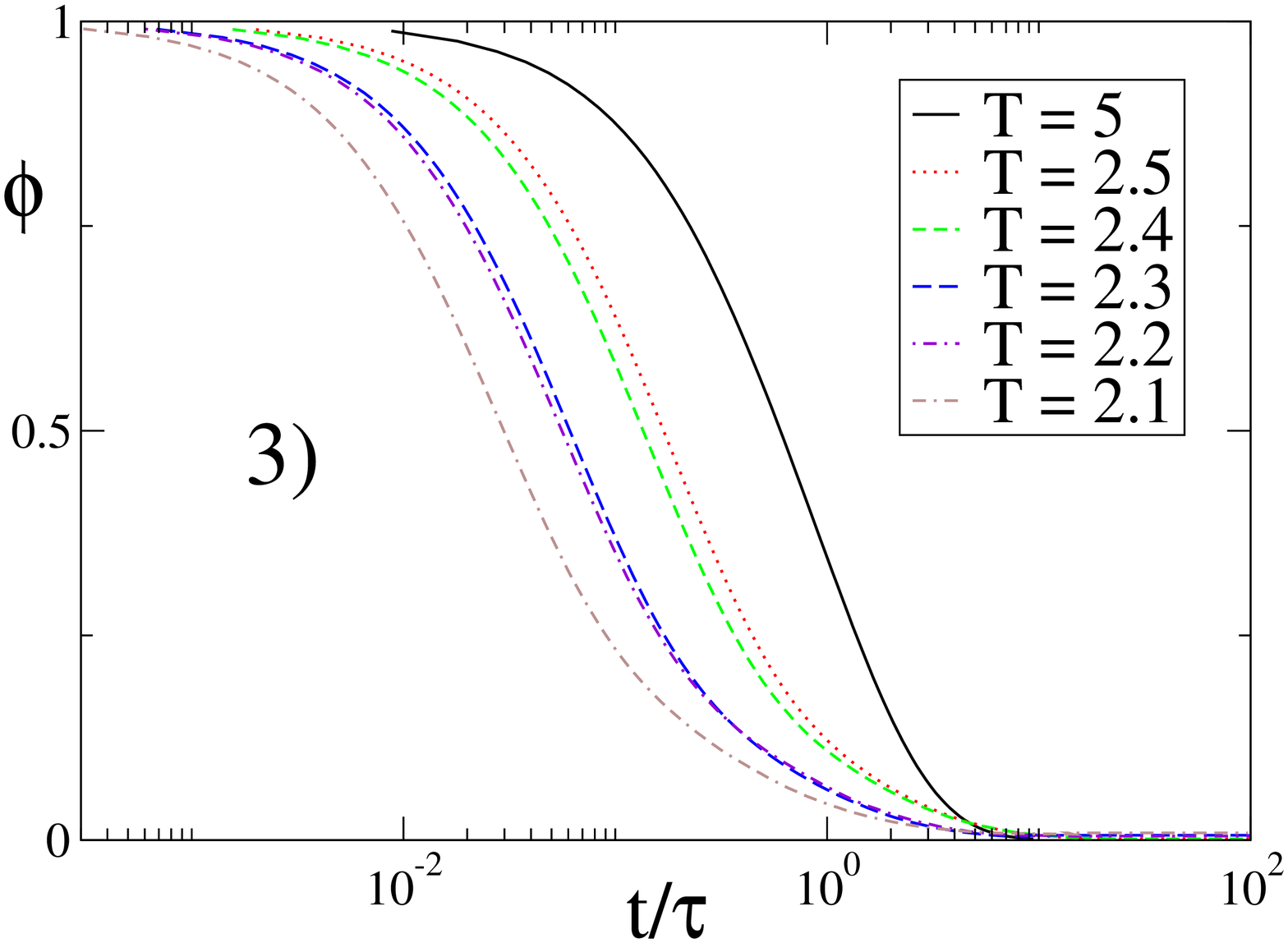}
\caption{Top: persistence curves $\phi(t)$ for $q_1= 0.25, 0.7, 0.85$ ($u=0.001$), respectively from left to right, for several temperatures.
Simulations for a graph of size $N=5 \cdot 10^4$, averages over $10$ realizations, plateaus from analitical results.
Bottom: Rescaled persistence curves}
\label{pers}
\end{center}
\end{figure}

\begin{figure}[h]
\begin{center}
\includegraphics*[width=0.3\textwidth]{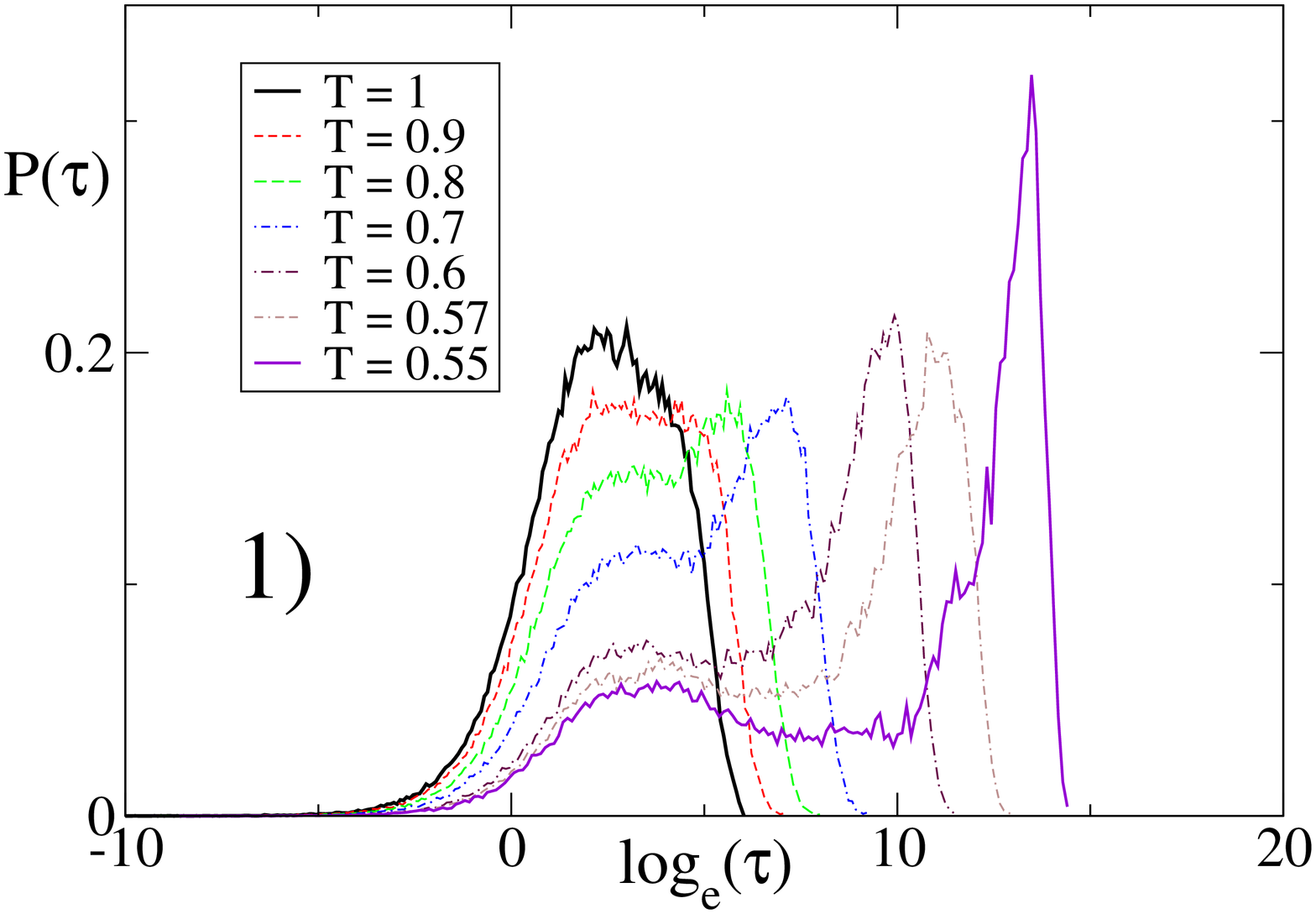}
\includegraphics*[width=0.3\textwidth]{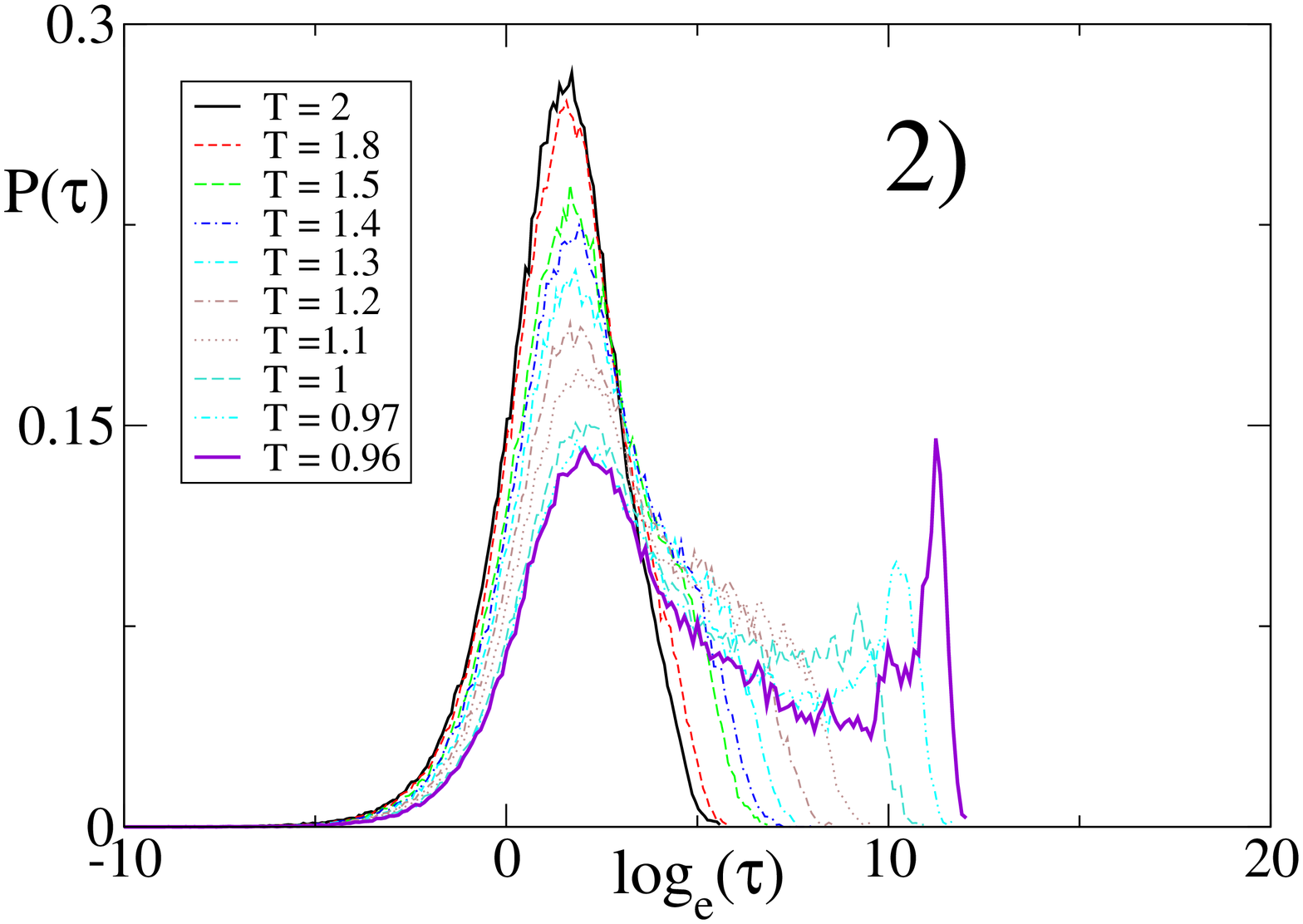}
\includegraphics*[width=0.3\textwidth]{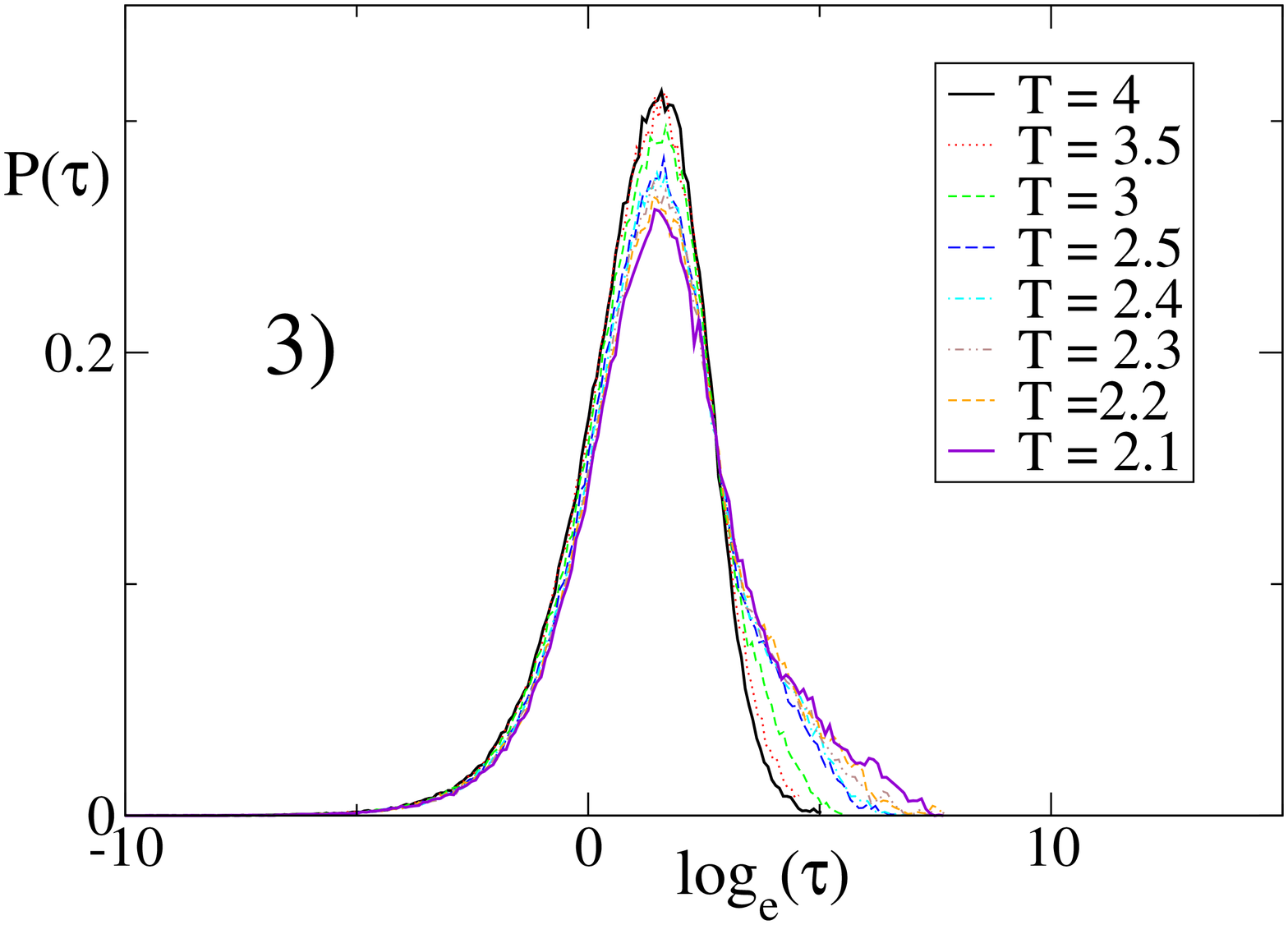}
\includegraphics*[width=0.3\textwidth]{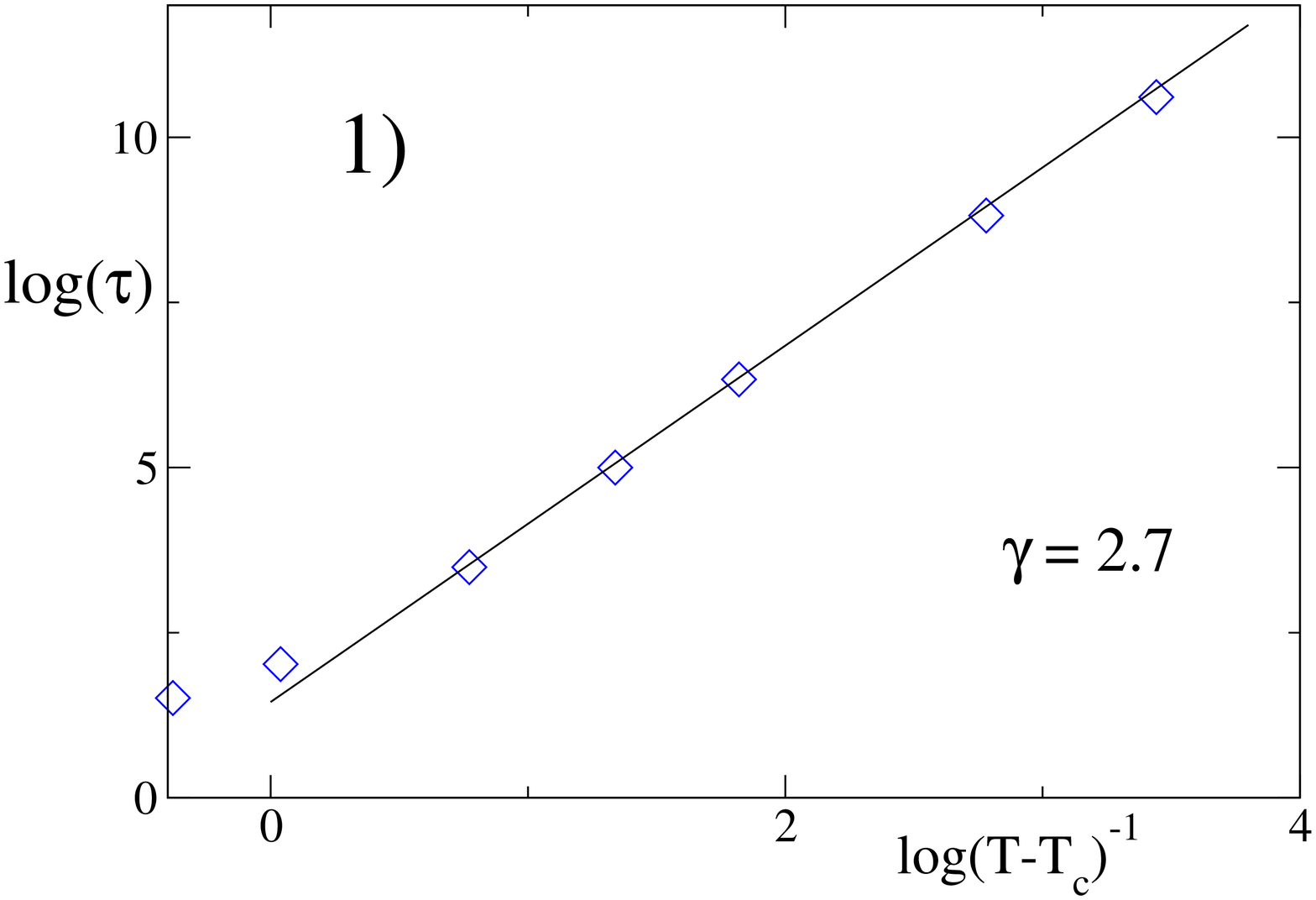}
\includegraphics*[width=0.3\textwidth]{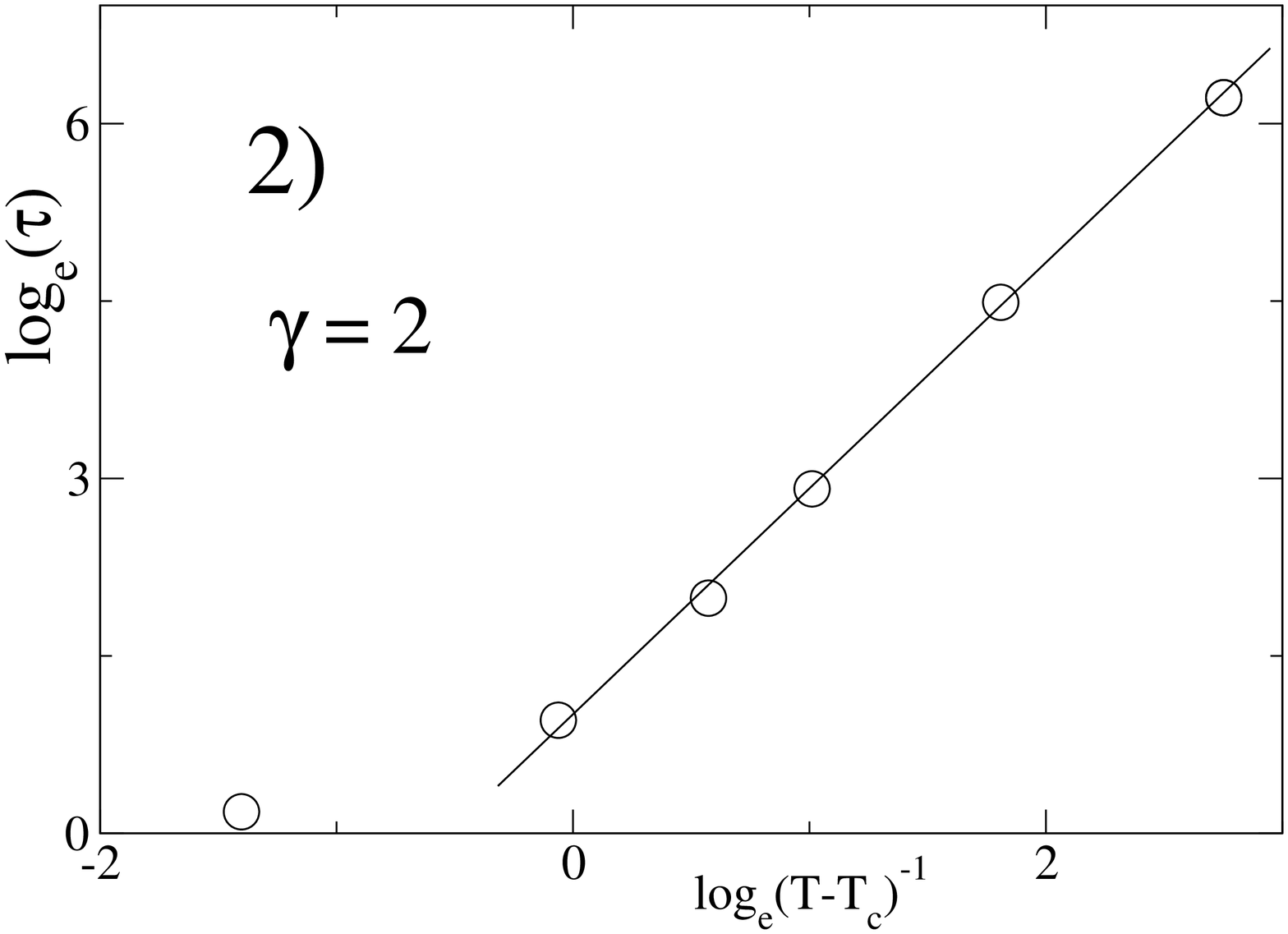}
\includegraphics*[width=0.3\textwidth]{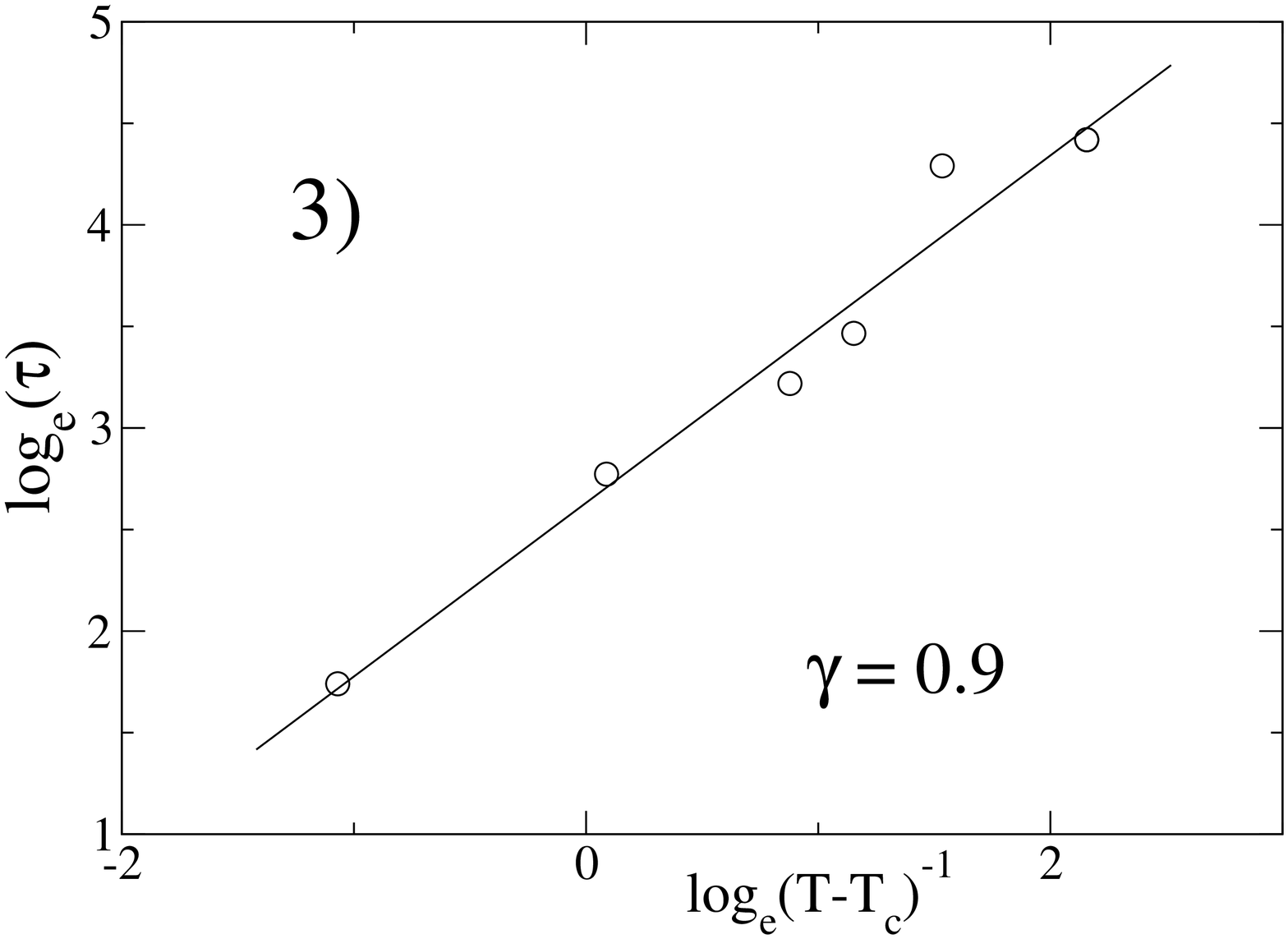}
\caption{Top: distribution of persistence times $P(\tau)$ for $q_1= 0.25, 0.7, 0.85$ ($u=0.001$), respectively from left to right, for several temperatures.
Simulations on top of a graph of size $N=2.5 \cdot 10^5$. Bottom: Average persistence times $\tau(T)$
}
\label{distr}
\end{center}
\end{figure}

\begin{figure}[h]
\begin{center}
\includegraphics*[width=0.3\textwidth]{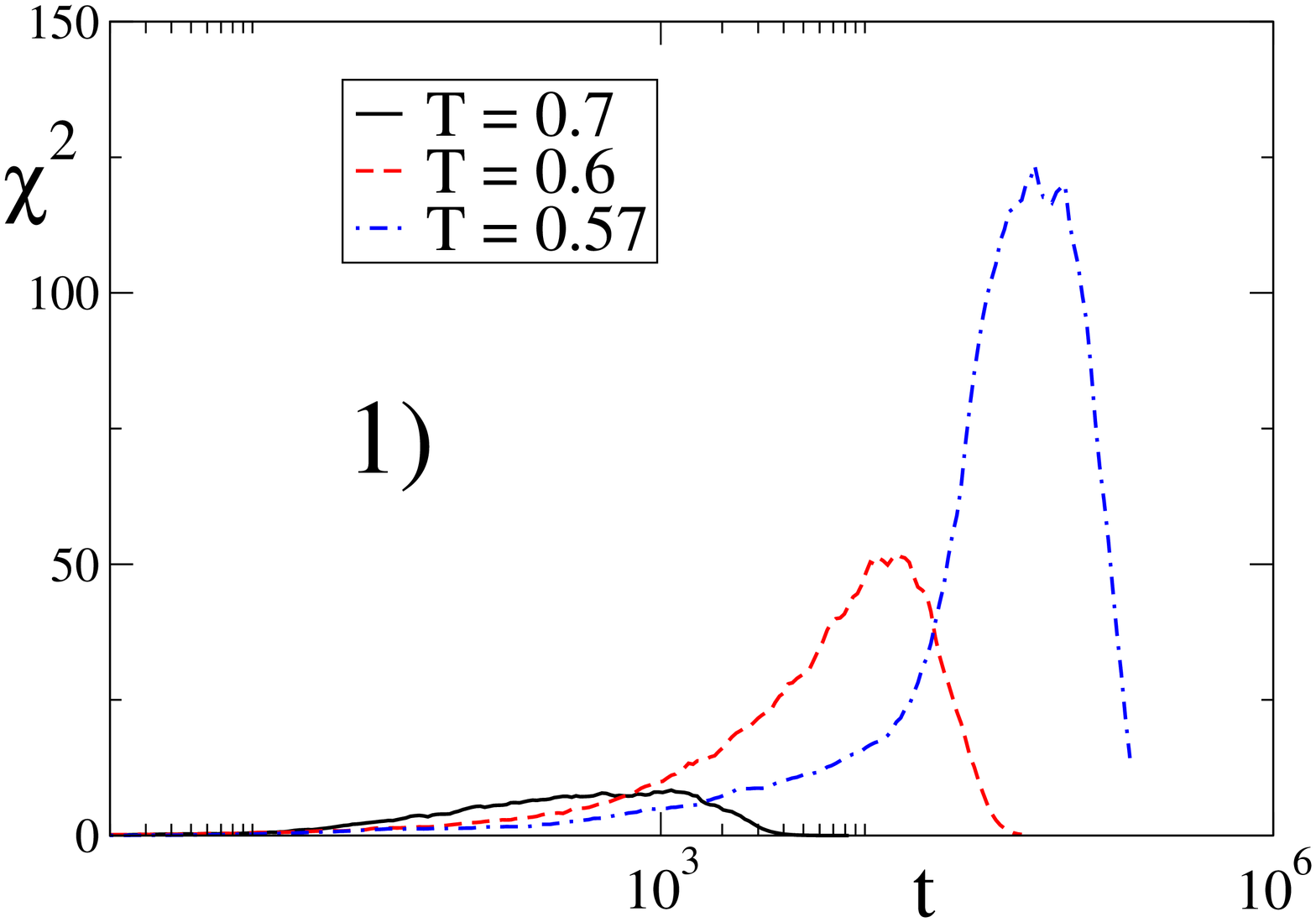}
\includegraphics*[width=0.3\textwidth]{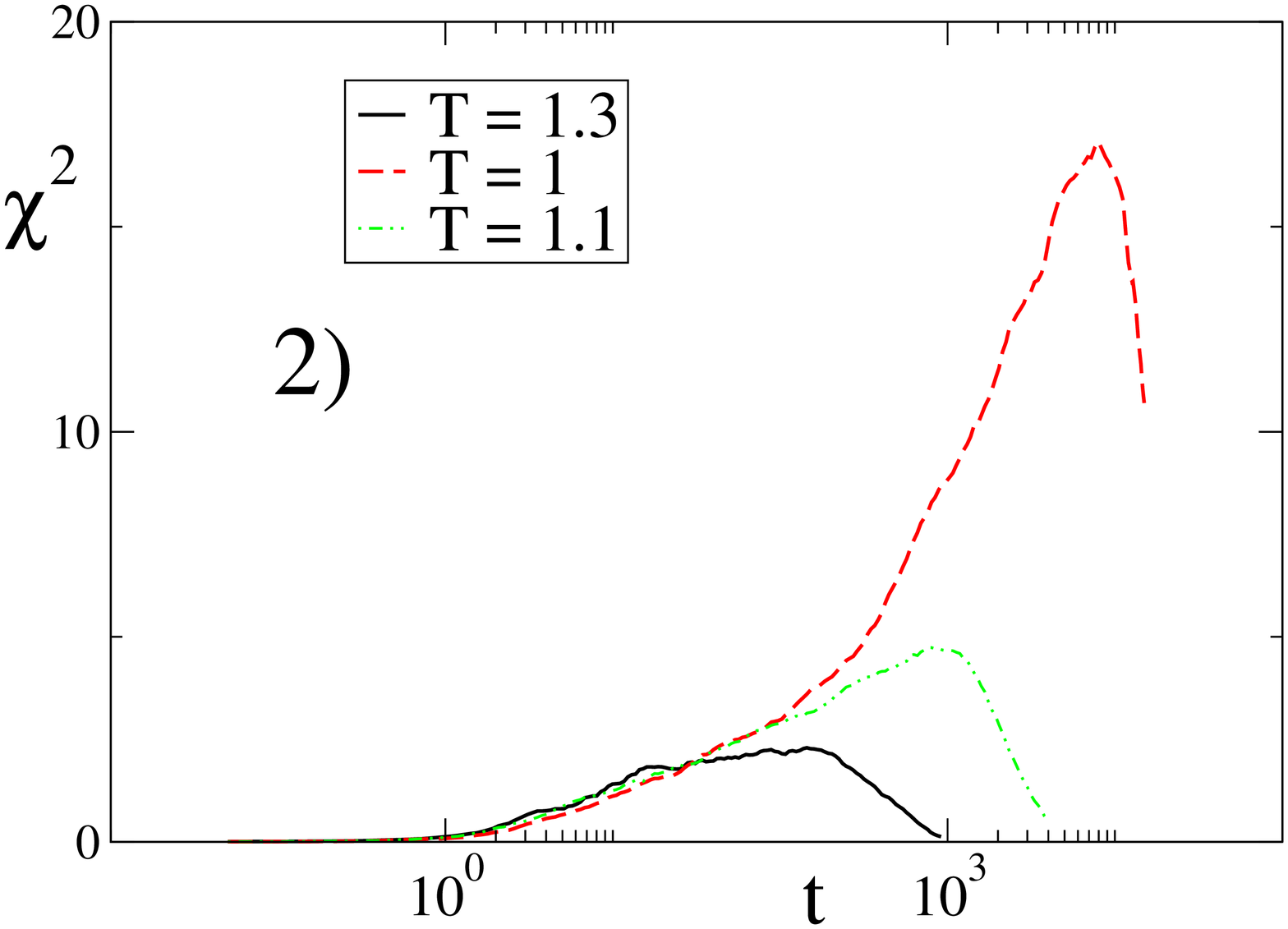}
\includegraphics*[width=0.3\textwidth]{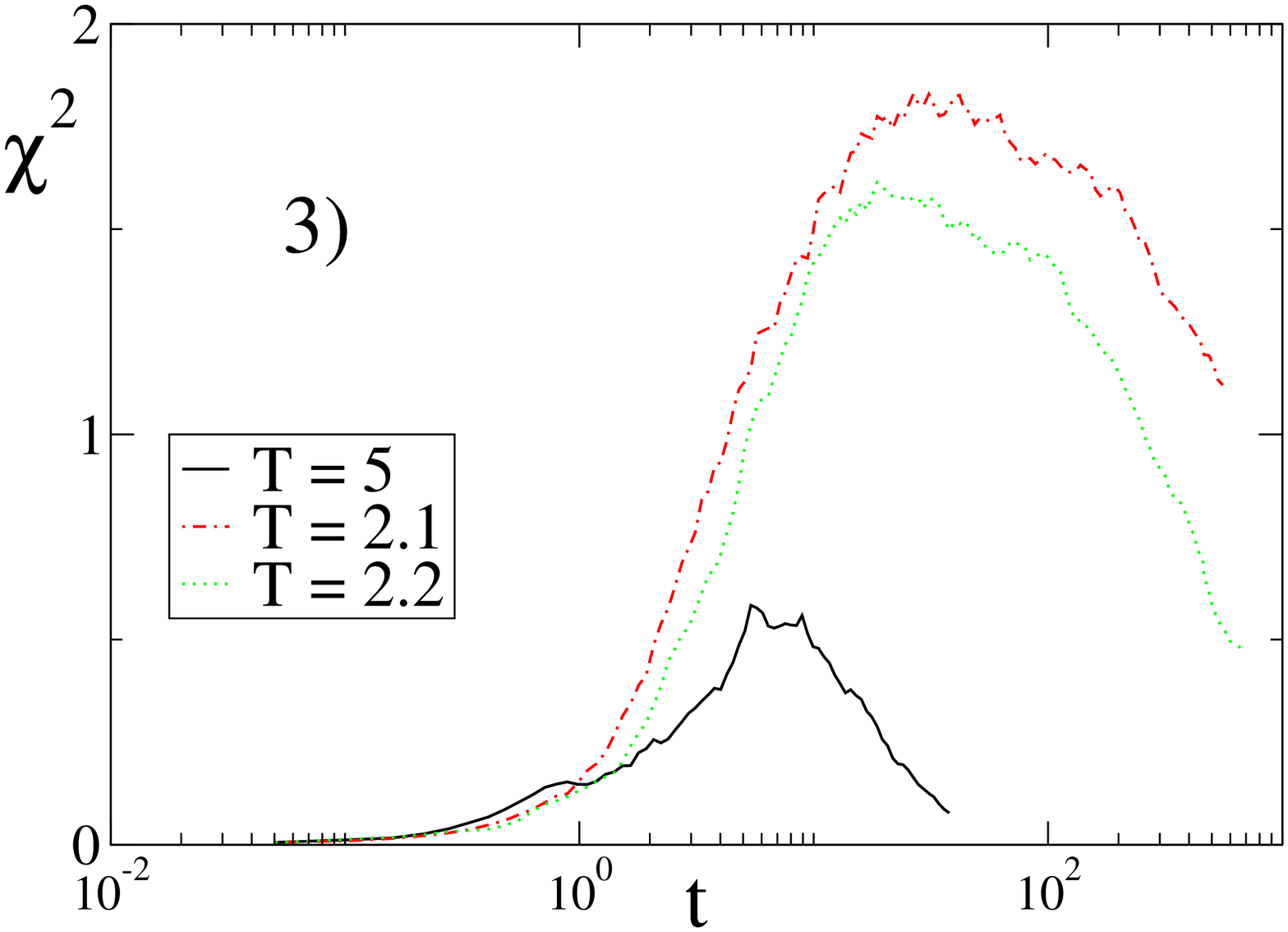}
\caption{Dynamical susceptibility  $\chi^2(t)$ for $q_1= 0.25, 0.7, 0.85$ ($u=0.001$), respectively from left to right, for several temperatures.
Simulations on top of a graph of size $N=10^4$, averages over $100$ realizations}
\label{susc}
\end{center}
\end{figure}

Fig.\ref{pers} shows the persistence $\phi(t)$, i.e. the fraction of spins that do not flip until time $t$, by simulations.
Upon decreasing $T$, the system starts to relax in a non-exponential way and then, it falls out of equilibrium, 
developing a plateau in the persistence, i.e the fraction of blocked spins  $\phi(\infty) = \Phi$.
As predicted from analitics, the plateau starts to develop discontinuosly and/or continuosly from the 
transition point, depending on the connectivity of the underlying graph (controlled by $q_1$). 
Nearby the transition temperature $T_c$, we can verify that there is a scaling law 
for the persistence of the type $\phi(t,T) = \phi(t/\tau(T))$, where 
$\tau$ is the integral time, i.e. simply the integral over time of the $\phi$.
This anomalous relaxation can be solved microscopically, looking 
at the distribution of persistence times $P(\tau)$, i.e. the time for the first spin-flip to occur.
In fig.\ref{distr} we can see that for both $q_1= 0.25, 0.7$, when decreasing the temperature, the distribution starts to develop another peak, instead, 
for $q_1=0.85$, when decreasing $T$, the distribution starts to develop a fat tail.    
Then, the dependence on temperature of typical relaxation times $\tau(T)$, calculated as  the average of the distribution 
is in a good agreement with a power law divergence at the critical temperature $\tau(T) \simeq (T-T_c)^{-\gamma}$, where
$\gamma$ depends on $q_1$.
Finally, we investigate the dynamical susceptibility $\chi^2(t) = N \langle(\phi(t)-\langle \phi(t)\rangle)^2\rangle$ (see fig.\ref{susc}). This increases with time
until it develops a maximum and/or a plateau, respectively for discontinuous and/or continuous transitions, whose height increases and diverges upon approaching $T_c$ 
However, it should be said that the divergence of relaxation time in the continuous case needs further investigations.

\subsection*{Facilitated spin mixtures on homogeneous graph}
An interesting point is that it is possible to obtain exactly the same results considering an homogeneous lattice and 
varying facilitation parameter from node to node. 
Lets' consider the FA model with a trimodal  $f=2,3,4$ distribution of facilitation parameter in a bethe lattice with degree $z=4$,
respectively a spin can have $f=2$ with probability $1-q$, $f=3$ w.p. $q-r$ and $f=4$ w.p. $r$
The self-consistent equation for $B$, the probability that, following a link, a spin is up, or can be in the next steps moving
the other spins on the top, is:

\begin{equation}
B = (1-p) + p ( B^3+3(1-q)(1-B)B^2 ) 
\end{equation}

We have for $x=1-B$, apart from the $x=0$ solution:
\begin{equation}
x = \frac{3(1-2q) + \sqrt{9p^2 -8p +12qp(1-p)}}{2p(2-3q)}
\end{equation}

This solution exists, for  $q < q_c = 1/2$, if $p > p_{c1} = \frac{4(3q-2)}{3(4q-3)}$, and, if $q > q_c = 1/2$,
it is positive until $p>p_{c2} =\frac{1}{3a}$.
Below $q_c$, we can expand around $p_{c1}$, $p=p_{c1}+\epsilon$, we have
\begin{equation}
x \simeq A_1 + A_2 \sqrt{\epsilon}
\end{equation}
where $A_1 = \frac{2q-1}{3q-2}$ and
$A_2=\frac{3\sqrt{2-3q}(4q-3)}{4(3q-2)^2}$.
The transition is discontinuos with exponent $1/2$  and $x_c = A_1$ at the critical point.
Above $a_c$, expanding around $p_{c2}$ we end up with
\begin{equation}
x \simeq A_3 \epsilon
\end{equation}
 where
$A_3 = \frac{3q^2}{1-2q}$, a continuos transition with exponent $1$.
The crossover is at the point $q_c=1/2$, $p_c=2/3$.

The fraction of blocked spins is:
\begin{eqnarray}
\phi =  p(x^3(4-3x)+6qx^2 (1-x)^2 + 4rx(1-x)^3)+\nonumber \\
(1-p)(h^3(4-3h)+6qh^2 (1-h)^2 + 4rh(1-h)^3)
\end{eqnarray}
where
\begin{equation} 
h = p(x^3 +3(1-q)x^2 (1-x) +3rx(1-x)^2)
\end{equation}
and x is the solution that we have discussed.
Below $q_c$ we have
\begin{equation}
\phi \simeq 4p_c r (1+3r(1-p_c)) A_3 \epsilon
\end{equation}
Above $a_c$, we have instead:
\begin{equation}
\phi-\phi(x_c)  \propto \sqrt{\epsilon}
\end{equation}
In synthesis, the picture is as follows:
\begin{itemize}
\item If $q<1/2$, above $ p_{c1} = \frac{4(3q-2)}{3(4q-3)}$
      there is a discontinuous jamming transition
      with exponent $1/2$
\item If $q>1/2$, above $p_{c2} = \frac{1}{3q}$
      there is a continuous transition with exponent $1$
\end{itemize}
\begin{figure}[h]
\begin{center}
\includegraphics*[width=0.5\textwidth]{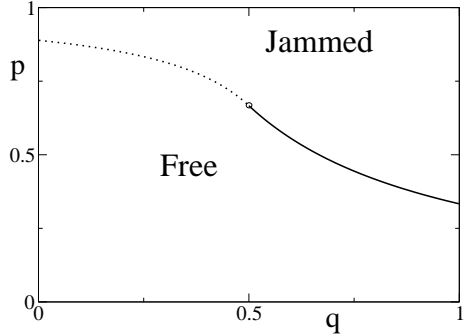}
\caption{Phase diagram of the system in the $(q,p)$ plane, at $(1/2,2/3)$ there is a critical point}
\end{center}
\end{figure}

\subsection{Conclusions}
In the cooperative case, the Frederickson-Andersen model can give a good mean-field microscopic description of the anomalous relaxation and dynamical crossover
of glass forming liquids in the range of temperature slightly above the crossover point. The relaxation of correlation functions 
proceedes by two steps, the height of the plateau starting discontinuosly from a finite value. 
The dynamical susceptibility grows in time till it reaches a maximum whose height increases upon  decreasing the temperature.
At odds of real liquids and similarly to the MCT there is a power law singularity at $T_c$, where the system starts to be jammed. 
Microscopically this correspond to a bootstrap percolation transition. 

I showed that this same model can change character on a diluted, heterogeneous structures. 
For high enough dilution the transition becomes continuous, within the class of simple percolation.
There are not two steps in the relaxation, that still shows stretched exponential dependence upon approaching the critical point.
The dynamical susceptibility develops a plateau whose height is slowly increasing when approaching the singularity.

This simple percolation dynamical arrest scenario is known to be the one of the sol/gel transition in polymer blends, and
recent numerical simulation studies shown that it is valid also for strongly confined fluids. 

It seems that the simple ingredient of a fixed heterogeneity, being enconded in the spatial structure or in the mobilities, 
can change qualitatively a dynamical arrest scenario, dividing systems in two classes from this point of view.

However, more detailed numerical investigations of this model in the continuous regime are needed, but
it should be  important to test the universality of such a mechanism applying it to other kinetically constrained models, like the one by Kob-Andersen\cite{KOB}
or others\cite{spiral}.

\chapter{Inverse phase transitions on heterogeneous graphs}
The relationship between model systems and the underlying topology is at the core of research in statistical mechanics.
It is a common belief that the distinctive equilibrium features of 
 simple model systems are affected only by the internal symmetries and by the dimensionality of the space.
In this chapter it is shown that a certain degree of heterogeneity in the underlying structure of network of interactions 
can trigger inverse phase transitions in tricritical model systems.

Inverse phase transitions are stricking phenomena in which an apparently more ordered phase becomes disordered by cooling.
In the first paragraph there is a  basic introduction to such phenomenon, 
with a special focus on inverse melting because of its relationship with some fundamental problems in statistical physics\cite{kauzrev}.
Then, there is a discussion about the simplest model system that shows inverse melting, 
i.e. the Blume-Capel model with higher degeneracy of interacting states\cite{schupper}.
Finally, I will show how inverse melting can emerge spontaneusly in tricritical model systems 
if the underlying graph has certain features, i.e. if sparse subgraphs are crucial for its connectivity.
I will work out many results\cite{nostopo} for the simple Blume-Capel model, and I will give some insights 
that the random field Ising model shares the same phenomenology.

\section{Experimental inverse transitions}
\begin{figure}[h]
\begin{center}
\includegraphics*[width=0.45\textwidth]{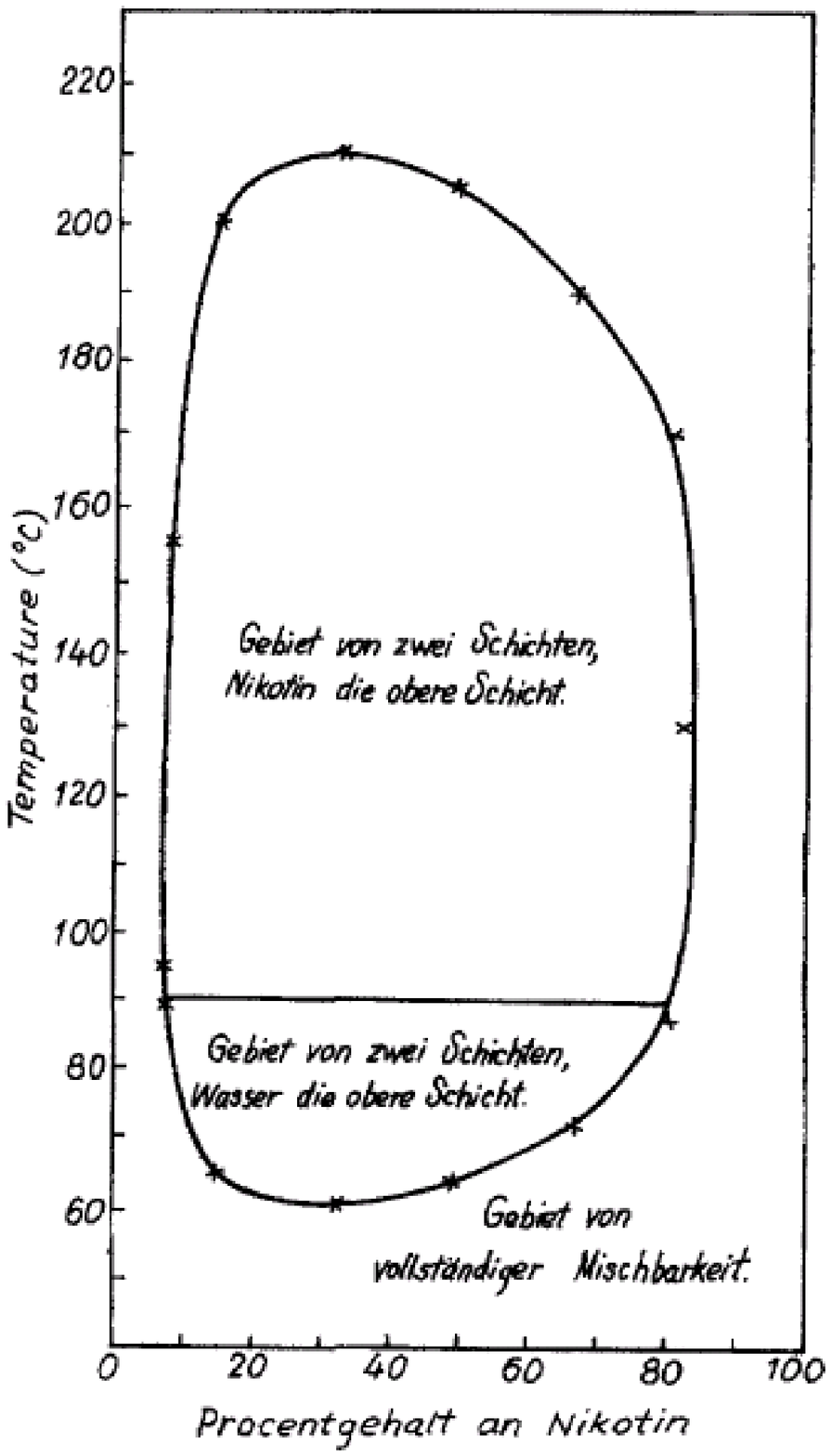}
\includegraphics*[width=0.45\textwidth]{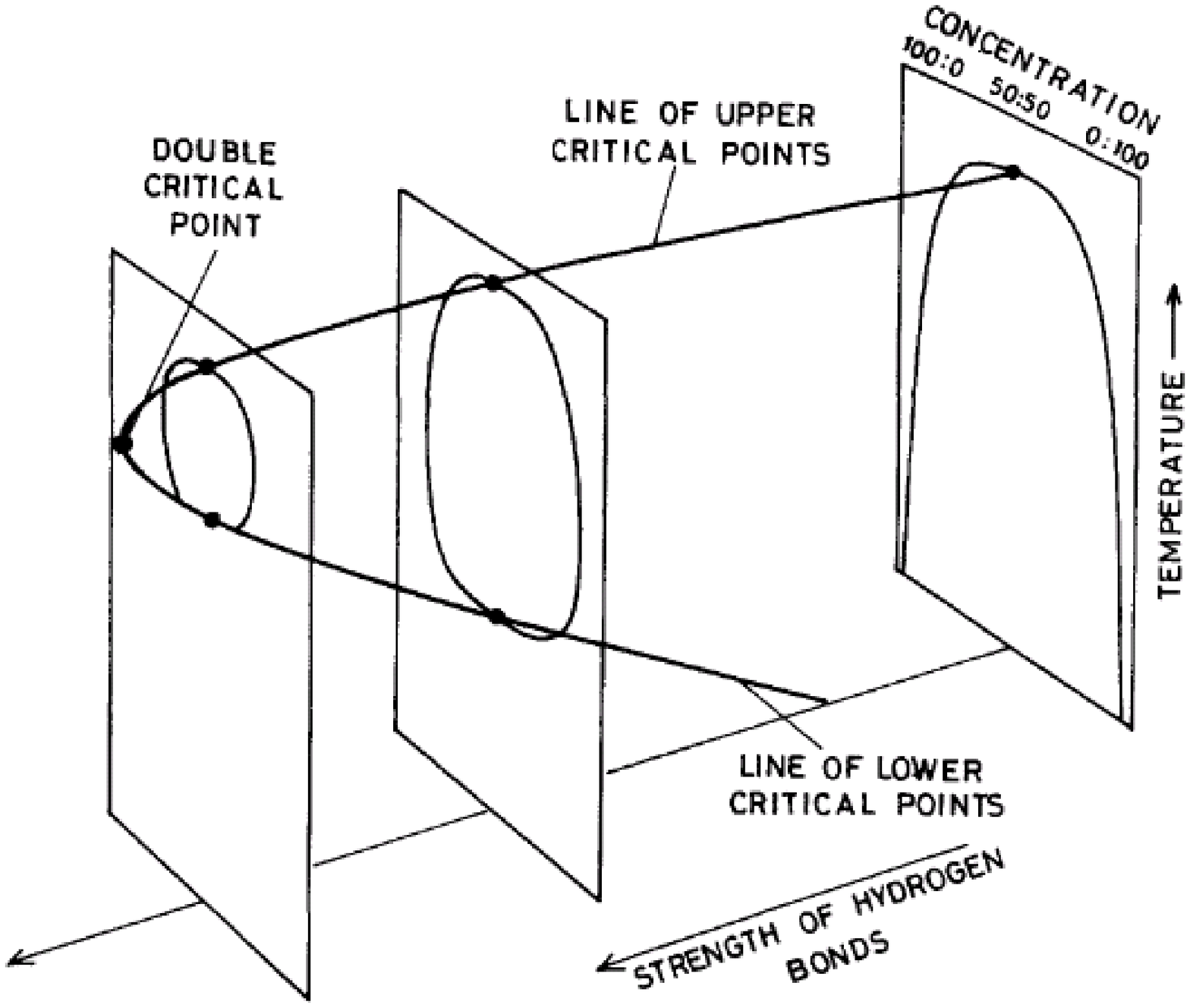}
\caption{Left: experimental sketch of the looped miscibility phase diagram $(T,c)$ of the solution Nicotine+$H_2 O$. From\cite{hudson}.
Right: sketch of the general miscibility phase diagram of a binary solution including the strenght of hydrogen bonds between unlike molecules as a third axis.
} \label{solutions}
\end{center}
\end{figure}

Inverse transitions in their most generic meaning have been detected
in a number of different materials and between phases of different nature (see \cite{schupper}and \cite{narayan} for a review).
The first experimentally seen inverse phase transition 
regards the miscibility properties of liquid mixtures\cite{hudson}.  
Fig \ref{solutions}(left) shows the loop-shaped phase diagram of the solution  Nicotine+$H_2 O$.
A reentrant phenomenon is evident. The solution is mixed at an high temperature, 
it demixes by cooling and it gets mixed again by further cooling. 
Many different multicomponent solutions show this behavior\cite{narayan}. 
The mechanism behind it relies on the 
strong directionality of hydrogen bonds between unlike molecules\cite{hirschfelder}.
In the low temperature mixed phase, when unlike molecules interact, they  form some complexes 
with a well defined orientation, thus freezing their internal rotational degrees of freedom. 
This in turn has the effect of lowering the total entropy with respect to the demixed phase.
Therefore in this case demixing is basically an entropy driven process and the demixed phase 
is counter-intuitively more disordered.

\begin{figure}[h]
\begin{center}
\includegraphics*[width=0.45\textwidth]{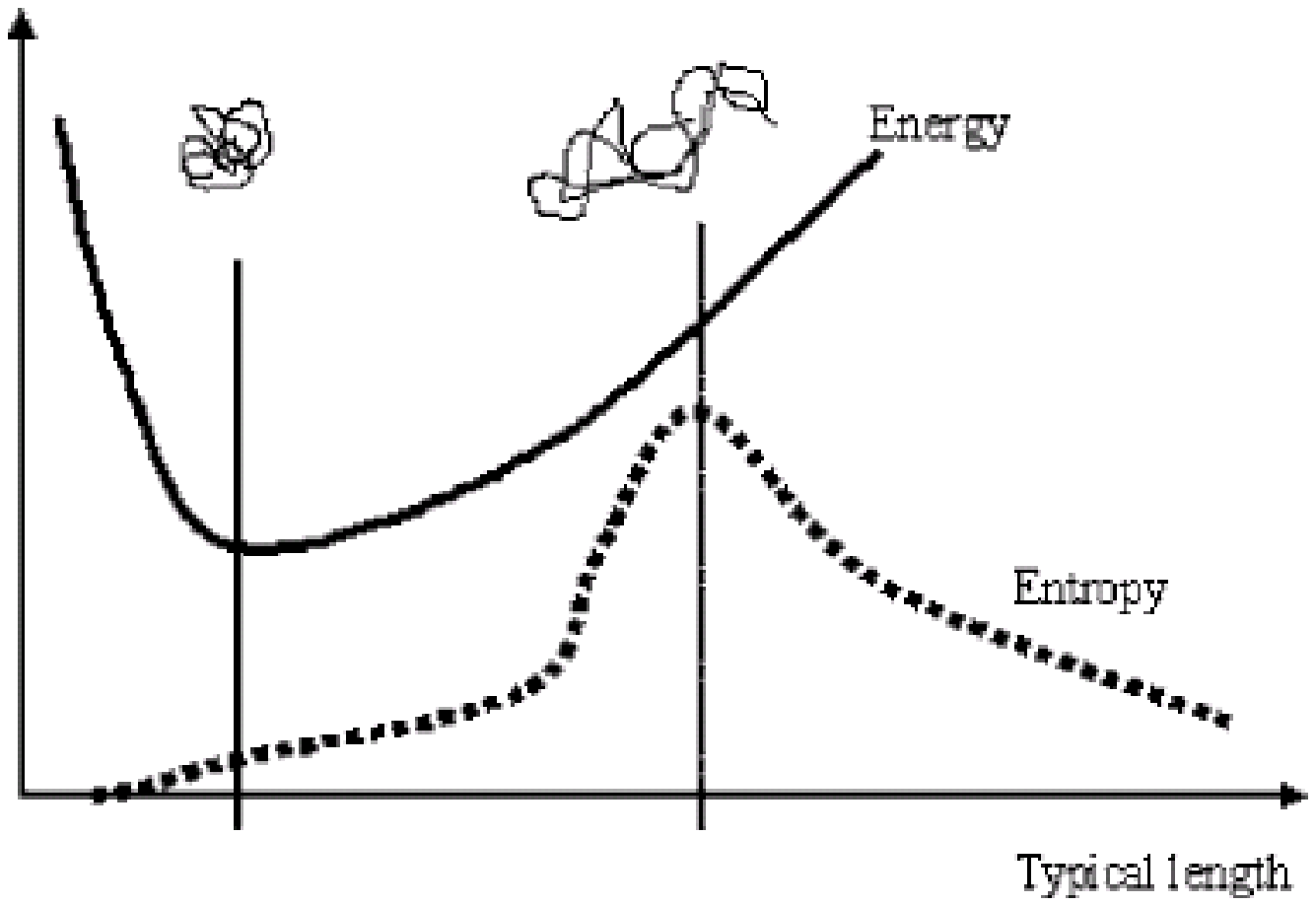}
\includegraphics*[width=0.45\textwidth]{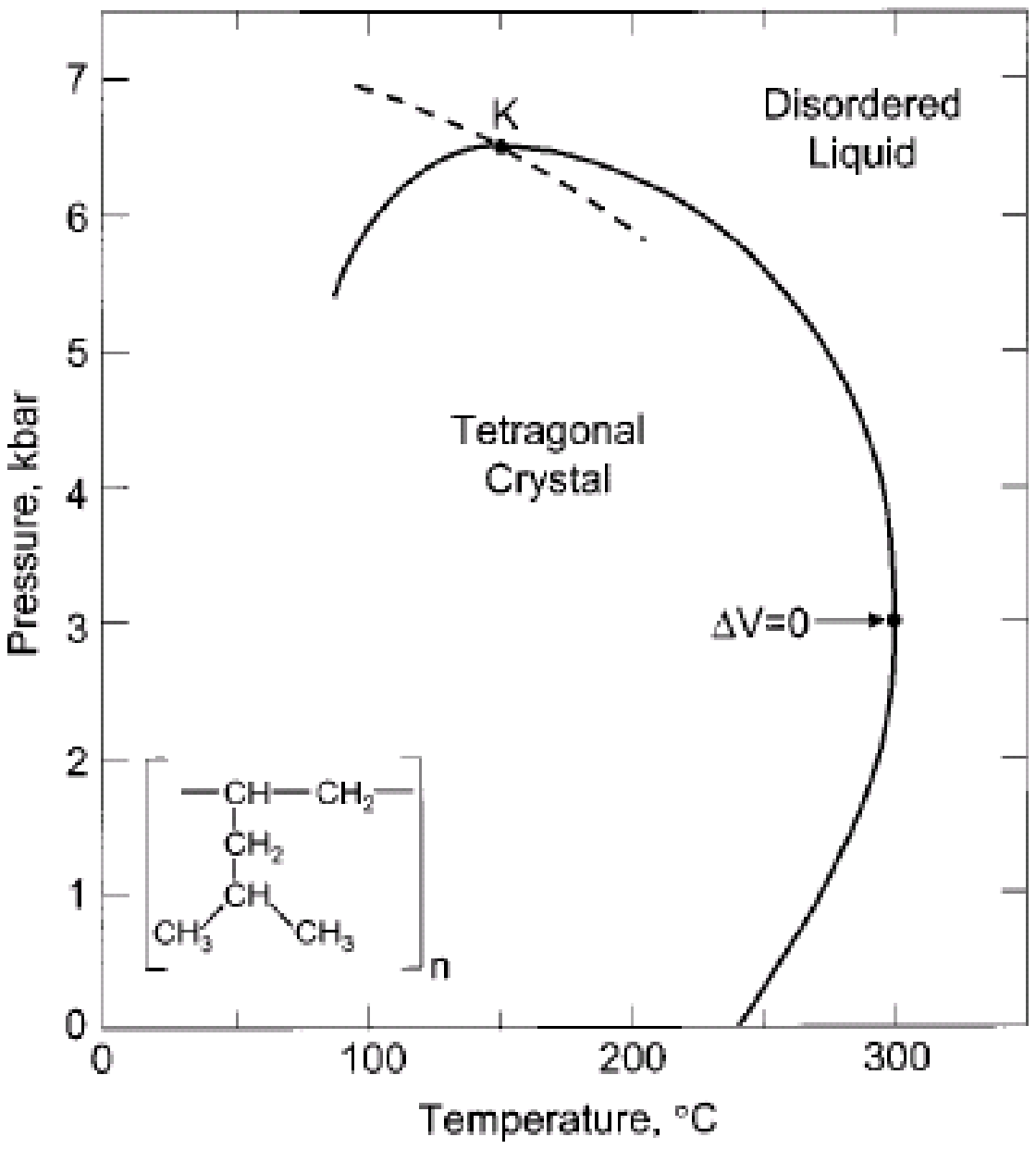}
\caption{Left: Sketch of the energy and entropy of a polymer chain as a function of the lenght.
Right: Sketch of the melting curve  in the $(T,p)$ plane for the polymer P4MP1.\cite{rastogi}
} \label{polimer}
\end{center}
\end{figure}

At the beginning of the last century\cite{tamman} speculations were put forward about the possibility of an {\em inverse melting}: 
a crystal that liquifies by cooling.
This is confirmed nowadays experimentally on many substances, 
the most famous examples being the inverse melting of $He_3$ and $He_4$ at high pressures\cite{kauzrev}.
The interesting point is that in this case 
the standard ratio of the entropies of the solid and liquid phases is inverted, the solid being more disordered.
In particular, at the point at which the inverse behavior starts, the entropies of the two phases are equal.
This is a practical realization of the Kauzmann scenario\cite{kauzrev} that I sketched in the second chapter.
The specific heat of many substances in the supercooled liquid phase is usually higher 
than the one of their crystalline phase. The  
decrease by cooling of the entropy of the supercooled liquid is steeper than the one of the crystal. 
Extrapolating it below the point at which the system
gets out-of-equilibrium, the glass transition point, 
there should be a temperature at which they are equal. 
It should be possible that the liquid below this point has a lower entropy.  
This was seen as a paradox because it was believed that the ground state of a physical system made
of identical objects should be a crystal. 
Many mechanisms were proposed to avoid this and some of them
are at the core of theoretical views on the glass transition.
However, the existence of inverse melting shows that in general this is not a paradox.
It is true that a crystal can have an higher entropy than a less interacting phase.
This can be explicitly pointed out in the polymer melts. 
A polymer can be in many microscopic configurations. 
The ground state is often  unique, non-interacting and looped(see fig\ref{polimer}). 
Thermal noise can unfold this structure, making the polymers interacting. All 
toghether they can form networks, i.e. a solid phase. 
A very well known case is the inverse melting of the crystal polymer 
made by the isotactic poly(4-methylpentene-1), P4MP1 (see fig\ref{polimer}\cite{rastogi}). 

\section{A simple model for inverse phase transition}
\begin{figure}[h]
\begin{center}
\includegraphics*[width=0.7\textwidth]{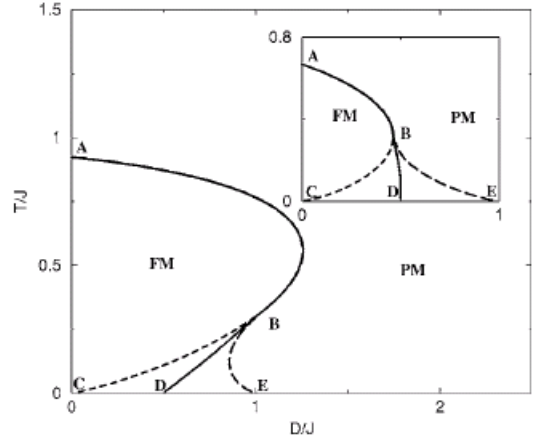}
\caption{Phase diagram of the Blume-Capel model in the $(\Delta,T)$ plane with higher degeneracy of
the interaction state $r=6$. Inset: the same for the normal case $r=1$.
} \label{rientro}
\end{center}
\end{figure}

Many mathematical models were proposed to explain how a phase transition can be inverted 
(see \cite{schupper}and \cite{wheeler} for a review).
In almost all of them the most interacting configuration 
of the units that made the system has by construction an higher degeneracy.
The simplest model that can encode this feature is the Blume-Capel model.
At first proposed to explain the occurrence of a first order magnetic transition in the $UO_2$\cite{blume}, it became 
the representative of tricritical systems. 
It consists of $N$ ferromagnetic interacting 1-spins $s_i=\pm1, 0$ 
that have a cost in energy to be present, i.e a chemical potential $\Delta$, the hamiltonian being:
\begin{equation}
H = - \sum_{\langle i,j \rangle} s_i s_j + \Delta \sum_i s_i^2 
\end{equation} 
Where the first sum is over the bonds of a given lattice.
Inspired by the already seen phenomenology of inverse transition in polymer melts 
we can think of the interacting phase as having more degeneracy than
the non-interacting one\cite{schupper}. 
That is, we imposed by hand that the states with $s_i=\pm1$ are $r\geq1$ times more present of the ones with  $s_i=0$.
We can recur to a mean field approximation, i.e. no spatial structure, by which every couple of spin is interacting. We rescale the
interaction by $2N$. 
Using standard gaussian integral techniques it is possible to find the expression of the free energy:
\begin{equation}
\beta f = \frac{\beta m^2}{2} - \log(1+2r \cosh(\beta m) e^{-2 \beta \Delta})
\end{equation}
Where $m$ is the order parameter, the average magnetization, that can be found by minimizing $f$.
This requires to solve the self consistent equation:
\begin{equation}
m = \frac{2r sinh(\beta m)}{e^{\beta \Delta} + 2r cosh(\beta  m)}
\end{equation}
There is always a solution $m=0$. We can expand in powers of $m$:
\begin{equation}
m = A m + B m^3 + \dots
\end{equation}
When $A=1$, another solution starts to occur, i.e.  when
\begin{equation}
\label{lambda}
\beta = 1+\frac{1}{2r} e^{\beta \Delta} 
\end{equation}
And the solution $m=0$ becomes unstable, i.e. a maximum for $f$.
The equation \ref{lambda} defines thus consistently a curve of second order, continuous critical points, till $B<0$.
When $B$ changes sign, at $T_c = \frac{1}{3}$, $\Delta_c = \log{4} T_c $ there is  a tricritical point,
after which the transition becomes discontinuous. 
However, the eq.\ref{lambda} after this point continues 
to be the line of stability of the $m=0$ solution, 
i.e. the spinodal curve of the paramagnets.
It is possible to study numerically the stability of the other solution, 
thus definying the spinodal curve of the ferromagnetic solution. 
In the region between the two spinodal curves both the paramagnetic and ferromagnetic solutions are minima of the free energy.
The transition is thus characterized by coexistence and hysteresis phenomena in this region.
It is possible to compare the free energies of both solutions to characterize which one is stable (absolute minimum).
In particular the Clasusius-Clapeyron equation is valid along the equilibrium curve :
\begin{equation}
\frac{d\Delta}{dT} = \frac{S_m-S_p}{\rho_m-\rho_p}
\end{equation}
Where $S$ is the entropy, $\rho=\langle s^2 \rangle$, and the labels $m$, $p$ refers to the ferromagnetic and paramgnatic phase respectively.
This equation shows that  $\frac{d\Delta}{dT}>0$ implies $S_m > S_p$.
In fig\ref{rientro} the phase diagram in the $(\Delta,T)$ plane is shown for $r=6$. There it is possible to see clearly 
the emergence of a reentrant phenomenon with respect to the normal case ($r=1$, in the inset)
Inverse phase transitions can emerge also {\em spontaneously} in tricritical model sytems, 
without the assumption of an  higher degeneracy
of the interacting states. For instance, the spin glass version of the Blume Capel model, with ferromagnetic and {\em antiferromagnetic} couplings,
shows inverse {\em freezing} \cite{leuzzi} between glassy and fluid phases.

I will show in the next paragraph that inverse phase transitions can emerge spontaneously also in the normal, 
ferromagnetic, Blume-Capel model on heterogeneous structures.

\section{Topology-induced inverse phase transition}
Let's consider the Blume-Capel model now on a general heterogeneous graphs\cite{nostopo}.
For random graphs of given degree distribution $P(k)$
it is possible to set up the following approximation scheme (Curie-Weiss).
We consider the following Hamiltonian function:
\begin{equation}
H = - \frac{1}{2N} \sum_{i\neq j} h_i h_j s_i s_j + \Delta \sum_i s_i^2 
\end{equation}
Where the $h_i$ are independently identically distributed quenched random variables
according to the distribution $P(k)$.
This is equivalent to consider the model on a fully connected geometry 
with link weights $a_{ij} = \frac{h_i h_j}{2N}$.
The calculation follows along the lines sketched in the previous paragraph.
Finally, we come up with self consistent equations for $m$, $m_v$, 
i.e. respectively the average magnetization of a randomly chosen node and 
that one of a node reached following a randomly chosen link:
\begin{eqnarray}
m_v = \sum_k \frac{kP(k)}{z} \frac{2 sinh(\beta k m_v)}{e^{\beta \Delta} + 2 cosh(\beta k m_v)} \\
m = \sum_k P(k) \frac{2 sinh(\beta k m_v)}{e^{\beta \Delta} + 2 cosh(\beta k m_v)}
\end{eqnarray}
Where $z$ is the average degree. The continuous critical line depends on the ratio between $\langle k^2\rangle$ and $z$:
\begin{equation}
\beta \frac{\langle k^2\rangle}{z} = 1+\frac{1}{2} e^{\beta \Delta} 
\end{equation}
Then, at $T_c = \frac{1}{3} \frac{\langle k^2\rangle}{z}$, $\Delta_c = \log{4} T_c $ there is  the tricritical point,
after which the transition becomes discontinuous. 
We can define rescaled variables $\delta=\Delta \frac{\langle k \rangle}{\langle k^2 \rangle}$,  $\tau=T \frac{\langle k \rangle}{\langle k^2 \rangle}$,
such that the $\lambda$-line collapses on a master function:
\begin{equation}
\delta = \tau \log[2(1/\tau -1)]
\end{equation} 
After the tricritical point, the line of first order  phase transitions
shows a striking difference between the homogenenous and heterogeneous case,
with the appearence of a reentrant phenomenon in the latter.

\begin{figure}[h]
\begin{center}
\includegraphics*[width=0.6\textwidth]{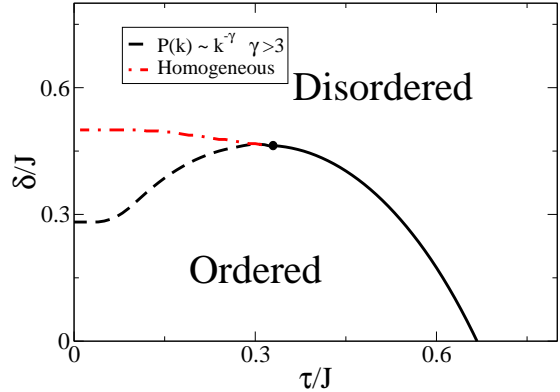}
\caption{Phase diagram $(\delta,\tau)$ of the Blume-Capel model within the Curie-Weiss approximation.
The first order branch can be reentrant for heterogeneous networks
} \label{diagrammi}
\end{center}
\end{figure}

\begin{figure}[h]
\begin{center}
\includegraphics*[width=0.45\textwidth]{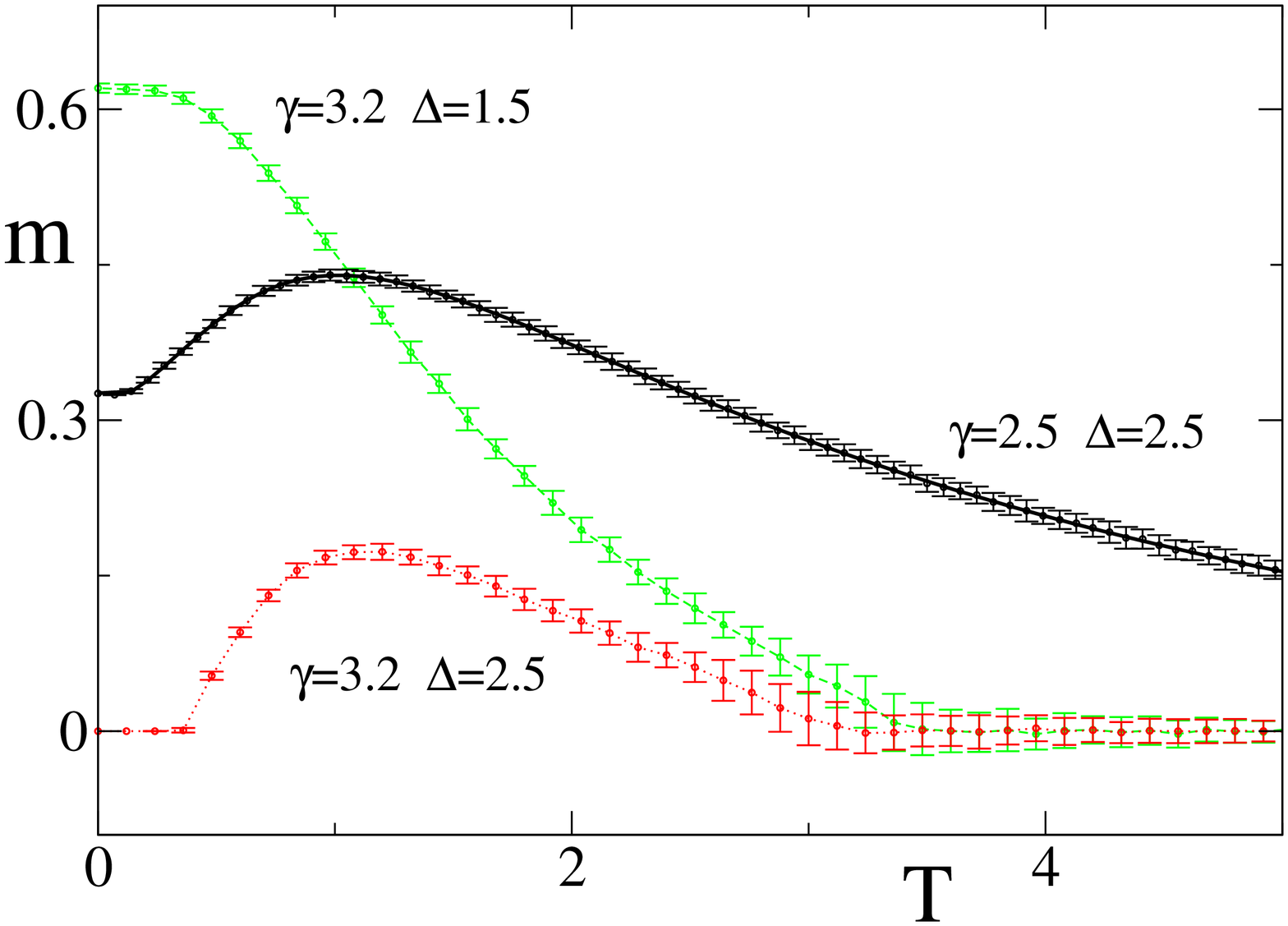}
\includegraphics*[width=0.45\textwidth]{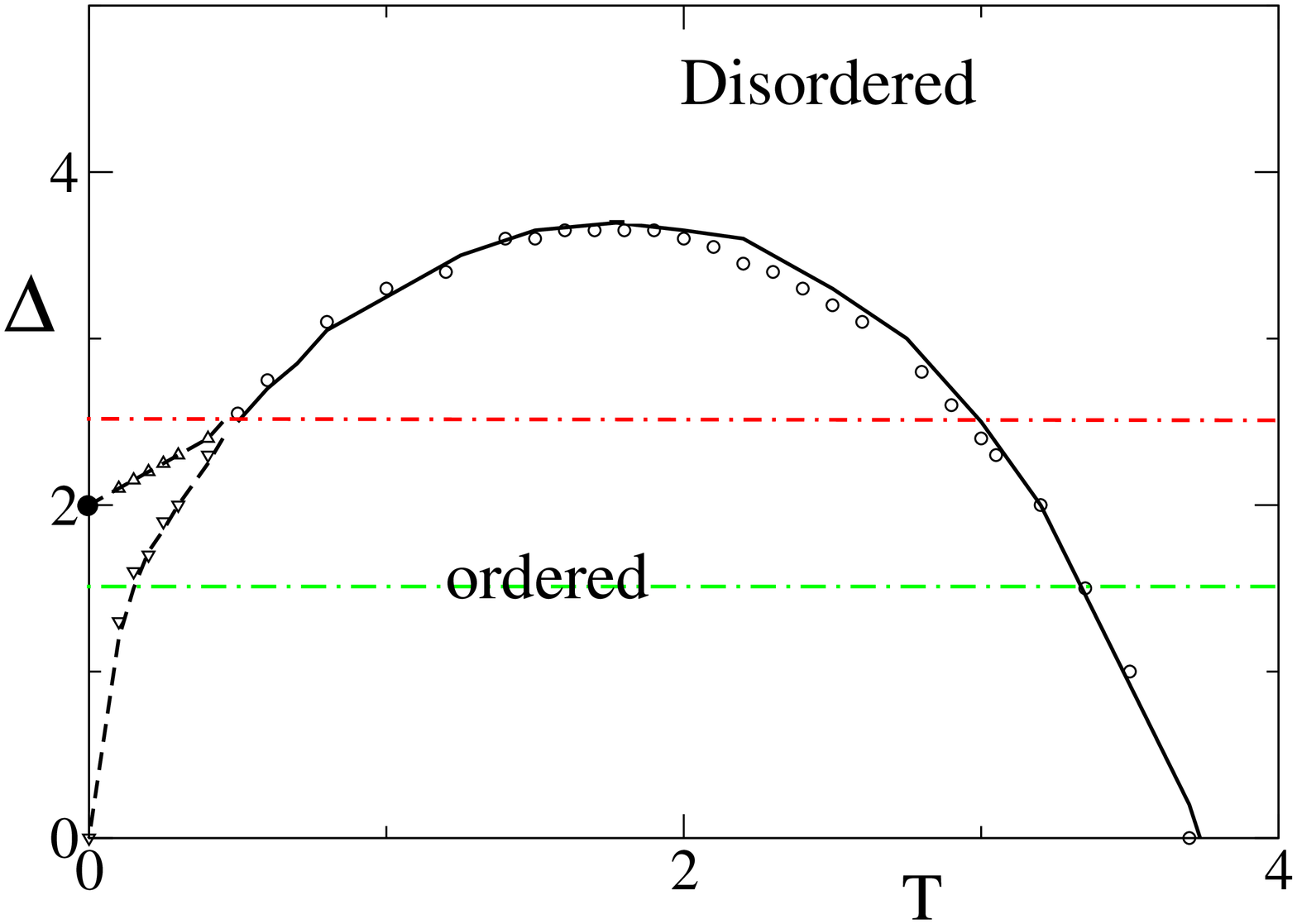}
\caption{Inverse phase transition and reentrant phenomena
in a heteorogeneous random graph of size $N=10^4$, with degree distribution $P(k) \propto k^{-\gamma}$, $\gamma=3.2$ and $k_{min}=2$.
Left: Transition curves $m(T)$. For large enough $\Delta$ there is a reentrant phase transition, 
that is dumped  for a different exponent $\gamma=2.5$. 
Right: Phase diagram. From Monte carlo simulations(points) and BP numerical calculations(lines).
} \label{powerlaw}
\end{center}
\end{figure}

However, this approximation is not exact. 
But given the tree-like nature of random graphs it is possible
to se up a better approximation (Bethe-Peierls).
In fact, we can write the partition function in a recursive way.
Let's select one node and write the partition function as a function of the ones  of the sub-branches from that node.
The approximation relies in the factorised form, that is, independent sub-branches (no loops).
\begin{eqnarray} 
\mathbb{Z} = \sum_{s_0} e^{-\beta \Delta s_0^2} \prod_{j\in N_0} g_{0j} (s_0) \\
g_{ij}(s_i) = \sum_{s_j} e^{\beta ( s_i s_j - \Delta s_j^2)} \prod_{k\in N_j, k \ne i} g_{jk} (s_j)
\end{eqnarray}
writing $g_{ij}(s_i) = A_{ij} e^{\beta(u_{ij} s_i - v_{ij} s_i^2)}$, 
we can solve for the $u_{ij}$, $v_{ij}$ from the equations

\begin{eqnarray}
x = \sum_{k\ne i }u_{jk} \nonumber \\
y = \Delta+\sum_{k \ne i} v_{jk} \nonumber \\
-2\beta v_{ij} = log \frac{(1 + e^{-\beta y } 2 cosh(\beta(x+1)))
                          (1 + e^{-\beta y } 2 cosh(\beta(x-1)))}
                          {1 + e^{-\beta y} 2 cosh(\beta x)}        \\
2 \beta u_{ij}  = log \frac{1 + e^{-\beta y} 2 cosh(\beta (x+1))}{1 + e^{-\beta y } 2 cosh(\beta (x-1)}
\end{eqnarray}
and get the magnetization per node
\begin{equation}
m_i = \frac{2 sinh(\beta \sum_{i} u_{0i}) }{e^{\beta(\Delta + \sum_{i} v_{0i})} + 2 cosh(\beta \sum_{i} u_{0i})}
\end{equation}
These equations can be solved for specific instances\footnote{At low temperatures is convenient to observe that
$ T \log(1+2cosh(\beta x) e^{-\beta y}) \to f(x,y)$, where $f(x,y) =0$ if $|x|<y$, $f(x,y) = |x|-y$ otherwise}.
Fig.\ref{powerlaw} shows for an heterogeneous random graph the transition curves $m(T)$ and the phase diagram  as well, from both simulations
and BP approximation scheme. The agreement is very good and the picture sketched previously by the CW is correct. 
A certain degree of heterogeneity for the graph can be responsible for reentrant phenomena and inverse phase transition in this model.

Is it possible to better characterize this ``certain degree of heterogeneity''?
Fig\ref{powerlaw} also shows how a change in the exponent of the degree distribution can suppress this inverse phase transition.

\begin{figure}[h]
\begin{center}
\includegraphics*[width=0.7\textwidth]{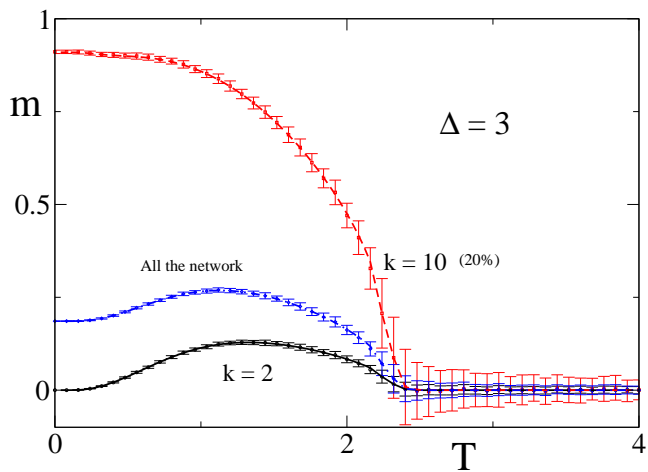}
\caption{
Transition curves $m(T)$ for the different components of a bimodal random graph. 
From Monte carlo simulations(points) and BP numerical calculations(lines).
} \label{bimodal}
\end{center}
\end{figure}

Once again we can turn to the CW approach to get useful insights. 
The zero-temperature self consistent equation takes the form
\begin{equation}
m_v = \sum_k \frac{k}{z} P(k) \theta(k m_v-\Delta)
\end{equation}
It is interesting to observe that in this approximation 
there is a degree $k^* = \frac{\Delta}{m_v}$  such that nodes with connectivity $k>k^*$ have $m_k=1$, while for the others $m_k=0$.
The fact that nodes with different connectivities can be in different phases can be easily checked for a bimodal random graph.
Fig.\ref{bimodal} shows the transition curves $m(T)$ at $\Delta=3$  of the components of a bimodal random graph with connectivity distribution 
$P(k) = 0.2 \delta_{k10} +0.8 \delta_{k2}$. The nodes with  degree $2$ show a reentrant phase transition, while the high degree nodes go to a value slighty
less then $1$ at zero temperature. 

\begin{figure}[h]
\begin{center}
\includegraphics*[width=0.7\textwidth]{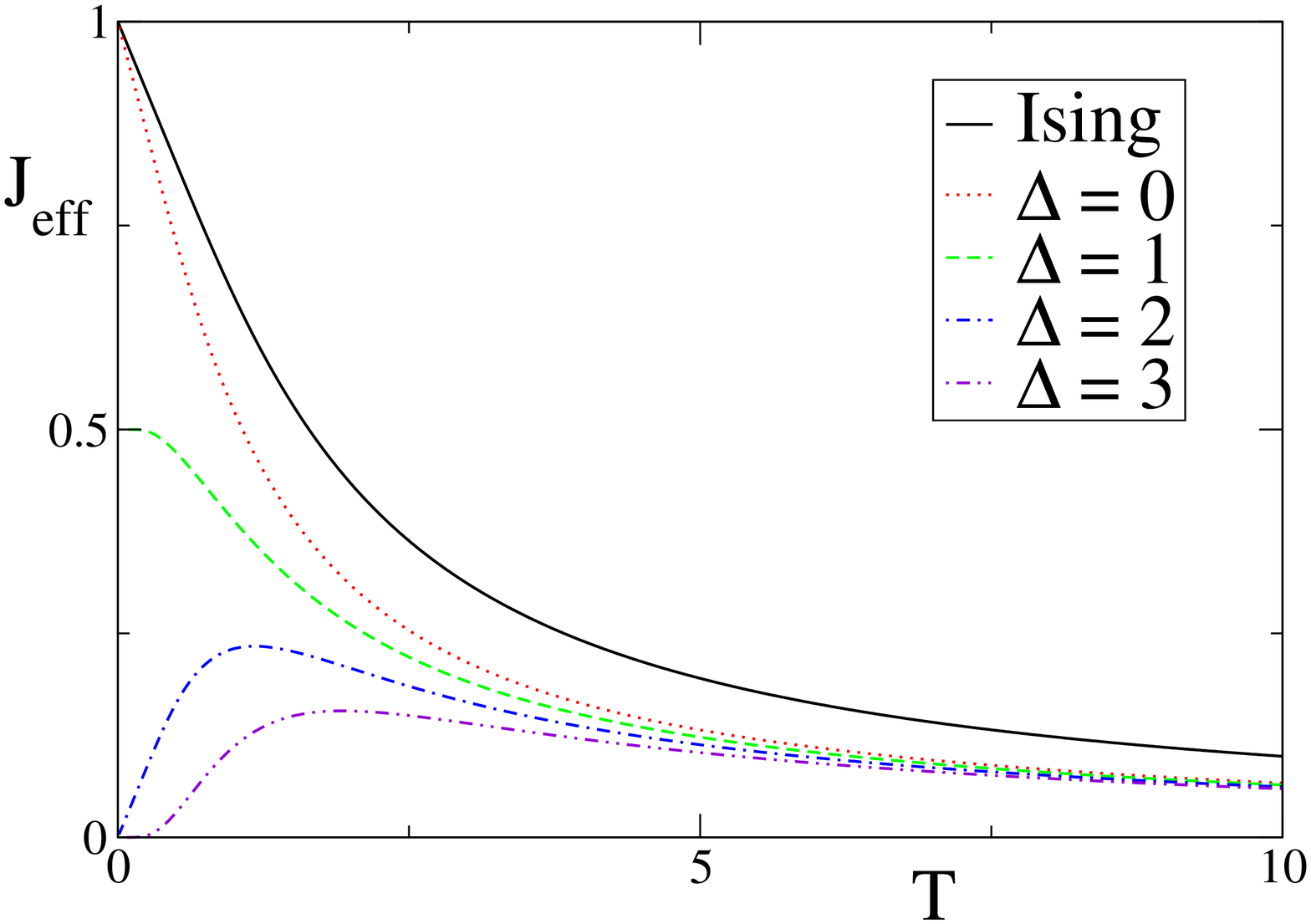}
\caption{Effective interaction $J_{eff}(T)$ between the ends 
of a node of connectivity $2$ as a function of the temperature for several $\Delta$.
} \label{inter}
\end{center}
\end{figure}

This suggests that the reentrant phenomenon can be ascribed to a mechanisms by which
low degree nodes are ``turned off'' at low temperature because they are frozen in the $s_i=0$ 
state by the effect of the chemical potential $\Delta$.
This in turn can lower the connectivity of their neighbours, with a cascade effect that can  
disconnect some parts of the graph. 
For nodes of degree $2$ this argument can be worked out rigorously.
The effective interaction transmitted by a node of degree $2$ between its ends depends on the temperature and 
it can be calculated with the use of the simplest renormalization group scheme:
\begin{equation}
2 \beta J_{eff} =  log(\frac{ 1+2 e^{-\beta\Delta} cosh(2\beta)}{1+2 e^{-\beta \Delta }})  
\end{equation}
Fig \ref{inter} shows its non-monotonous behavior for $\Delta>2$.

This argument suggests that  also the role of degree-degree correlations of the graph can be crucial for the collective behavior
of such model system.  
The degree correlations of a network can be quantified by the assortativity:
\begin{equation}
r = \frac{\langle k k' \rangle_l - \langle (k+k')/2 \rangle^{2}_l }{\langle (k^2+k'^2)/2 \rangle_l - \langle (k+k')/2 \rangle^{2}_l}
\end{equation}
Where $\langle \rangle_l$ denotes an average over the links, and $(k,k')$ denotes  the degree of the nodes at either end of links.

\begin{figure}[h]
\begin{center}
\includegraphics*[width=0.45\textwidth]{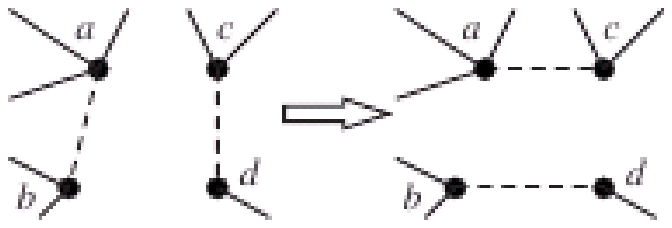}
\includegraphics*[width=0.45\textwidth]{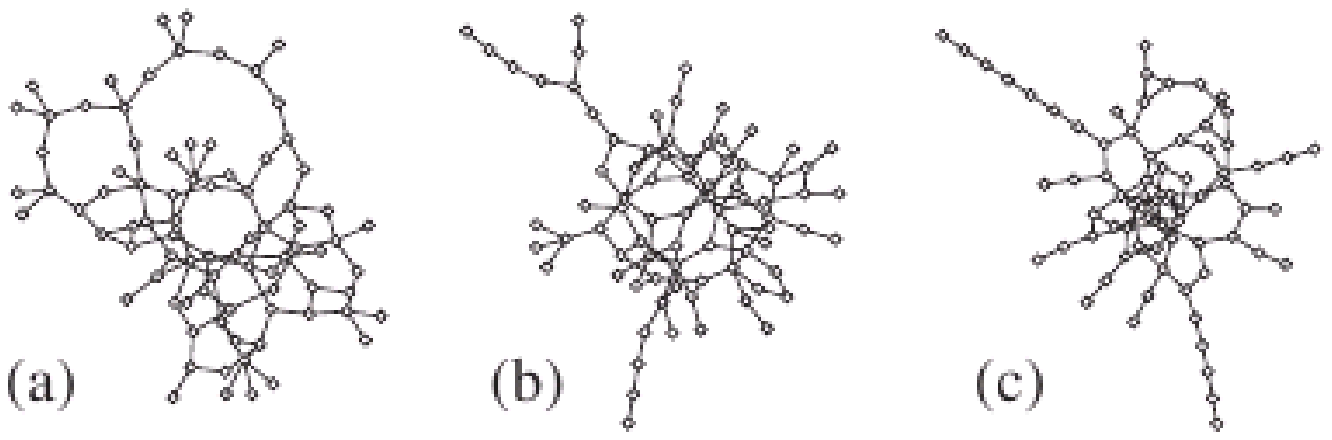}
\caption{Left: a sketch of the rewiring procedure. 
Right: Its application onto a network of $100$ nodes. $a$ is disassortative ($\theta=-1$), $b$ uncorrelated and $c$ is assortative ($\theta=1$). 
From \cite{statnet}.
} \label{rewiring}
\end{center}
\end{figure}

It is possible to obtain a graph with a given degree distribution and assortativity $r$ 
along the lines of the following exponential random graph model\cite{statnet}.
Let's Suppose that we want to construct a network model specified by an observable $x$. 
We can think of an ensemble in which the probabilistic weight 
of a given network $G$ is $P(G)\propto e^{-H(G)}$, 
where $H(G)= \theta x(G)$ and  $\theta$ should be such that $x$ is equal to the desired value.
Then a suitable monte-carlo scheme has to be adopted to sample the network ensemble.
In our case $x=r$ and $H = -\theta/2 \sum_{\langle i,j \rangle} k_i k_j$, if the degree distribution is fixed.
We can think of the following mixing procedure (see fig\ref{rewiring}). 
Two links are randomly drawn $(a,b)$ and $(c,d)$ and are sobstituted by the new links $(a,c)$ and $(b,d)$ with probability
$P = min\{1,exp(-\theta (k_a-k_d)(k_c-k_b)\}$. 
This update rule  verifies the detailed balance and doesn't change the degree of the nodes. 
The effects of this procedure are shown graphically in \ref{rewiring} for a small network.

\begin{figure}[h]
\begin{center}
\includegraphics*[width=0.7\textwidth]{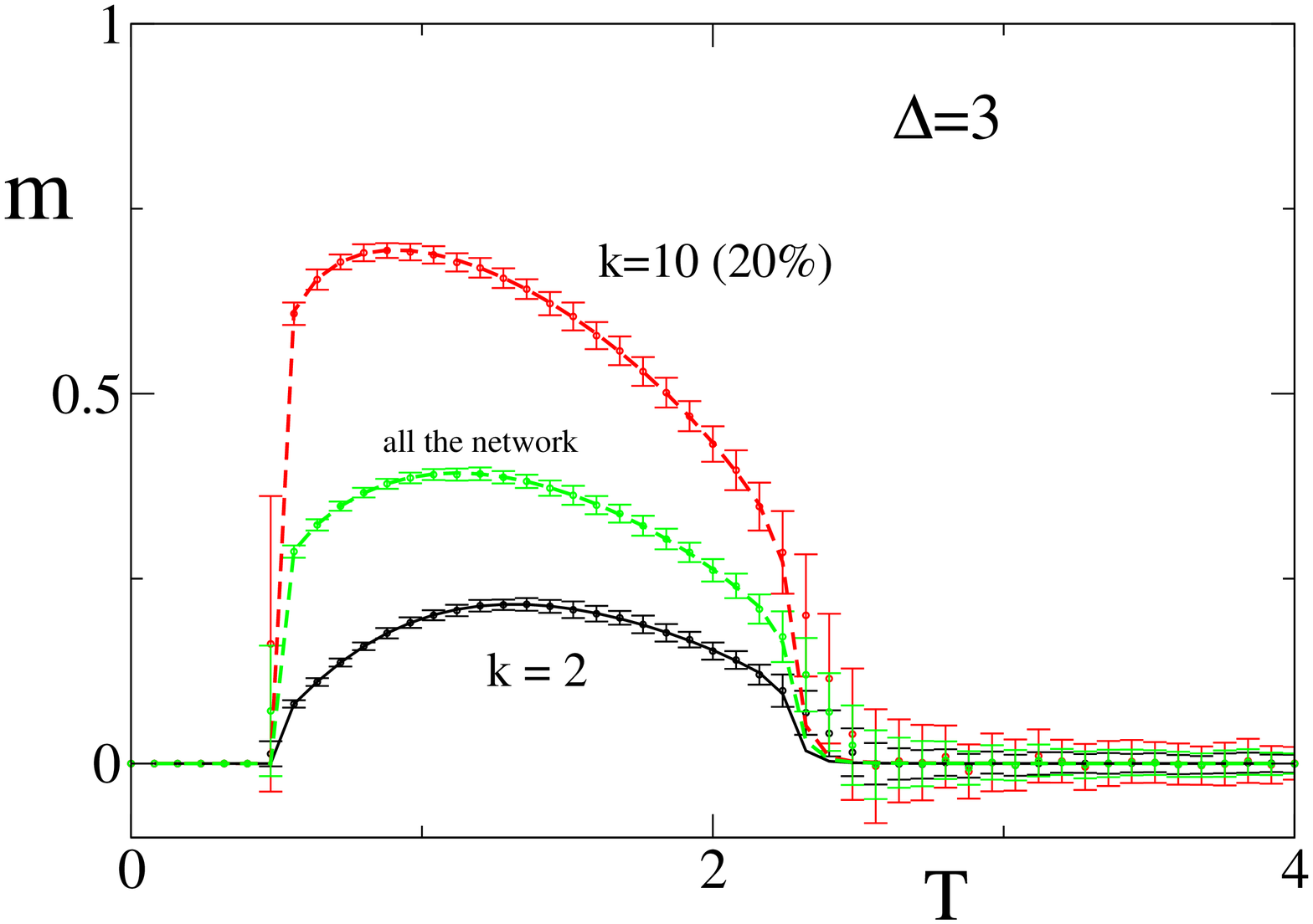}
\caption{Transition curves $m(T)$, $\Delta=3$ onto a disassortative bimodal random graph of size $N=10^4$. 
From Monte carlo simulations(points) and BP numerical calculations(lines).} 
\label{correla}
\end{center}
\end{figure}

Fig.\ref{correla} shows the transition curves $m(T)$ at $\Delta=3$ 
of the different components of the previous bimodal graph 
after this mixing procedure ($\theta=-1$, disassortative mixing).
This time the fact that the low degree nodes are turned off at decreasing temperature is enough to disconnect the whole graph of interactions, tuning 
a reentrant phase transition.
This mechanism for reentrance based on the  freezing of sparse subgraphs 
can give an explanation of the different behaviors observed in the left part of fig.\ref{powerlaw}. 
It is the case that random graphs with a power law degree distribution 
have qualitatively different structures if the value of the exponent $\gamma$ is above or below $3$.
In fact, the  number of short loops in a network with $\gamma<3$ is big. These networks are more clustered and 
sparse subgraphs should not be crucial for their connectivity, as it should be the case if $\gamma>3$\cite{matteo}.

\section{Conclusions}
In this chapter I showed how an inverse phase transition can emerge spontaneously
in the Blume-Capel without recurring to an higher degeneracy of the interacting state.
I showed a mechanism that trigger this phenomenon, based on the freezing of sparse subgraphs.
If they are crucial for the connectivity, the overall graph of interactions can be disconnected by cooling.
It should be the case that this picture is correct in general for tricritical model system and
I will give some hints about this mechanism at work also for the random field ising model.

\subsection*{The random field ising model}
In this model, ferromagnetic interacting Ising spins $s_i = \pm 1$ are subject to 
local quenched fields, the hamiltonian being:
\begin{equation}
H = -\sum_{<i,j>} s_i s_j + \sum_i h_i s_i  
\end{equation}
The $h_i$ are i.i.d. random variables, distributed in a bimodal fashion $p(h_i) = 1/2 (\delta_{h_i,h} + \delta_{h_i,-h})$.
This model was introduced to study disordered magnets and it is the minimal model
to describe phase transition in systems that show {\em crackling noise}\cite{setna}.
The response of this model to time-varying  external field has three regimes:
\begin{itemize}
\item At high $h$ the system responds as a paramagnet, 
      i.e. its magnetization follows the external field in a continuous way.
\item At low $h$ the system responds as a magnet. It responds as a paramagnet for high $T$ and as a ferromagnet
      for low $T$, i.e. the magnetization jumps discontinuosly depending on the sign of the external field.
\item At intermediate $h$ the magnetization follows the external field with little jumps. 
      The spin flip dynamics is characterized by avalanche events whose size in scale free\cite{shukla}.
\end{itemize}
The dynamics of each spin is the result of a possible competition between the local field and the effective field coming 
from the interaction with the neighbors.

\begin{figure}[h]
\begin{center}
\includegraphics*[width=0.4\textwidth]{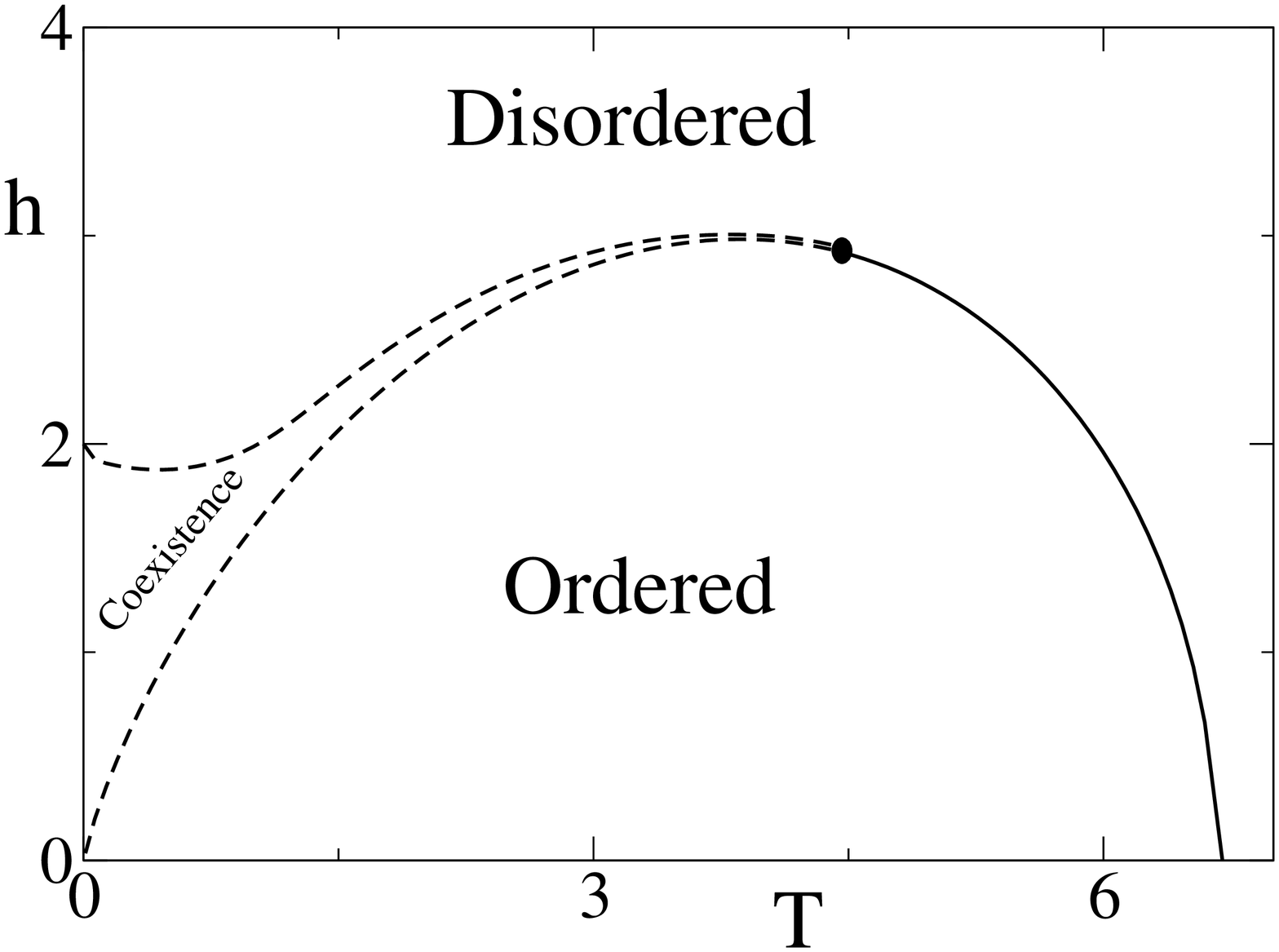}
\includegraphics*[width=0.4\textwidth]{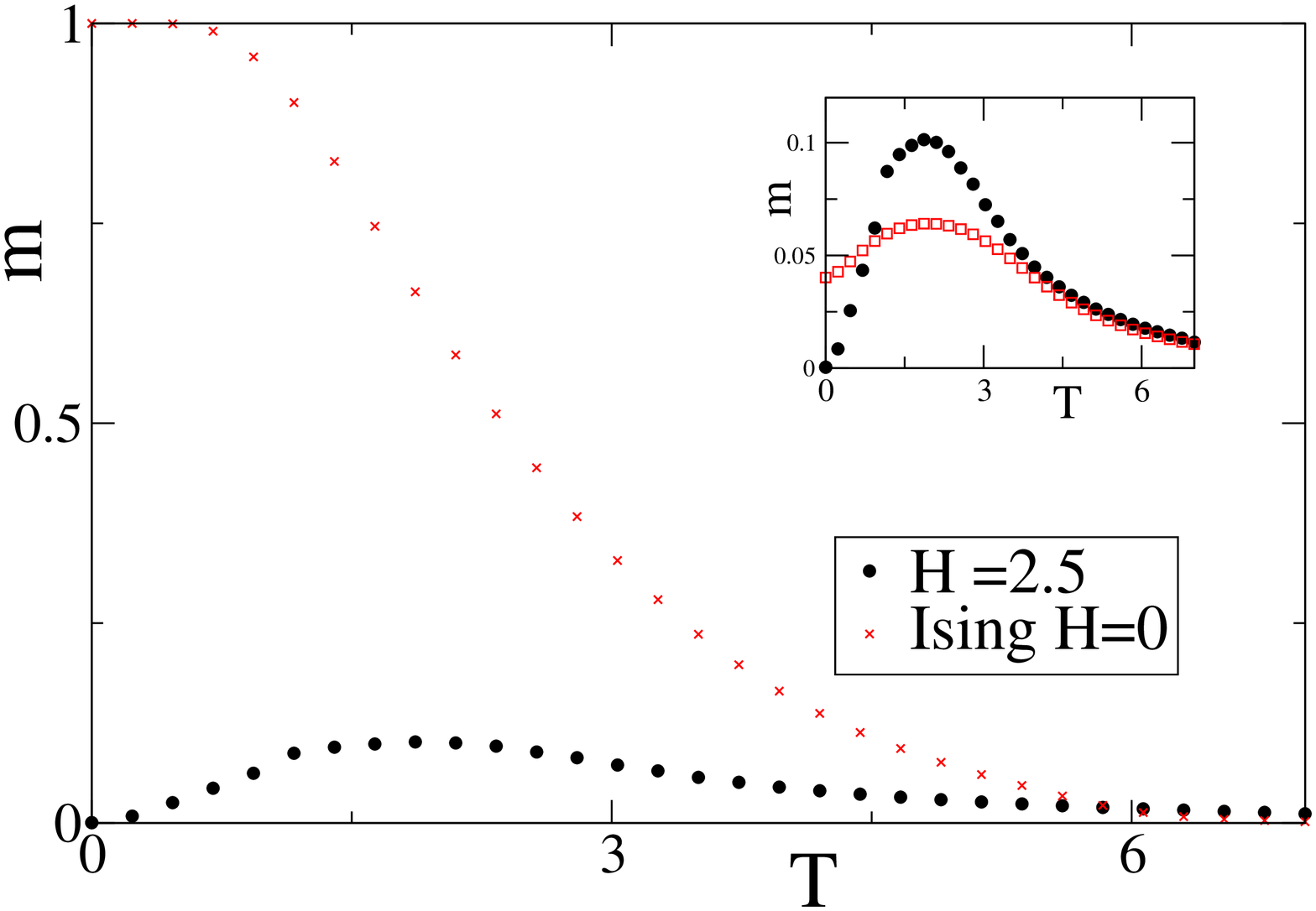}
\caption{
Left: phase diagram $(T,h)$ for the random field ising model on a random graph with degree distribution $P(k) \propto k^{-3.2}$, $k_{min} = 2$ $k_{max} = 100$. 
From CW approximation. Right: transition curves from BP calculations. The inset shows $m(T)$ ($H=2.5$)  
for different microscopic realizations of the quenched local fields.
}
\label{rfim}
\end{center}
\end{figure}

The study of the  model  on a general random graph with degree distribution $P(k)$ 
follows the same lines of the Blume-Capel model.
Hence, in the Curie-Weiss scheme, we have the self consistent equations for $m$, $m_v$, i.e. respectively the average magnetization
of a randomly chosen node and that one of a node reached following a randomly chosen link:
\begin{eqnarray}
m_v = \frac{1}{2} \sum_k \frac{kP(k)}{z} ( tanh(\beta(k m_v + h)) + tanh(\beta(k m_v - h)) ) \\
m = \frac{1}{2} \sum_k P(k) ( tanh(\beta(k m_v + h)) + tanh(\beta(k m_v - h)) )
\end{eqnarray}
In general, expanding in power of $m_v$ the rhs, we can find the critical second order $\lambda$-line $\beta \frac{<k^2>}{z} = cosh^2(\beta h)$,
until $tanh^2(\beta h) < \frac{1}{3}$. Then, at $T_c = 2/3 \frac{<k^2>}{z}$, $h_c = T_c tanh^{-1}(1/\sqrt{3})$ there is  a tricritical point,
after which the transition becomes first-order. Again, for a power law degree distribution a reentrant phase diagram is found (fig.\ref{rfim}, left).

It is possible to set up a BP approach to improve this approximation.
We have the recursive equations for the partition function:
\begin{eqnarray} 
\mathbb{Z} = \sum_{s_0} e^{\beta h_0 s_0} \prod_{j\in N_0} g_{0j} (s_0) \\
g_{ij}(s_i) = \sum_{s_j} e^{\beta ( s_i s_j + h_j s_j)} \prod_{k\in N_j, k \ne i} g_{jk} (s_j)
\end{eqnarray}
Then we can write $g_{ij}(s_i) = A_{ij} e^{\beta u_{ij} s_i}$ and we have the equations:
\begin{equation}
2\beta u_{ij} = log ( \frac{cosh(\beta(1+h_j +\sum_{k \ne i} u_{jk}))}{cosh(\beta(-1+h_j +\sum_{k \ne i} u_{jk}))}) 
\end{equation}
from which we can get the magnetization per node $m_0 = tanh(\beta(h_0 + \sum_i u_{0i})$.
Fig.\ref{rfim} (right) shows the transition curves $m(T)$ for a random graph with degree distribution $P(k) \propto k^{-3.2}$, $k_{min} = 2$ $N = 10^4$. 
At $H=2.5$ there is a weak reentrant phenomenon, but the form of the 
curves $m(T)$ is strongly dependent on the microscopic realization of the quenched local fields(inset).  
However the phase diagram of this model has a complex structure of singularities even on homogeneous graph\cite{bruinsma}
and further investigations are needed.


\chapter{Volatility and evolution of social networks}
The idea of applying methods and concepts from natural science, in particular statistical physics, 
to the study of social systems has a long history\cite{ball}.

Many social and economic phenomena have an inherent network dimension\cite{vegaredo}. 
The question of embedness of such phenomena in {\em social networks} has been  addressed directly only in recent times,
because of the recent technological development in storing and hadling large dataset of informations.

Social networks have complex structures that are evolving in time.
In particular, the same kind of network can show qualitatively different structures, 
sparse and disconnected {\em vs} dense and connected.

After a brief introduction on the structure and evolution of social networks, 
we will present the mechanism of coevolution to model their formation trough 
a class of models proposed by G.Ehrhardt et al\cite{14}. 
In these, the evolution of the network is coupled with the dynamics defined on top of it. 
A feedback effect can trigger the apparence of different phases, 
disordered and disconnected in clusters or ordered and connected, respectively, in a discontinuous way. 

Within the simplest model of this class, I will show how the kind of volatility, 
i.e. the rate at which nodes and/or links disappear, affects the evolution of the network\cite{nostro}.
It is found that when the volatility is mostly node-based the emergence of an ordered phase is definitively suppressed.

\section{Complex social networks}
The recent surge of interest of the worldwide public opinion on social networks perhaps 
has its motivations in the spreading of virtual settings like Facebook. 
Anyway, the network dimension is important in many concrete aspects of social and economic life. Examples range from informal contacts in labour market \cite{Topa} 
and peer effects in promoting (anti-)social behaviors \cite{6} to inter-firm agreement for R$\&$D \cite{8}.
This issue of embeddedness was addressed in early times\cite{5}, but only recently it is possible to deal
with it in practice thanks to the recent technological development in informatics.

Let's consider as an example the question of the social dimension in science research. 
Some insights about it can be gained from the study of coathourship networks. 
Here the nodes are scholars, that are connected by a link if they wrote a paper togheter.
Ref.\cite{newman} reports on the analysis of:
\begin{itemize}
\item A network of coauthorship of papers in the Medline bibliographical database from 1995 to 1999.
Medline is a widely used and compendious database for covering biomedical research.
\item A networks of coauthorships of physicists assembled 
from papers posted on the widely used Physics E-print archive at Cornell university between 1995 and 1999.
\item A collaboration network of mathematicians compiled from databases mantained by the journal  {\em Mathematical Reviews},
covering the period from 1940 to the present without break, from \cite{grossman}. 
\end{itemize}

\begin{figure}[h]
\begin{center}
\includegraphics[width=0.7\textwidth]{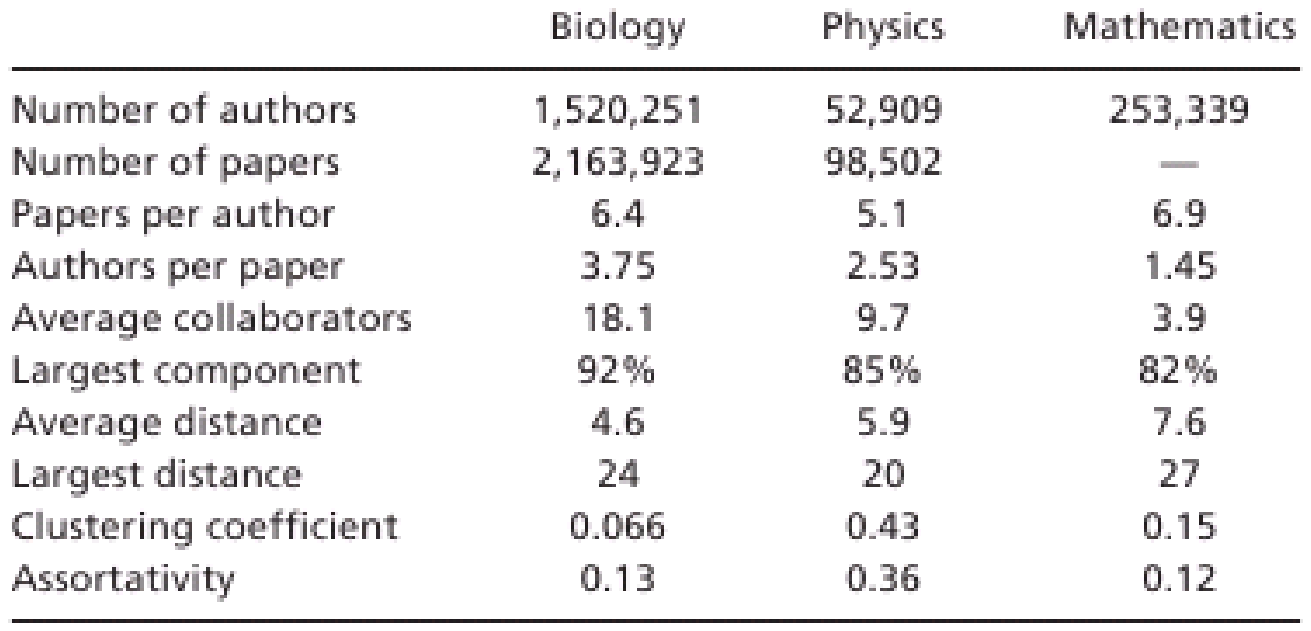}
\includegraphics[width=0.7\textwidth]{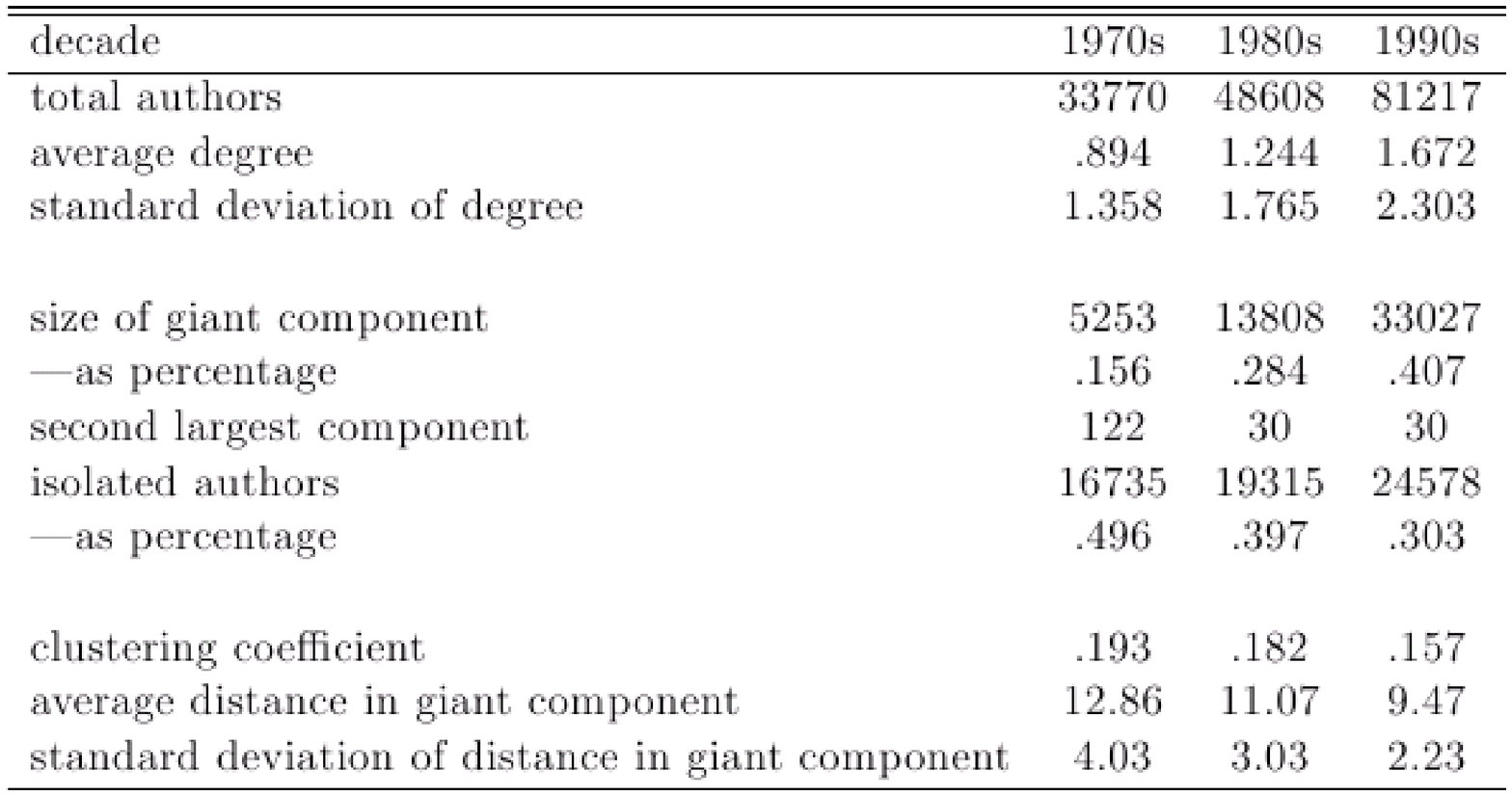}
\caption{Top: statistical features of the coauthorship networks analized in \cite{newman}. 
Bottom: Evolution of the statistical features of the coauthorship networks of economists analized in \cite{7}} 
\label{table}
\end{center}
\end{figure}

\begin{figure}[h]
\begin{center}
\includegraphics[width=0.7\textwidth]{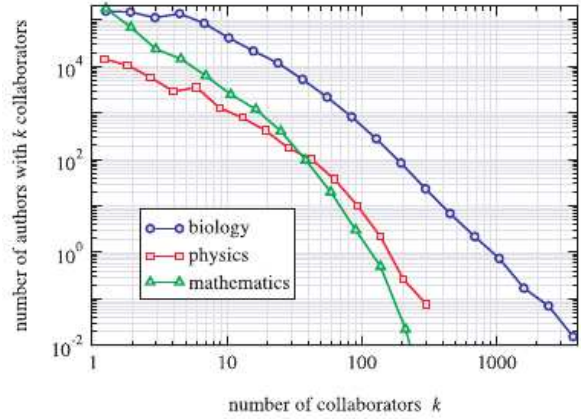}
\caption{Distribution of the number of collaborators per scientists in the coauthorship networks analized in \cite{newman}}                                                                     
\label{colla}
\end{center}
\end{figure}
\begin{figure}[h!!]
\begin{center}
\includegraphics[width=0.6\textwidth]{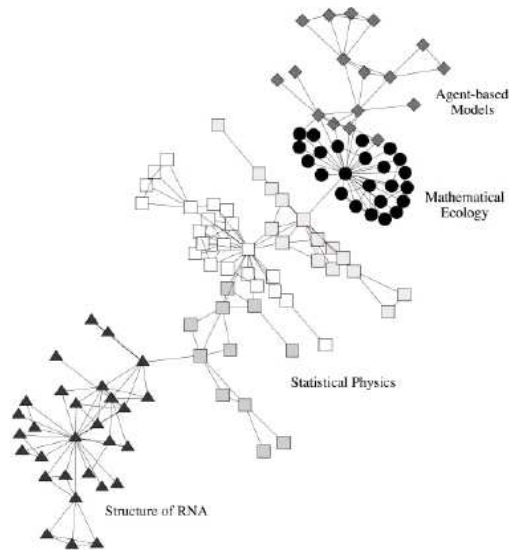}
\caption{The largest component in the network of collaboarations in the Santa Fe institute. From\cite{newman2}}
\label{santa}
\end{center}
\end{figure}

A summary of the basic statistics of these networks is given in table \ref{table}
\footnote{There are no precise data about the number of papers in the mathematics database. 
However, according to \cite{grossman} they should be around 1.6 million.}. 
These reveal many interesting features about academic communities. 
All the three communities have a largest connected component that cover the great part of the graph,
with rather small average distances and diameters.
The statement of scientific research as a common and collective enterprise immediately comes up in a graphical way.
The average number of authors per paper and the average number of collaborators is bigger in biology and smaller in mathemathics, with physics in between.
This should be presumably a result of different methods of research. Biological research in fact consists mostly  of experimental 
work by large groups of laboratory scientists. Mathematical research  instead consists of theoretical work done primarly by individuals alone or by pairs of collaborators.
Physics should be a combination of the two. 
The distribution of collaborators per scientists is shown in fig.\ref{colla}. 
They all show fat-tails, showing the presence of few scientists with a lot of collaborations and probably having in them a leading role. 
The average clustering coefficients, i.e. the probability that two collaborators of a researcher are collaborators themselves
are also rather different in different fields. This is very high in physics $0.43$, and quite low in biology $0.06$. 
The finer details of these networks show morevoer interesting community structures\cite{newman2}(see e.g. fig \ref{santa}).

Finally, these networks are subject to a dynamical evolution. 
The evolution of the coauthorhip networks of economists from the Econlit database is analyzed in \cite{7}.
The basic statistical features are shown again in fig.\ref{table}(bottom). 
The largest connected component is growing,  with emerging features similar to the previous cases, 
and the authors state of the research in economy as an {\em emerging  small world}, 
as it is called a compact network, whose diameters and average distances are small with respect to the size.

If the analysis of the structure of large and complex social networks 
needs statistical methods, the analysis of their evolution can gain insights from statistical mechanics modeling.

\section{Models of co-evolving networks}
The competition between order and disorder is by no means restricted to physics\cite{loreto}.
Also economies and societies 
-- as systems of many interacting individuals -- 
organize themselves in different (macroscopic) states, 
with different degrees of order -- informally interpreted as coordination on social norms, 
compliance with laws or conventions \cite{PYoung}. 
Besides all its inherent complexity, one important element of additional richness is that the relation between 
the degree of order in a society and the cohesion of the underlying social network is not unidirectional as in physics, 
where the topology of interactions is fixed.
The structure of the networks in these phenomena is dynamically shaped by incentives of agents (nodes), be they individuals or organizations, 
who establish bilateral interactions (links) when profitable.
In addition, this interplay typically takes place in a volatile environment.
That is, the favourable circumstances that led at the same point to the formation of a particular link may later on deteriorate, 
causing the removal or rewiring of that link. 
This combination of factors raises a number of interesting issues in statistical physics, as the collective behavior 
of the interacting degrees of freedom may radically change 
when they are coupled to the dynamics of the network they are defined on.

I will review here  the results of ref.\cite{14}.
In all the models defined there, the feedback between nodes and networks dynamics arises from assuming that 
the formation of a link requires some sort of similarity or proximity of the two parties. 
This captures different situation. For example, in cases where trust is essential in the establishment of new relationships
(e.g. in crime or trade networks), linking may be facilitated by common acquaintances or by the existence of a chain of acquaintances
joining the two parties. In other cases (e.g. in R$\&$D or scientific networks) a common language, methodology, or comparable level of technical
competence may be required for the link to be feasible or fruitful to both parties.
This class of models reveals a generic behavior characterized by a
discontinuous transition from an uncoordinated state characterized by
a sparse network, to a coordinated state on a dense network. 
As discussed in Ref. \cite{14}, this agrees with anecdotical evidence
on the observation of sharp transitions\cite{7},\cite{8} and resilience properties\cite{9},\cite{10} of some social networks. 
However, our focus here will be mostly on the statistical phenomenon, than on
its interpretation in socio-economic terms, given that the phenomenology bears a formal similarity with the liquid-gas transition. 

Consider a population of $N$ agents. They form the nodes $i=1,...,N$ of a network that is described by an undirected graph.
The formation and destruction of links proceedes by the following steps:
\begin{itemize}
\item Each node $i$ attempts to establish a new link with a randomly chosen node $j$ at rate $\eta$. 
\item Given a notion of a {\em social distance} $d_{ij}$ between nodes $i$ and $j$ , 
       if $d_{ij} \leq \bar{d}$ the link is formed, otherwise it is formed with probability $\epsilon$. 
\item Links are destroyed at rate $\lambda$
\end{itemize}

It is possible to set up a mean field approximation. 
It consists in neglecting degree correlations between neighbouring nodes, i.e.
we approximate the network with a random graph. 
Random graphs  are characterized only by their degree distribution $P(k)$. 
We make the hypothesis that the $p(k)$ satisfies a master equation whose rates are:
\begin{eqnarray}
w(k \to k-1) = \lambda k \\
w(k \to k+1) = 2\eta(\epsilon +(1-\epsilon) P(d_{ij}\leq \bar{d})) 
\end{eqnarray}

The factor $2$  comes because each node can either initiate or receive a new link.
The definition of the social distance should depend upon the specific phenomena we are looking at.
 
\begin{figure}[hb!!!]
\begin{center}
\includegraphics[width=0.7\textwidth]{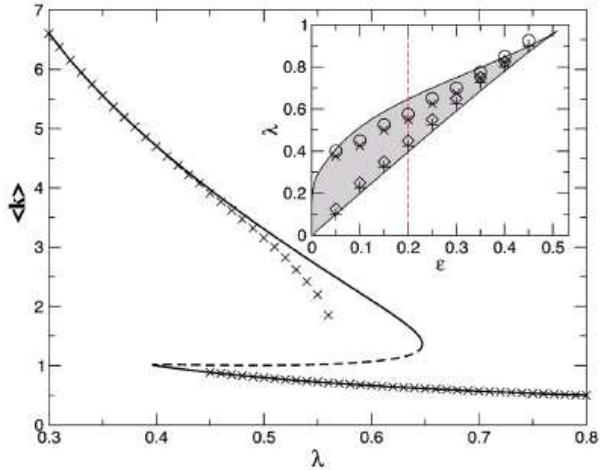}
\caption{Mean degree $\langle k\rangle$ as a function of $\lambda$ for $\epsilon=0.2$,$\eta=1$, when $d_{ij}$ is the distance on the graph and $\bar{d}=N-2$. 
Lines are from  mean field theory, points from simulations, starting from both low and high connected phases. Inset: phase diagram from the mean field. Coexistence occurs within the shaded region
, whereas above(below) only the sparse(dense) state is stable. Numerical simulation agree qualitatively The high(low) density state is stable up (down) to the points marked
with  $X$ ($\diamondsuit$) and is unstable at points marked with $\bigcirc$($+$). From \cite{14}}
\label{socialnet}
\end{center}
\end{figure}

A first simple specification can be with $d_{ij}$ being the geodetic distance on the graph, and $\bar{n} =N-2$.
This describes a situation in which the formation of new links is strongly influenced by proximity on the graph.
If $i$ and $j$ are in different components the rate of link formation is $2\epsilon \eta$, otherwise is $2\eta$. In the large $N$ limit the latter only occurs
if the graph has a giant component  with a finite fraction $\gamma$ of the nodes.
For random graphs (see e.g. \cite{vegaredo})    
$\gamma = 1-\phi(u)$, where $\phi(s) =\sum_k p(k) s^k$ is the generating function, and $u$ is the probability that a link,
followed in one direction, does not lead to the giant component.
This latter satisfies
\begin{equation}
 u = \phi'(u)/\phi'(1) 
\end{equation}
Hence $u^k$ is the probability that an agent with degree $k$ is not in the giant component, and then
\begin{equation}
w(k \to k+1) = 2\eta(\epsilon +(1-\epsilon) \gamma (1-u^k)) 
\end{equation}
The stationary state condition brings  the equation for $\phi$:
\begin{equation}
\lambda \phi'(s) = 2\eta (\epsilon +(1-\epsilon)\gamma)\phi(s) -2\eta(1-\epsilon)\gamma\phi(us)
\end{equation}
which can be solved numerically to the desidered accuracy.
The solution of this equation is summarized in fig.\ref{socialnet}. 
Either one or three solutions are found, depending on the parameters. In the latter case
the intermediate solution (dashed line in fig.\ref{socialnet}) is unstable and it separates the basins of attraction of the two stable solutions within this mean field approach.
The solution is exact when there is no giant component, and numerical simulations show 
that the approach is very accurate away from the phase transition. 

\begin{figure}[h]
\begin{center}
\includegraphics[width=0.75\textwidth]{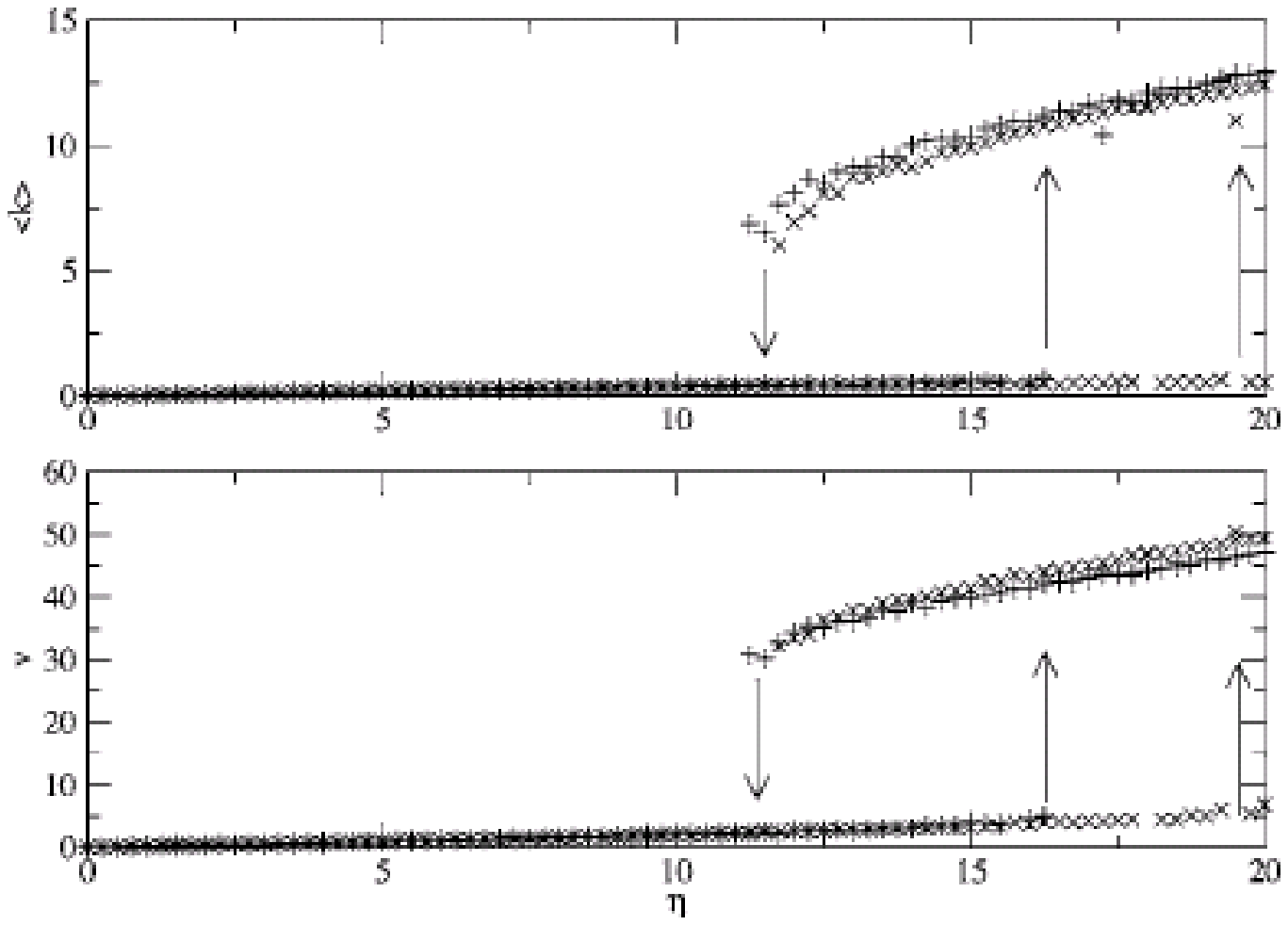}
\caption{Mean degree $\langle k \rangle$ and growth rate $v$ as a function of $\eta$ from numerical simulations of the BP model.
Shown are simulations with $N=500$ (plusses) and $1000$ (crosses). Arrows denote the approximate point at which the system jumps
from one phase to the other. Here $\epsilon=0.001$, $\Delta=0.1$ and similarity treshold $\bar{d}=2$. From \cite{14}}
\label{socialnet2}
\end{center}
\end{figure}

Next let's consider a setup in which $d_{ij}$ reflects the proximity of nodes in terms of some continuous, non negative attributes $h_i$.
In short, the attributes could represent the level of technical expertise of two firms involved in a  R$\&$D partnership, or the competence
of two researchers involved in a joint project. Each agents updates his attribute $h_i$ with a rate $\nu$, that we suppose much larger than $\lambda$ and $\eta$.
Let's explore a setting of {\em best practice} imitation (BP) 
where individuals aim at improving in the direction of increasing $h_i$ by on site efforts and by learning from
their neighbors.
We posit that $h_i(t^+) = max_{N_i} \{ h_j(t) \} +\eta(t)_i$, 
where $\eta_i$ are i.i.d. gaussian random variables with zero mean and variance $\Delta$, that capture 
idyosincratic change of expertise due to $i$'s own (say research) efforts.
We set the distance $d_{ij} =|h_i-h_j|$. Fig \ref{socialnet2} reports typical results of simulations of this model.
As in the previous model, there is a discontinuous transition between a sparse and dense network state, characterized by hysteresis effects.
In the stationary state $h=\langle h_i\rangle$ grows linearly in time with velocity $v$. 
Notably the growth is much faster in the dense state that in the sparse one.
This model exihibits an interplay between the process on the network, i.e. the dynamics of the $h_i$ and the network evolution. It is this interdependence
and the corresponding positive feedback that produces the discontinuous transition and phase coexistence.

The last specification I consider is such that link formation requires  some form of coordination or compatibility.
For example, a profitable interaction may fail to occur if the two parties do not speak the same language and/or do not adopt 
compatible technologies or standards.
We can characterize each agent $i$ with a variable $\sigma_i$ which represents the social
norm (convention or technological standard) adopted.
There are $q$ possible social norms, i.e. $\sigma_i\in\{1,\ldots,q\}$. 
I will call them colors.We impose that the formation 
of a new link between $i$ and $j$ requires that $\sigma_i=\sigma_j$.
The color of a node is updated with rate $\nu$ to the
color of any of its neighbors, unless the node is isolated.
In the latter case the nodes takes a random color.
In terms of statistical physics, the model can be
thought of as a $q$ state Potts model defined on a graph of $N$ nodes,
with $T=0$, that evolves in a coupled fashion to the dynamics of the system.
This model is solved exactly in \cite{14}, but given its simplicity,  
we will see in the next paragraph how to generalize it to take into account
the possibility of agents'turnover, i.e. a node-based volatility.

\section{Node-based volatility}
Indeed, the effect of volatility was up to now limited to link
removal, but the turnover of agents (i.e. node
removal and arrival) may be an important factor in many real systems.
In order to investigate this question, we concentrate on the
simplest model for which a full analytic treatment is possible. In the
concluding section, we argue that this qualitative change is expected
in a wider class of model, and it can have much stronger effects.
I consider the model of coordination sketched in the end of the last paragraph,
but I generalize it, considering the possiblity that with rate $\alpha$ all the links of a node disappear.
This will show how the alternative assumptions of link or node based volatility have profound
effects on the dynamics of network formation\cite{nostro}.
In the following I set for sake of simplicity $\lambda=1$ and I rescale by a factor  $2$ the link creation rate $\eta$.  

Therefore, the parameter $\alpha$ interpolates between two kinds of volatility. For $\alpha=0$ volatility only affects links and for
$\alpha\gg 1$ it mostly affects nodes. As observed in Ref. \cite{14}, the color update rule is
effective only for isolated nodes, in the long run, and in that case the color is drawn at random. Since only links between same
type agents are created, after a transient all nodes are either isolated, or connected to nodes of the same color. Therefore the
particular way in which the neighbor is chosen is immaterial. For example, both a majority rule (most frequent color) or a voter-type
rule (random neighbor) would give the same dynamics.  The model can be generalized to a probabilistic update rule for the colors
introducing a finite temperature $T$ (see \cite{14}). Results do not change considerably as long as $T$ is small enough, so we shall
confine ourselves to the $T=0$ case.

Ref. \cite{14} has shown that for $\alpha=0$, the system shows an
hysteretic transition in $\eta$ from a symmetric to an asymmetric
state\footnote{I will recover it as a special case}. 
The symmetric state is characterized by a sparse network, with
average degree $\langle k\rangle<1$, with a symmetric distribution of
colors. In the asymmetric state, instead, a dense network with
$\langle k\rangle>1$ arises, along with a dominant color, which is
adopted by agents more frequently than the others. All the colors
are a priori equivalent and the fact that only one is selected
is a simple example of a {\em spontaneous simmetry breaking}.
The dominant component is selected by random fluctuations and is
stabilized by the feedback mechanism between link formation (which
is more frequently successful for nodes of the dominant component) and
the freezing of the colors of connected nodes (akin to the
ferromagnetic interaction in Potts models).
In this sense, the
model shows how order and disorder are intimately related with the
dynamics of the social network in a volatile environment.

In what follows, we solve the model in the stationary state for
$N\to\infty$, for all the values of $\alpha$. We find that the
$\alpha=0$ behavior is generic for all $\alpha<1$, but the transition
is softened as $\alpha$ increases. For $\alpha>1$ instead we show that
the system is always in the symmetric phase. Hence, in terms of
statistical mechanics, $\alpha=1$ is a second order critical point
separating a phase with spontaneously broken symmetry from a symmetric
phase.

If we call $n_{k,\sigma}$ the density of nodes with $k$ links and color $\sigma=1,\dots,q$ we have the following rate equations:
\begin{eqnarray}
\dot{n}_{k,\sigma} & = &
(k+1)n_{k+1,\sigma}-kn_{k,\sigma}-\alpha n_{k,\sigma}+ \nonumber\\
 & & {} + x_\sigma(n_{k-1,\sigma} - n_{k,\sigma}) \\
\dot{n}_{0,\sigma} & = &  \alpha\sum_{k>0} n_{k,\sigma}
+ n_{1,\sigma}-x_\sigma n_{0,\sigma}+ \nonumber\\
 & & {} + \frac{\nu}{q}\sum_{\sigma'=1}^{q}(n_{0,\sigma'}-n_{0,\sigma} )
\end{eqnarray}
where, for future convenience, we have introduced the dynamical variables
\begin{equation}
x_\sigma=\eta\sum_{k=0}^\infty n_{k,\sigma}.
\end{equation}
Making the sum over all $k$ of these equations and multiplying by $\eta$ we find
\begin{equation}
\label{dotnsig}
\dot x_\sigma=\frac{\eta\nu}{q}\sum_{\sigma'=1}^{q}(n_{0,\sigma'}-n_{0,\sigma} )
\end{equation}
which implies that, in the stationary state, each component has the same fraction $n_{0,\sigma}=n_0/q$ of disconnected ($k=0$) nodes. 
it is straightforward to derive an equations for the characteristic functions $\pi_\sigma(s)$ of the degree distribution 
$p_{\sigma}(k)=n_{k,\sigma}/\sum_q n_{q,\sigma}$ of the component $\sigma$. In the stationary state this reads:
\begin{equation}
(1-s)\frac{d\pi_{\sigma}}{ds}
=[\alpha+x_{\sigma}(1-s)]\pi_{\sigma}(s)-\alpha .
\end{equation}
The stationary solution is found by direct integration:
\begin{equation}
\label{charp}
\pi_\sigma(s)=\alpha\int_0^1\!dz z^{\alpha-1}e^{-x_\sigma(1-s)(1-z)}
\end{equation}
It is easy to see that this interpolates between a Poisson
distribution, $\pi_\sigma(s)=e^{x_\sigma(s-1)}$ for $\alpha\to
0$, which coincides with the result of Ref. \cite{14}, and an
exponential distribution
$\pi_\sigma(s)=\alpha/[\alpha+x_\sigma(1-s)]$ for $\alpha\to\infty$.
The latter limit is derived upon changing variables to $y=z^\alpha$ in
Eq. (\ref{charp}) and expanding $1-y^{1/\alpha}\simeq
-\frac{1}{\alpha}\log y$ in the argument of the exponential.
Notice also that the average degree in component $\sigma$ is $\langle
k\rangle_\sigma=\pi_\sigma'(1)= x_\sigma/(1+\alpha)$. This is
precisely what one expects from balance of link creation and
destruction of links in component $\sigma$.

Observing that $\pi_\sigma(0)=\eta\frac{n_{0,\sigma}}{x_\sigma}=\frac{\eta n_0}{q x_\sigma}$ we find an equation for $x_\sigma$ in the stationary state, which reads
\begin{equation}
\label{xsig}
G_\alpha(x_\sigma) \equiv \alpha x_\sigma\int_0^1\!du u^{\alpha-1}e^{x_\sigma(u-1)} = \frac{\eta n_0}{q}.
\end{equation}
Notice that the r.h.s. of Eq. (\ref{xsig}) is independent of $\sigma$. The variables $x_\sigma$ are determined by Eq. (\ref{xsig}) and the normalization condition, which takes the form
\begin{equation}
\label{norm}
\sum_{\sigma=1}^qx_\sigma =\eta.
\end{equation}
The properties of the solutions of Eqs. (\ref{xsig},\ref{norm}) depend on the behavior of the function $G_\alpha(x)$, which are discussed in the appendix,
and can be classified in symmetric and asymmetric solutions.

\subsection{$\alpha>1$: The symmetric solution}

For $\alpha>1$ the function $G_\alpha(x)$ is a monotone increasing function. 

The function $G_\alpha(x)$ can be written as
\[
G_\alpha(x)=\alpha\int_0^x\!dz\left(1-\frac{z}{x}\right)^{\alpha-1}e^{-z}
\]
For $\alpha>1$ we have
\[
\frac{dG_\alpha}{dx}=\frac{\alpha(\alpha-1)}{x^2}
\int_0^x\!dzz\left(1-\frac{z}{x}\right)^{\alpha-1}e^{-z}>0
\]

Hence Eq. (\ref{xsig}) 
has a single solution and Eq. (\ref{norm}) implies that $x_\sigma=\eta/q$ for all components $\sigma$. 
Notice also that
\[
\frac{d}{dx}\frac{G_\alpha(x)}{x}=-\alpha
\int_0^1\!duu\left(1-u\right)^{\alpha-1}e^{-ux}
\]
i.e. $n_0=G_\alpha(\eta/q)/(\eta/q)$ in the symmetric solution is a decreasing function of $\eta/q$. In addition $G_\alpha(x)\simeq x$ for $x\ll 1$, i.e. $n_0\to 1$.
Hence Eq. (\ref{xsig}) yields the total fraction of disconnected nodes
\[
n_0=\frac{q}{\eta}G_\alpha\left(\eta/q\right)
\]
as a function of the parameters $q,\alpha$ and $\eta$. We can analyze
the stability of the symmetric solution recalling that
$\eta n_{0,\sigma}= G_\alpha(x_\sigma)$. Then Eq. (\ref{dotnsig})
becomes a dynamical equation for $x_\sigma$
\begin{equation}
\label{dotxsig}
\dot x_\sigma = \frac{\nu}{q}\sum_{\sigma'=1}^{q}\left[G_\alpha(x_{\sigma'})-G_\alpha(x_{\sigma} )\right].
\end{equation}
Linear stability of the symmetric solution is addressed by setting $x_\sigma=\eta/q+\epsilon_\sigma$, with $\sum_\sigma \epsilon_\sigma=0$. Then to linear order
\begin{equation}
\label{stabsymm}
\dot\epsilon_\sigma= \frac{\nu}{q}G_\alpha'(\eta/q)
\sum_{\sigma'=1}^{q}[\epsilon_{\sigma'}-\epsilon_\sigma]=
-\nu G_\alpha'(\eta/q) \epsilon_\sigma.
\end{equation}
Hence, as long as $G_\alpha(x)$ is an increasing function of $x$, the symmetric solution is stable. This is always the case for $\alpha>1$, as we shall see, it fails to hold for $\alpha<1$.

\subsection{$\alpha<1$: The asymmetric solution}

For $\alpha<1$ the symmetric solution still exists.
However the function $G_\alpha(x)$ now has a maximum for some $x_0(\alpha)$  
and $G_\alpha(x)\to \alpha$ from above as $x\to\infty$.
Therefore the symmetric solution becomes unstable when $\eta>\eta_+$ where
\begin{equation}
\eta_+\equiv q x_0(\alpha)
\end{equation}
because beyond that point $G_\alpha'(\eta/q)<0$.

The occurrence of a maximum in $G_\alpha$ also implies that Eq.
(\ref{xsig}) admits solutions with $x_\sigma =x_-<x_0(\alpha)$ for
some $\sigma$'s and $x_\sigma=x_+>x_0(\alpha)$ for the other
components. Since $x_\sigma$ is related to the density of a component
$\sigma$, we shall call a component dense if $x_\sigma=x_+$ and
diluted if $x_\sigma=x_-$.  As in Ref. \cite{14}, all solutions with
more than one dense component are unstable. Indeed, by the same
argument used to analyze the stability of the symmetric solution, a
perturbation with $\epsilon_\sigma=0$ for all diluted components would
grow as $\dot\epsilon_\sigma=-\nu G_\alpha'(x_+)\epsilon_\sigma$ on
all dense components. These unstable modes correspond to density
fluctuations across dense components. Once one of these components
acquires slightly more mass, the density of links in it increases,
which makes it less likely for nodes in this component to become
isolated. At the same time, this component will recruit isolated nodes
at a slightly faster pace, due to its larger density. It is then
intuitively clear that the initial density perturbation will grow
unboundedly.

\begin{figure}[h!!]
\includegraphics*[scale=1.3, angle=0]{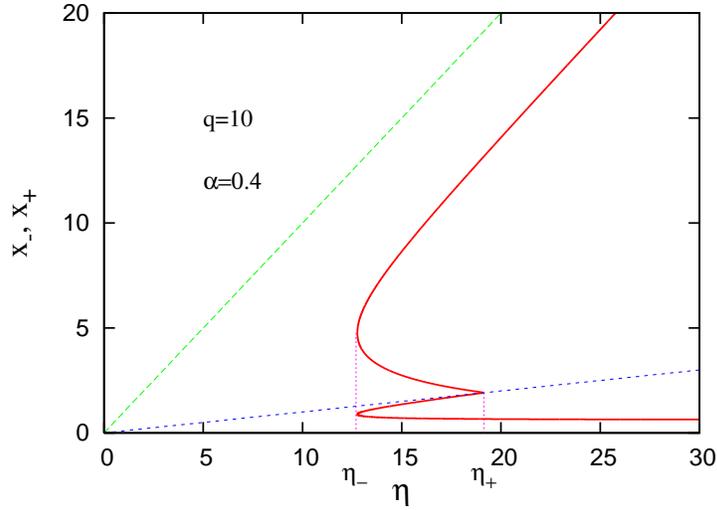}
\caption{Solutions $x_\pm$ as a function of $\eta$ for $q=10$ and $\alpha=0.4$. The dashed line $x=\eta/q$ separating the two curves is the symmetric solution.}
\label{figxb}
\end{figure}

These unstable modes ($G_\alpha'(x_+)>0$ implies $\dot\epsilon_+>0$) 
are clearly absent in the solution with only one
dense component. These are the asymmetric solutions we shall focus on
in what follows. There are $q$ of them, depending on which color is
associated with the dense component. The variables $x_\pm$ are
determined by the system of equations
\begin{eqnarray}
G_\alpha(x_+) & = & G_\alpha(x_-) \\
x_++(q-1)x_- & = & \eta \label{norma}
\end{eqnarray}
This solution is shown in Fig. \ref{figxb}.

Actually, of the two asymmetric solutions the one with $x_+$ decreasing with $\eta$ is clearly unphysical as this would have a connected component 
with an average degree $\langle k\rangle_\sigma=x_+/(1+\alpha)$ which decreases with the rate $\eta$ with which links are formed. 
As in ref. \cite{14}, it is easy to see that only solutions with $x_+$ increasing in $\eta$ are  stable. Indeed, regarding $\eta$ and $x_-$ as functions of $x_+$ in Eq. (\ref{norma}) we find
\[
\frac{d\eta}{d x_+}=1+(q-1)\frac{d x_-}{d x_+}=\frac{G_\alpha'(x_-)+(q-1)G_\alpha'(x_+)}{G_\alpha'(x_-)}.
\]
Consider perturbations of the form $x_\sigma=x_++\epsilon$ for the dense component and $x_\sigma=x_--\epsilon/(q-1)$ for the others. Then by a derivation analogous to that leading to Eq. (\ref{stabsymm}), we find
\[
\dot\epsilon=-\frac{\nu}{q}\left[G_\alpha'(x_-)+(q-1)G_\alpha'(x_+)\right]\epsilon
=-\frac{\nu}{q}G_\alpha'(x_-)\frac{d\eta}{d x_+}\epsilon.
\]
Given that $G_\alpha'(x_-)>0$, this implies that on solutions with $x_+$ decreasing with $\eta$, the perturbation $\epsilon$ grows unboundedly.

The asymmetric solution ceases to exist for $\eta<\eta_-$\footnote{We note, in passing, that the condition $d\eta/dx_+=0$ provides an equation which allows to determine $\eta_-$.}.
In the region $\eta\in [\eta_-,\eta_+]$ both the symmetric and the asymmetric solutions co-exist. The coexistence region, in the $\alpha,\eta$ plane is reported in Fig. \ref{figphasediag}.

\begin{figure}[hb!!!]
\includegraphics*[scale=0.35,angle=270]{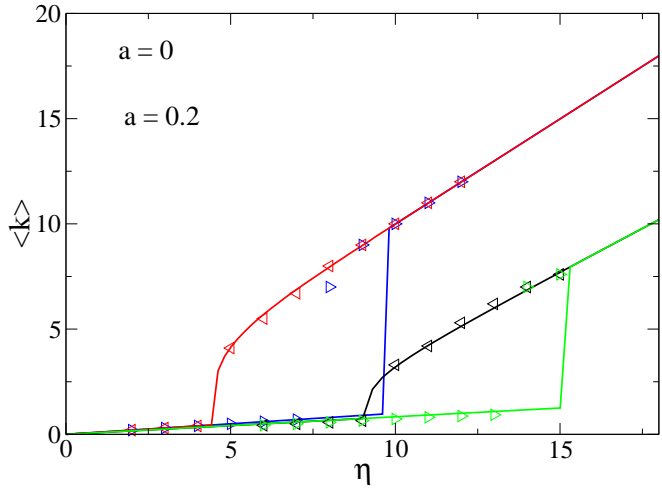}
\caption{Mean degree $<k>$ as a function of $\eta/\lambda$  for a system with $q=10$ colors, for $\alpha=0$ and $0.2$, simulations are
for systems of $1000$ nodes.}\label{figsweep}
\end{figure}

\begin{figure}[h!!!]
\includegraphics*[scale=1.2,angle=0]{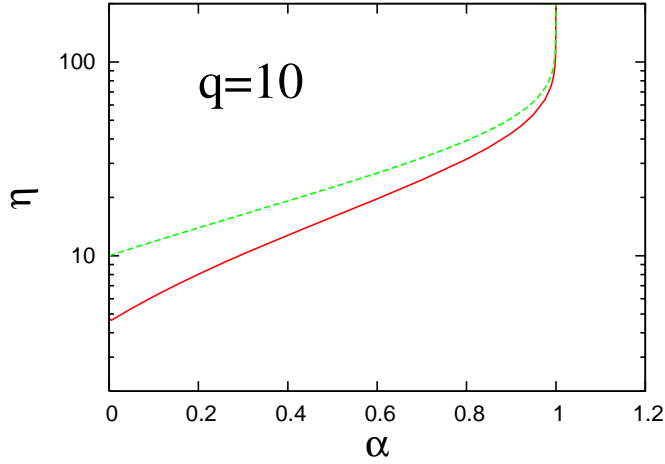}
\caption{Phase diagram for $q=10$. The symmetric phase extends below and to the right of the (full) line $\eta_-(\alpha)$ whereas above the (dashed) line $\eta_+(\alpha)$ only the asymmetric phase is stable. The coexistence region, where both phases are stable, is delimited by the two curves.}\label{figphasediag}\end{figure}

The practical relevance of the results derived so far is best discussed introducing an order parameter
\begin{equation}
m=\frac{x_+-x_-}{\eta}
\end{equation}
which is the difference in the density of the dense and diluted components. This
vanishes in the symmetric phase and is non-zero in the asymmetric one.
In Fig. \ref{figsweep} where we report the behavior of the average degree of the network
\begin{equation}
\langle k\rangle=\sum_{k,\sigma} n_{k,\sigma}  k=\frac{\eta}{1+\alpha}\frac{1+(q-1)m^{2}}{q}.
\end{equation}

Fig. \ref{figsweep} shows that as $\eta$ sweeps through the coexistence region the system undergoes an hysteresis loop: the degree jumps from low to high values at $\eta_+$ as $\eta$ is increased whereas when $\eta$ decreases from large values, the network collapses back to the symmetric phase when $\eta_-$ is crossed. In the case $\alpha=0$ \cite{14}, the symmetric phase is characterized by sparse networks, with a vanishing giant component. This is no more true when $0<\alpha<1$, specially close to $\eta_+$ \footnote{Indeed the condition for the presence of a giant component is $\langle k(k-1)\rangle_\sigma>\langle k\rangle_\sigma$ which, by a straightforward calculation, reads $\eta\ge q(1+\alpha/2)$. At the critical point $\eta_+=qx_0(\alpha)$ this reads $x_0(\alpha)\ge 1+\alpha/2$ which holds true for all $\alpha>0$.}. Numerical simulations fully confirm this picture, even though for finite systems the symmetric (asymmetric) phase is meta-stable close to $\eta_+$ ($\eta_-$) and therefore the transition occurs for lower (larger) values of $\eta$.

\subsection{The critical region: $\alpha\approx 1$}

\begin{figure}[hb!!]
\includegraphics*[scale=0.35,angle=270]{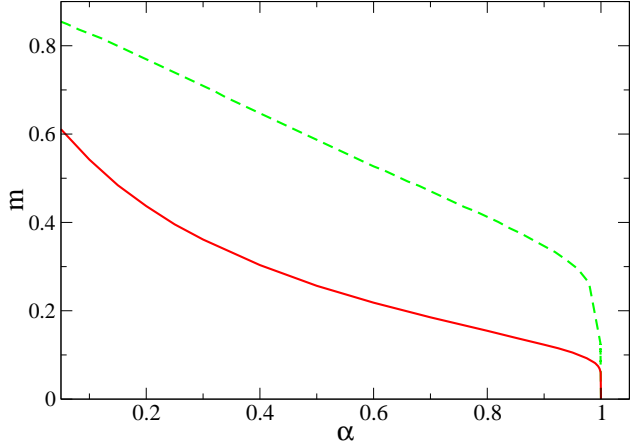}
\caption{Order parameter $m$ on the boundary of the coexistence region
  $\eta_-$ (full line) and $\eta_+$ (dashed line) as a function of
  $\alpha$ for $q=10$.}\label{figm}
\end{figure}

The behavior of the order parameter $m$ on the critical lines which confine the coexistence region is shown in Fig. \ref{figm}. 
This shows that the transition is continuous but with a peculiar critical behavior.
For $\alpha=1-\epsilon$ we can approximate
\[
   G_{\alpha}(x)\simeq(1-\epsilon)\int_0^x du\left[1-\epsilon\log\left(1-\frac{u}{x}\right)\right]e^{-u}
\]
\[
     = (1-\epsilon)(1-e^{-x}+\epsilon(E_{i}(x)-\gamma))
\]
where $E_{i}(x)$ is the exponential integral function, and for $\epsilon \to 0$ we have $x_{\pm} \to \infty$ and $E_{i}(x)\simeq \frac{e^{x}}{x}$i.
Hence
\[
        G_{\alpha} \simeq (1-\epsilon)(1-e^{-x}+\epsilon/x)
\]
Exactly at the critical point $\eta = qx_{0}$, where $x_{0}$ is such that $G'_{\alpha}(x_{0})=0$.
We have  $\epsilon \frac{e^{x_{0}}}{x_{0}^{2}}=1$  
from which we can get
$x_{0}\simeq -\log\epsilon+2\log|\log\epsilon|$.

From $G_{\alpha}(x_{+})=G_{\alpha}(x_{-})$ and the expressions of $x_-$ and $x_+$  with respect to $x_0$ and $m$,  
we finally have


\begin{equation}
m\sim c/x_0(\alpha)\sim |\log(1-\alpha)|^{-1}.
\end{equation}
where c is given self-consistently by:
\[
 c=e^{c}(1-e^{-qc})/q
\]


In terms of the usual description of critical phenomena, where $m\sim |1-\alpha|^\beta$, this model is consistent with an exponent $\beta=0^+$.
Indeed, the singular behavior of $m$ is very close to that of a first order phase transition.

\subsection{Conclusions}

The introduction of node volatility, in the simple model discussed here, makes the transition from a symmetric 
(disordered) diluted network to an asymmetric (ordered) dense network less sharp. 
Indeed when node volatility dominates ($\alpha>1$) the transition disappears altogether, and the symmetric (disordered) state prevails. 
The phenomenology is strongly reminiscent of that of first order phase transitions 
(e.g. liquid-gas or paramagnet-ferromagnet) though the critical behavior is highly non-trivial.

The virtue of the particular model studied is that it allows a detailed analytic approach which allows one to gain insight on all aspects of its behavior. 
This model belongs to a general class of models which embody a generic feedback mechanism between the nodes and the network they are embedded in, which can be expressed in the following way: while the network promotes similarity or proximity between nodes, proximity or similarity enhances link formation.
This feedback allows the system to cope with environmental volatility, which acts removing links at a constant rate. 
Interestingly, the emergence of an ``ordered'' state plays a key role in this evolutionary struggle.
\begin{figure}[h!!!]
\includegraphics*[scale=0.25,angle=270]{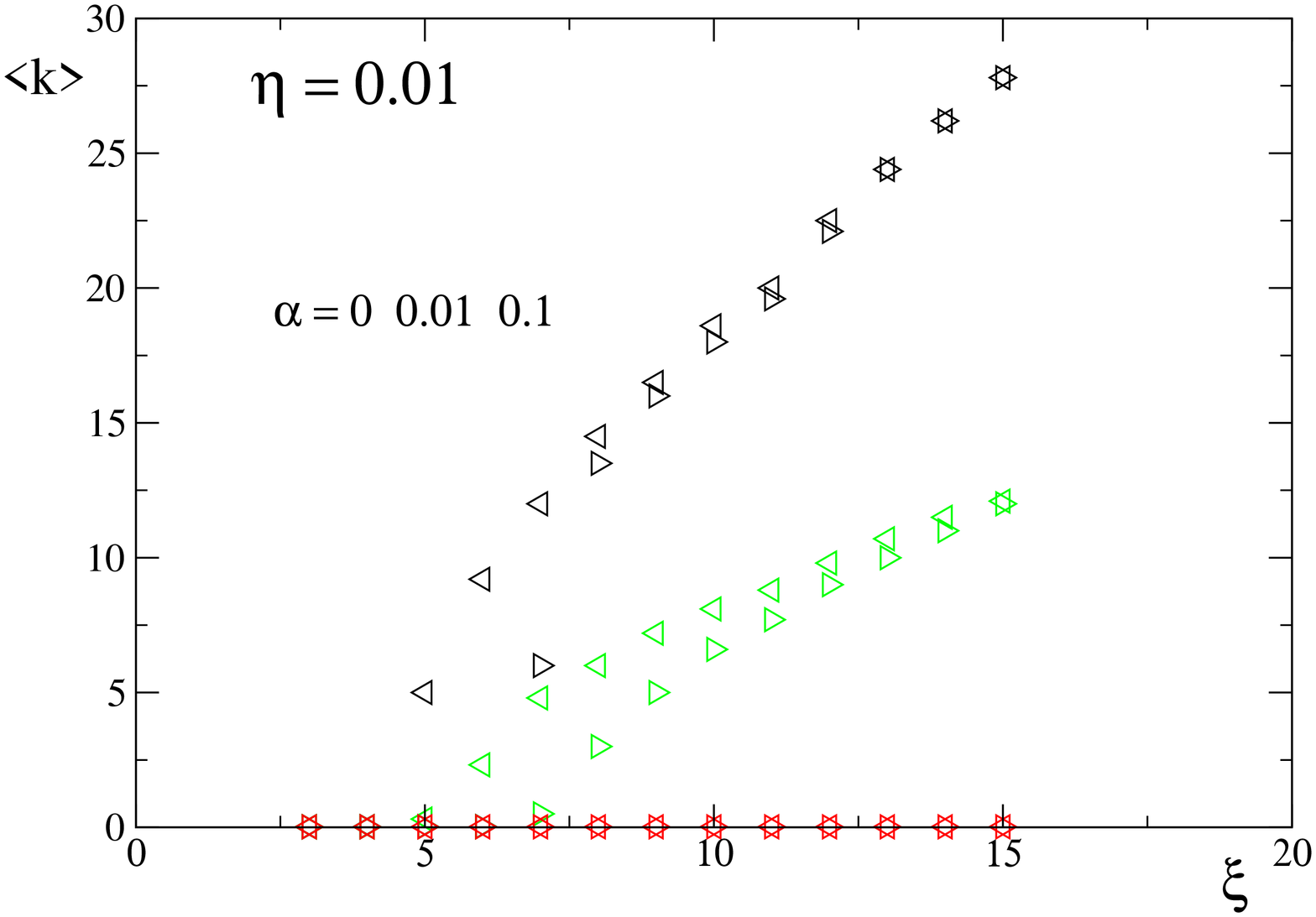}
\includegraphics*[scale=0.25,angle=270]{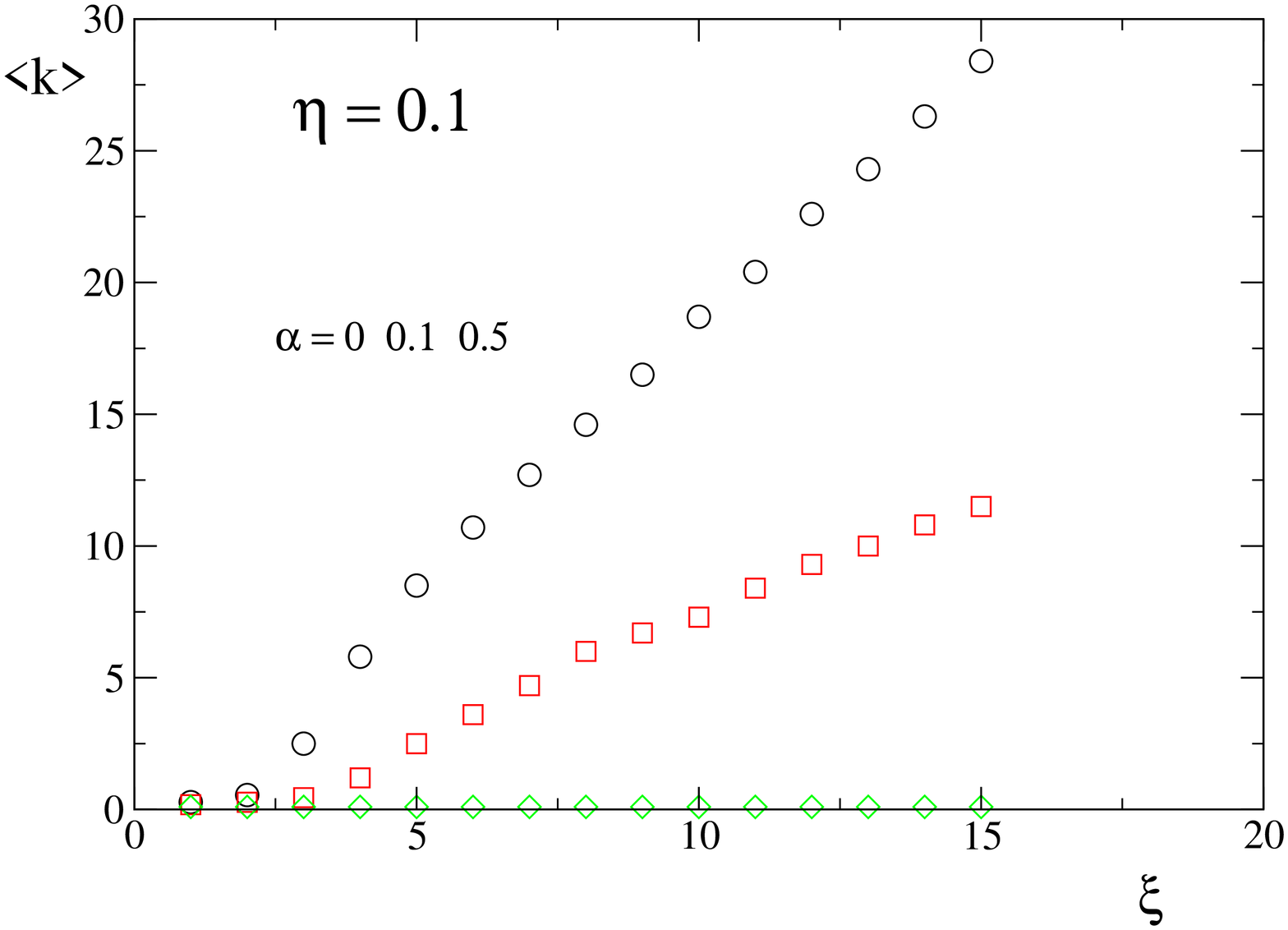}
\caption{Mean degree as a function of the rate $\xi$ of formation of
  links with neighbours of neighbours, for $N=1000$ ($\lambda=1$).
  Top: $\eta=0.01$, Bottom: $\eta=0.1$.}\label{vic}
\end{figure}
We believe the general findings discussed here will extend to the general class of models of Ref. \cite{14}. 
In particular, we expect the phase transition to be blurred by the effect of node volatility and to disappear when the latter exceed a particular threshold.

Actually, Fig. \ref{vic} shows that this is the case even for the model of Ref. \cite{13}. 
This is a model where link creation occurs either by long distance search at rate $\eta$ (as in the model discussed here) 
or through local search (on second neighbors) at rate $\xi$. Again links decay at unit rate. We refer the interested reader to Ref. \cite{13} 
for further details, for the present discussion let it suffice to say that the effects of (link) volatility are contrasted by the creation of 
a dense network with small-world features (a somewhat similar model with node volatility has been considered in Ref. \cite{12}). Fig. \ref{vic} 
shows that the effects of node volatility are very strong. Indeed, even a very small $\alpha$ reduces considerably the size 
of the coexistence region and the value $\alpha_c$ at which the latter disappears is also relatively small.

These results suggest that node volatility is indeed a relevant effect
in the co-evolution of socio-economic networks, as it may affect in
dramatic ways the ability of the system to reach a dense and/or
coordinated state.

\chapter{Conclusions}
In this thesis each chapter is mostly independent and self-contained.
Each of them is in fact referring to a specific phenomenology.
Is there something in common among phase transitions in physical materials, 
congestion  phenomena in informatic systems and the evolution of social networks?

I studied all of them using interacting dynamical models on heterogeneous 
graphs with the use of statistical mechanics techniques and concepts.
The wide applications of lattice models from statistical mechanics  
has its main reason in the fact that network based rappresentations
are widely used to describe many real complex systems.

This thesis is about how a certain degree of heterogeneity 
in the underlying topology can affect the collective statistical 
behavior of a system.
It is not intended as a review on it, rather 
I showed practically that this question
emerges spontaneously and gives insights in specific instances.

In the first chapter I showed how to give a 
statistical mechanics perspective to the problem of 
congestion in large communication networked system. 
This  comes out from a natural
extension of queuing network theory  
to large systems and to congested states. 
In order to do it, the use of statistical networks' ensembles and
the concept of congestion as a phase transition were really important.
In this chapter we have a specific example of the fact that the 
collective behavior of a system can depend 
crucially on the underlying structure,
e.g. traffic control is uneffective in homogeneous networks.

Usually it is believed that
the equilibrium distinctive features of statistical physics systems
should depend on the simmetries of the internal degrees of freedom
and only on the dimension of the spatial structure.
However the results on congestion phenomena
suggest that in out of equilibrium systems
the collective behavior could depend 
even on the details of the structure.
I showed in the second chapter through 
a simple model of kinetically constrained spins
subjected to a dynamical arrest,
how a simple dilution of the underlying graph can change 
completely the nature of this transition.
In the context of mean field models for 
the dynamical glass transition
we can have bootstrap or simple percolation transition 
on homogeneous or heterogeneous structures, respectively.

Given their general nature, the dependence of the collective properties 
of simple interacting models on the underlying  
graph is interesting per se, without referring to a specific phenomenology.
In the third chapter, I showed that the equilibrium features 
of tricritical model systems can change dramatically if the underlying graph has a certain degree of heterogeneity. 
The usual entropy ratio of the ordered and the disordered phase can invert on some graphs. 
There is an inverse phase transition, by which tricritical model systems becomes disordered upon cooling.
I showed a mechanism that trigger this phenomenon, based on the freezing of sparse subgraphs. 
If they are crucial for the connectivity, the overall graph of interactions can be disconnected by cooling.

Up to this point I discussed cases in which the behavior of a system
is affected by the underlying structure, in particular this last is fixed.
But where a  given structure of interactions comes from?
In  social networks the graph itself changes in time with an evolution 
that can be coupled with the dynamics defined on top of it.
In the last chapter we exploited a nice analogy with statistical physics in this context.
Models of coevolving social networks show
discontinuous transitions between sparse and dense structures, 
with hysteresis and coexistence phenomena.
I showed a mechanism that dump this transition, up to a critical point like 
in the Van der Waals picture of the liquid-gas transition.

Then, apart from the specific insights that can be obtained 
from applying statistical mechanics to such specific instances, 
is there something {\em more general} that we can get?

In the introduction I gave the examples of epidemic spreading processes and of the Ising model
as model systems that show a  phase transition whose behavior is 
affected in a non-trivial way by the heterogeneity of the underlying graph.
The critical point of these models scales with the system size in heterogeneous graphs in a 
way such that large enough systems are always in practice in the non-trivial phase.
In the SIS model the nodes of a network can be susceptible (S) or infected (I), respectively.
An infected node can infect its neighbors with rate $\nu$. Infected nodes recover from the infection with rate $1$.
These rules define a dynamical process of spreading that, for a given $\nu$ and initial conditions, can
end up in a state that can be characterized by  the fraction of infected nodes $\phi_I$.
If $\phi_I=0$ the spreading of the infection is stopped (this is an absorbing state). For large enough $\nu$
we have a final state $\phi_I>0$, there is a continuous transition whose transition point, 
within a mean field approximation on random graphs is\cite{epidemic}
\begin{equation}
\nu_c = \frac{\langle k\rangle}{\langle k^2 \rangle}.
\end{equation} 
The critical temperature on random graphs\cite{ising} of the Ising model is
\begin{equation}
\frac{1}{T_c} = \frac{1}{2} \log(\frac{\langle k^2 \rangle}{\langle k^2 \rangle-2\langle k \rangle}).
\end{equation}
These equations show that the collective behavior 
of such model systems can be  ruled by the tails of the degree distribution of the underlying graph. 
 
This is not the case in general.
Here, in the context of the queuing network theory and of tricritical spin models I showed an analogous parallel.
For these models, that show continuous and {\em discontinuous} transitions, there is a rather different behavior.
Different parts of the same system can be in different phases.
In fact the mean field analysis on heterogenenous graphs of both models is 
characterized by a cut-off degree such that only nodes whose degree is higher than it
are in the non-trivial phase, congested or magnetized, respectively.
In the range of parameters such that the transition is continuous, the behavior of the system is still ruled by high degree nodes. 
But, when the transition is discontinuous, their behavior is ruled by the central body of the degree distribution. 
In queuing network theory on random graphs, the critical inserction rate of packets per node $p_c$ scales according to:
\begin{equation}
p_c \simeq \left\{
 \begin{array}{rl}
 \frac{\mu \langle k \rangle/k_M}{\mu+(1-\mu) \langle k \rangle/k_M} & \textrm{continuous, low traffic control}.\\
 \mu k_m/\langle k \rangle & \textrm{discontinuous, high traffic control}.
 \end{array} \right.
\end{equation}
Where $k_m$ and $k_M$ are the minimum and the maximum of the degree distribution.
The  chemical potential at the transition point $\Delta_c$ of the Blume-Capel model on a random graph scales, 
within a mean field approximation, according to:
\begin{equation}
\Delta_c \simeq \left\{
 \begin{array}{rl}
 2  T \log (\beta \frac{\langle k^2 \rangle}{\langle k \rangle}-1) & \textrm{for the continuous branch}.\\
 \langle k \rangle & \textrm{around T=0}.
 \end{array} \right.
\end{equation}

This difference in turn can trigger highly non-trivial phenomena, like inverse phase transitions
in tricritical spin models or mixed phase transitions in the queuing networks.
It should be interesting to test the general validity of this picture.



\backmatter

\chapter*{Acknowledgements}
I would like to thank the SISSA and ICTP institutions for their stimulating environments.
I acknowledge G.Mussardo for the idea and concrete realization of this phd curriculum in statistical physics.
Among the many interesting people that I met during this period, I would like to thank
M.Sellitto, J.J.Arenzon, G.Bianconi and D.Helbing.
I thank M.Barthelemy that accepted to be the external referees of this thesis.
I thank all my collegues and friends in SISSA and ICTP, in particular Serena, Fabio, Raffo, Elena, Pier and Jacopo .
I would like to thank my boss Matteo that 
was always showing me the ``right spirit'' in doing this job.
A special thank to Luca Dall'Asta for his really valuable collaboration and patience all along this period. 
I thank A.Barato and L.Leuzzi for partial reviews of this manuscript.

Un abbraccio ed un saluto alla mia famiglia, in particolare a Ludovica,
un abbraccio ed un saluto a tutti gli amici e alla ``famiglia'' di Trieste ed infine, un bacio a Maya!

\end{document}